\documentclass[a4paper,11pt]{article}
\pdfoutput=1 % if your are submitting a pdflatex (i.e. if you have
\RequirePackage{snapshot}  % required by bundledoc

\usepackage[colorlinks,citecolor=blue,urlcolor=blue,linkcolor=blue]{hyperref}
\usepackage{bm,amsmath,amssymb,slashed,graphicx,%
            enumerate,alltt,xspace,multirow,xcolor,mathrsfs}
\usepackage{fancyvrb}
\usepackage{booktabs,tabularx}
\usepackage{graphicx}
\usepackage{subcaption}
\usepackage{xspace}
\usepackage[utf8]{inputenc} 
\usepackage[compat=1.0.0]{tikz-feynman}
\usepackage{scalerel}
\usepackage{layouts}
\usepackage{url}
\usepackage{lineno}

\RequirePackage[numbers,sort&compress]{natbib}
\makeatletter
\g@addto@macro\bfseries{\boldmath}
\makeatother

\definecolor{labelkey}{rgb}{0,0.5,0.0}

\usepackage{listings}
\definecolor{darkgreen}{rgb}{0,0.4,0}
\definecolor{grey}{rgb}{0.5,0.5,0.5}
\definecolor{semiblue}{rgb}{0.3,0.3,0.8}
\newcommand{\logbook}[2]{}
\newcommand{\order}[1]{\mathcal{O}\left(#1\right)}
\newcommand{\as}{\alpha_s}

\newcommand{\ptilde}{{\widetilde p}}

\newcommand{\qbar}{{\bar q}}

\newcommand{\itilde}{{\tilde \imath}}
\newcommand{\jtilde}{{\tilde \jmath}}

\tikzstyle{block} = [rectangle, minimum width=1.0cm, minimum height=0.75cm, thin, draw=black]
\tikzstyle{blob} = [circle, minimum width=0.5cm, thin, draw=black]
\tikzset{blackarrow/.style={-stealth, semithick, draw=black}}
\tikzset{connection/.style={inner sep=0,outer sep=0}}

\newcolumntype{C}{>{\centering\arraybackslash}X}
\begin{document}
\title{\textbf{Spin correlations in final-state parton showers and jet
  observables}}
\date{}
%\preprint{OUTP-21-09P}

%\newcommand{\OXaff}{Rudolf Peierls Centre for Theoretical Physics, Clarendon Laboratory, Parks Road,
%  University of Oxford, Oxford OX1 3PU, UK}
%\newcommand{\ASCaff}{All Souls College, Oxford OX1 4AL, UK}
\author{
Alexander Karlberg$^{1\,}$,
Gavin P.~Salam$^{1,2\,}$,
Ludovic Scyboz$^{1\,}$,
Rob Verheyen$^{3\,}$
\\[9mm]
{\small\it $^1$Rudolf Peierls Centre for Theoretical Physics, University of Oxford,}\\
{\small\it Clarendon Laboratory, Parks Road, Oxford OX1 3PU, United Kingdom} \\[3mm]
{\small\it $^2$All Souls College, Oxford OX1 4AL, United Kingdom}\\[3mm]
{\small\it $^3$ Department of Physics and Astronomy, University College London,} \\ %
{\small\it London, WC1E 6BT, United Kingdom}\\[3mm]
        }
\maketitle

%\author[a]{Alexander Karlberg,}%
%\author[a,b]{Gavin P.~Salam}%
% \author[a]{Ludovic Scyboz,}%
% \author[c]{Rob Verheyen}%
%
%
%\affiliation[a]{\OXaff}
%\affiliation[b]{\ASCaff}
%\affiliation[c]{Department of Physics and Astronomy, University College London, London, WC1E 6BT, UK}
%
%\date{Received: date / Accepted: \today}

\begin{abstract}
  \noindent
  As part of a programme to develop parton showers with controlled
  logarithmic accuracy, we consider the question of collinear spin
  correlations within the PanScales family of parton showers.
  We adapt the well-known Collins-Knowles spin-correlation algorithm
  to PanScales antenna and dipole showers, using
  an approach with similarities to that taken by Richardson and
  Webster.
  To study the impact of spin correlations, we develop
  Lund-declustering based observables that are sensitive to
  spin-correlation effects both within and between jets and extend the
  MicroJets collinear single-logarithmic resummation code to
  include spin correlations.
  Together with a 3-point energy correlation observable
  proposed recently by Chen, Moult and Zhu, this provides a powerful
  set of constraints for validating the logarithmic accuracy of our
  shower results.
  The new observables and their resummation further open the pathway
  to phenomenological studies of these important quantum mechanical
  effects.
\end{abstract}
\thispagestyle{empty}
\vfill

\newpage
%\keywords{QCD, Parton Shower, Resummation, LHC}

%\tableofcontents 

%======================================================================
\section{Introduction}
\label{sec:intro}

One of the most striking properties of Quantum Mechanics is that the
spin angular momentum of two or more particles can be created in an
entangled state~\cite{Einstein:1935rr}.
As a consequence, when measuring the spin of the individual particles,
or more generally the angular distributions of particle decays and
branchings, long-distance correlations will be found depending on the degree of
entanglement.
At colliders, spin correlations are most widely studied in the context
of heavy particle decays (see e.g.\ Refs.~\cite{Mahlon:1995zn,Bernreuther:2001rq,Bernreuther:2010ny,Mahlon:2010gw,Melnikov:2011ai,Artoisenet:2012st,ATLAS:2012ao,Aad:2013ksa,Chatrchyan:2013wua,Aad:2014pwa,Aad:2014mfk,Bernreuther:2015yna,CMS:2018jcg,Behring:2019iiv,Afik:2020onf,Fabbrichesi:2021npl}).
However they play a significant role also in the pattern of QCD
branchings that occur in jet fragmentation, as studied for example at
LEP~\cite{Barate:1997ha,Moretti:1998zc}.
The quantum mechanical nature of the problem is reflected in the need
to sum coherently over the spin states of intermediate particles in
the jet fragmentation, similarly to the need to sum coherently over
the spins of an electron-positron pair in an Einstein--Podolsky--Rosen
(EPR) experiment~\cite{Einstein:1935rr}.
It was recognised long ago~\cite{Webber:1986mc,Collins:1987cp,Knowles:1987cu} that
the core tools for simulating jet fragmentation, i.e.\ parton showers,
should incorporate such effects.

In this article we consider collinear spin correlations in the context
of the PanScales
programme~\cite{Dasgupta:2018nvj,Dasgupta:2020fwr,Hamilton:2020rcu} of
QCD parton-shower development.
The core aim of the programme is to develop parton showers with
well-understood logarithmic accuracy.
A first step on that path is to achieve so-called next-to-leading
logarithmic (NLL) accuracy.
There are many senses in which a shower can be NLL accurate and we
choose two broad criteria~\cite{Dasgupta:2020fwr}.
The logarithmic phase-space for QCD branching involves two
dimensions, corresponding to the logarithms of transverse momentum and
of angle, which can conveniently be represented using Lund
diagrams~\cite{Andersson:1988gp}.
To claim NLL accuracy, we firstly require a shower to reproduce the correct
matrix element for any configuration of emissions where all branchings
are well separated from each other in the Lund diagram.
Secondly, for all observables where suitable resummations exist (e.g.\
event shape distributions), the shower should reproduce the
resummation results up to and including terms $\as^n L^n$.
Here $L$ is the logarithm of the value of the observable and terms
$\as^n L^n$ are NLL in a context where $\as^n L^{n+1}$ terms
exponentiate~\cite{Catani:1992ua}.

Until now, the PanScales shower development has been based on
unpolarised splitting functions, as is common in dipole and antenna
showers.
According to our accuracy criterion, however, spin correlations are a
crucial part of NLL accuracy, because in configurations with
successive branchings at disparate angles, they are required in order
to reproduce the correct azimuthal dependence of the matrix elements.
The core purpose of this article, therefore, is to start the
implementation of spin correlations, specifically as concerns nested
collinear emissions.
The algorithm that we will adopt is based on the well-established
proposal by Collins~\cite{Collins:1987cp}, used notably in the Herwig
series of angular ordered
showers~\cite{Corcella:2000bw,Bahr:2008pv,Bellm:2015jjp,Bellm:2019zci}.
Our adaptation for the PanScales antenna and dipole showers
bears similarities with the implementation for the {Herwig7}
dipole shower by Richardson and Webster~\cite{Richardson:2018pvo,Webster:2019cwq} (for
work in other shower frameworks, see
Refs.~\cite{Nagy:2007ty,Nagy:2008eq,Forshaw:2019ver,Forshaw:2020wrq}).
One class of configuration that is not addressed by the Collins
algorithm is that of two or more commensurate-angle energy-ordered
soft emissions followed by a collinear splitting of one or more of
them.
Strictly these configurations should also be addressed for NLL
accuracy, however we defer their study to future work.

Whilst the techniques to implement parton-shower spin correlations are
relatively well established, our philosophy is that a parton shower
implementation is only complete if one has a set of observables,
resummations and associated techniques to validate the implementation.
The PanScales shower development in
Refs.~\cite{Dasgupta:2020fwr,Hamilton:2020rcu} has been able to draw on
a rich set of event shapes and associated resummed calculations.
However for testing spin correlations there was no such set of
observables.
Recently, while our work was in progress, Chen, Moult and
Zhu~\cite{Chen:2020adz} introduced and resummed a 3-point
energy-energy correlator that is spin-correlation sensitive. 
Here we introduce a new set of spin-correlation sensitive observables
based on Lund declustering.
We extend the {MicroJets} collinear resummation
code~\cite{Dasgupta:2014yra,Dasgupta:2016bnd} to allow for the
treatment of azimuthal structure, so as to obtain $\as^n L^n$
numerical predictions for all of these observables (in the process,
confirming the analytic results of Ref.~\cite{Chen:2020adz}).
These new observables, and our study of their properties, are of
potential interest also in their own right for practical measurements
of spin correlations in jets.

The paper is structured as follows. In section~\ref{sec:spinalg} we
give details of the algorithm used to implement spin correlations in
the PanScales showers.
We introduce an azimuthal observable based on the Lund plane picture
in section~\ref{sec:spin-Lund-observable}, and discuss general
features of spin correlations in the strictly collinear
limit.\footnote{Ref.~\cite{Dasgupta:2020fwr} also used azimuthal
  observables for testing showers, however for the case of emissions
  with commensurate $k_t$ that are widely separated in angle.
  That is a region dominated by independent soft emission and is free
  of spin correlations.
  The recoil effects in traditional dipole showers that cause that
  region to be incorrect at NLL accuracy can also introduce azimuthal
  correlations that obfuscate the genuine spin correlations that we
  discuss here, as is visible in Appendix~\ref{sec:collinear-cuts},
  though an understanding of the logarithmic structure associated with
  this effect would require further work.
  This issue was discussed also in Ref.~\cite{Richardson:2018pvo}. }
In
section~\ref{sec:tests} we validate our implementation and show that it achieves
NLL accuracy. In section~\ref{sec:concl} we conclude.

\section{The Collins-Knowles algorithm and its adaptation to dipole showers}
\label{sec:spinalg}

An efficient algorithm to include spin correlations in angular-ordered
MC generators was proposed by Collins~\cite{Collins:1987cp} for
final-state showers and subsequently extended by
Knowles~\cite{Knowles:1987cu,Knowles:1988vs,Knowles:1988hu} to include initial-state radiation and
backwards evolution.
It is based on the factorisation of the tree-level matrix element in
the collinear limit, and can be generalised to include spin
correlations in the matching of the parton shower to a hard matrix
element, between initial and final state radiation, as well as in
decays~\cite{Richardson:2001df}.

The Collins-Knowles algorithm can be readily applied to showers that first
generate a full branching tree with intermediate virtualities and
momentum fractions at each stage of the splitting, and then only
subsequently assign azimuthal angles and reconstruct the full event
kinematics.
However, in dipole showers the azimuthal angle needs to be chosen at
each stage of the branching, because it affects the phase space for
subsequent branchings.
This requires a reordering of the steps in the Collins-Knowles
algorithm.
Furthermore, in an angular-ordered shower, it is straightforward to
identify the mapping between the shower kinematics and the azimuthal
angle as needed in the Collins-Knowles algorithm.
In dipole (and antenna) showers the corresponding mapping can be less
straightforward.
In this section we therefore introduce a modified version of the
Collins-Knowles algorithm that is applicable to any parton shower,
including those of the dipole or antenna kind.
We first develop a collinear branching formalism in terms of boost
invariant spinor products and then show how to apply these in the
context of dipole and antenna showers.

Our work here is not the first to adapt the Collins-Knowles algorithm
to showers with dipole-like structures, see for example
Refs.~\cite{Richardson:2018pvo,Webster:2019cwq,Nagy:2007ty,Nagy:2008eq}.
Our approach is inspired by and bears a number of similarities with
that by Richardson and
Webster~\cite{Richardson:2018pvo,Webster:2019cwq}, as used in the
Herwig7~\cite{Bellm:2019zci} framework.
That algorithm continuously boosts between the lab frame and frames
specific to each individual collinear splitting, where individual
Collins-Knowles steps may be applied directly.
In our implementation of the algorithm no such boost is necessary, as
our expressions are formulated in terms of boost-invariant spinor products.

\subsection{Collinear branching amplitudes}
\label{sec:coll-branching}
To determine the appropriate azimuthal distribution of shower
branchings, the Collins-Knowles algorithm makes use of collinear branching amplitudes
$\mathcal{M}_{a \to b c}^{\lambda_a \lambda_b \lambda_c}$ for a
splitting $a \to bc$ with spin labels $\lambda = \pm 1$.
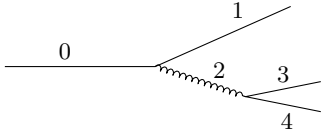
\begin{figure}
  \centering
  \begin{tikzpicture}
    \draw[black] (0.0,0.0) -- (2.0,0.0) -- (3.8, 0.8);
    \draw[black,decorate,decoration={coil,amplitude=1.0pt,segment length=2.5pt}] (2.0,0.0) -- (3.2,-0.4);
    \draw[black] (4.2,-0.2) -- (3.2,-0.4) -- (4.2,-0.6);

    \node at (0.8,0.2) {\footnotesize $0$};
    \node at (3.1,0.75) {\footnotesize $1$};
    \node at (2.85,-0.05) {\footnotesize $2$};
    \node at (3.7, -0.1) {\footnotesize $3$};
    \node at (3.75, -0.73) {\footnotesize $4$};
  \end{tikzpicture}
  \caption{Two subsequent collinear splittings $0 \to 1 2$, $2 \to 3 4$. }
  \label{fig:intro-example}
\end{figure}
As an example, consider the situation of Fig.~\ref{fig:intro-example}, where an unpolarised parton $0$
emits a gluon $2$, which subsequently splits into a $q'\bar{q}'$
pair, $2 \to 34$. In the collinear limit, the azimuthal distributions are determined by the factorised matrix element
\begin{equation} 
|M|^2 \propto \mathcal{M}_{0\to 12}^{\lambda_0
\lambda_1 \lambda_2} \mathcal{M}_{0 \to 12}^{* \lambda_0 \lambda_1 \lambda_2'} \mathcal{M}_{2 \to 3
4}^{\lambda_2 \lambda_3 \lambda_4} \mathcal{M}_{2 \to 3 4}^{* \lambda_2' \lambda_3
\lambda_4} \;, 
\label{eq:ME-factorization-example}
\end{equation}
where summation over repeated spin indices is implied.\footnote{Note
  the separate $\lambda_2$ and $\lambda_2'$ indices in the amplitude
  and its complex conjugate: this reflects the independent coherent
  sums over intermediate particle spins in the amplitude and its
  conjugate.}
\begin{table}
  \centering
  \begin{tabular}{  c | c | c | c | c | c  }
  \toprule
  $\lambda_a$  & $\lambda_b$  & $\lambda_c$  & $q \to qg$ & $g \to q\bar{q}$ & $g \to gg$    \\
  \midrule
  $\lambda$ &      $\lambda$        &      $\lambda$       & $\frac{1}{\sqrt{1-z}}$ & $0$ & $\frac{1}{\sqrt{z(1-z)}}$ \\
  $\lambda$ &      $\lambda$        &      $-\lambda$      & $\frac{z}{\sqrt{1-z}}$ & $-z$ & $\frac{z^{3/2}}{\sqrt{1-z}}$ \\ 
  $\lambda$ &      $-\lambda$       &      $\lambda$       & $0$ & $1-z$ & $\frac{(1-z)^{3/2}}{\sqrt{z}}$ \\
  $\lambda$ &      $-\lambda$       &      $-\lambda$      &           $0$               &$0$                  &                 $0$    \\
  \end{tabular}
  \caption{The helicity-dependent Altarelli-Parisi splitting
    amplitudes $\mathcal{F}_{a \to bc}^{\lambda_a \lambda_b
      \lambda_c}(z)$, where $z = E_b/E_a$.} 
\label{tab:hel-dep-fcns}
\end{table}
Ultimately we will test our approach in the PanScales shower framework, which
currently only supports massless particles in its kinematic maps.
Accordingly, we work in the massless quark limit, though the extension
to massive quarks is straightforward, as discussed in
Ref.~\cite{Richardson:2018pvo}.
In the massless quark limit, the branching amplitudes may be written as 
\begin{equation}
\mathcal{M}_{a \to bc}^{\lambda_a \lambda_b \lambda_c} = \frac{1}{\sqrt{2}} \frac{g_s}{p_b{\cdot}p_c} \mathcal{F}_{a \to bc}^{\lambda_a \lambda_b \lambda_c}(z) S_{\tau}(p_b,p_c)\;,
\label{eq:branch-amp-spinor-prod}
\end{equation}
where the functions $\mathcal{F}_{a \to bc}^{\lambda_a \lambda_b
  \lambda_c}(z)$, listed in Table~\ref{tab:hel-dep-fcns}, are the
(colour-stripped) helicity-dependent Altarelli-Parisi splitting
amplitudes that depend on the collinear momentum fraction $z$ carried
by parton $b$.  They are related to the usual unregularised splitting functions
through\footnote{In order to get the full unregularised splitting functions an appropriate factor of $C_A$, $C_F$ or $T_R$ has to be included.}
\begin{equation}
  P_{a \to bc}(z) \propto \frac{1}{2}\sum_{\lambda_a, \lambda_b, \lambda_c}\left[\mathcal{F}^{\lambda_a \lambda_b \lambda_c}_{a \to bc}(z)\right]^2.
  \end{equation}
The function $S_{\tau}(p_b,p_c)$ is a spinor product, where the label $\tau = \pm1$ indicates the 
sign of the complex phase associated with the spinor product. 
It is given by
\begin{equation}
\tau = \tilde{\lambda}_b + \tilde{\lambda}_c - \tilde{\lambda}_a \mbox{ where } \tilde{\lambda} = 
\begin{cases}
\lambda/2 \mbox{ for a quark}, \\ 
\lambda \mbox{ for a gluon}.
\end{cases}
\end{equation} 
The derivation of these branching amplitudes can be found in
Appendix~\ref{sec:app:spinor-product-der}, together with our
convention for spinor products.

Inserting an explicit expression for the spinor product,
Eq.~\eqref{eq:branch-amp-spinor-prod} may alternatively be written as  
\begin{equation}
\mathcal{M}_{a \to bc}^{\lambda_a \lambda_b \lambda_c} = \tau \frac{g_s}{\sqrt{p_b{\cdot}p_c}} \mathcal{F}_{a \to bc}^{\lambda_a \lambda_b \lambda_c}(z) e^{i \tau \phi}\;,
\label{eq:branch-amp-phase}
\end{equation}
where $\phi$ is the azimuthal angle as defined in a reference frame
where the parent is along a specific (e.g.\ $z$) direction.
Eq.~\eqref{eq:branch-amp-phase} is useful in the context of a parton
shower only when this azimuthal angle is used to parameterise its
phase space directly.
This can be the case when the azimuthal variable $\phi$ for the
shower's phase space generation is defined with respect to a fixed
azimuthal reference direction.
However, in dipole or antenna showers, $\phi$ usually
represents the azimuthal direction of the transverse momentum with
respect to the plane of the dipole or antenna parent partons.
As the PanScales showers are of the dipole/antenna type, we will use
Eq.~\eqref{eq:branch-amp-spinor-prod} directly, and evaluate the
spinor products numerically using
Eq.~\eqref{eq:spinor-product-explicit-form}.
This choice guarantees that the implementation of the algorithm
remains independent of the type of parton shower it is applied to.
However one could also choose to track the relation between the shower
azimuths and the collinear azimuths, as done in
Refs.~\cite{Richardson:2018pvo,Webster:2019cwq}.

\subsection{The algorithm}
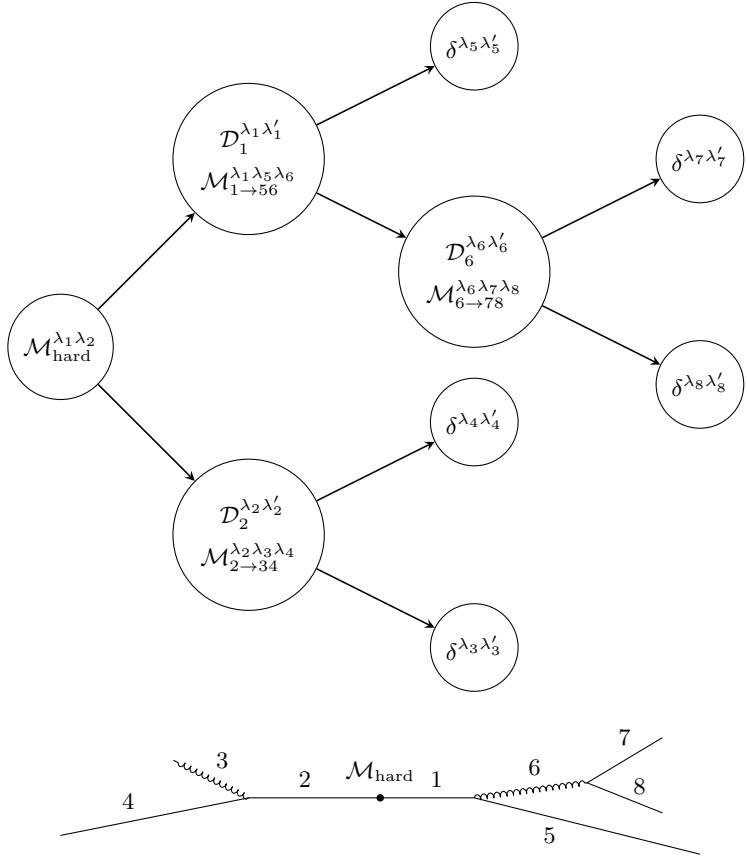
\begin{figure}
  \centering
  \begin{tikzpicture}
    \node[blob] at (0.0,0.0)(ME) {\footnotesize ${\mathcal{M}_{\mathrm{hard}}^{\lambda_1 \lambda_2}}$};
    \node[blob] at (2.5,2.5)(one) {\footnotesize $\begin{aligned} \mathcal{D}_{1}^{\lambda_1 \lambda'_1} \,\,\, \\ \mathcal{M}_{1\rightarrow 5 6}^{\lambda_1 \lambda_5 \lambda_6}\end{aligned}$};
    \node[blob] at (2.5,-2.5)(two) {\footnotesize $\begin{aligned} \mathcal{D}_{2}^{\lambda_2 \lambda'_2} \,\,\, \\ \mathcal{M}_{2\rightarrow 3 4}^{\lambda_2 \lambda_3 \lambda_4}\end{aligned}$};
    \node[blob] at (5.5, -4)(three) {\footnotesize $\delta^{\lambda_3 \lambda'_3}$};
    \node[blob] at (5.5, -1)(four) {\footnotesize $\delta^{\lambda_4 \lambda'_4}$};
    \node[blob] at (5.5, 4)(five) {\footnotesize $\delta^{\lambda_5 \lambda'_5}$};
    \node[blob] at (5.5, 1)(six) {\footnotesize $\begin{aligned}\mathcal{D}_{6}^{\lambda_6 \lambda'_6} \,\,\, \\ \mathcal{M}_{6\rightarrow 7 8}^{\lambda_6 \lambda_7 \lambda_8}\end{aligned}$};
    \node[blob] at (8.5, 2.5)(seven) {\footnotesize $\delta^{\lambda_7 \lambda'_7}$};
    \node[blob] at (8.5, -0.5)(eight) {\footnotesize $\delta^{\lambda_8 \lambda'_8}$};

    \draw[blackarrow] (ME) to (one);
    \draw[blackarrow] (ME) to (two);
    \draw[blackarrow] (two) to (three);
    \draw[blackarrow] (two) to (four);
    \draw[blackarrow] (one) to (five);
    \draw[blackarrow] (one) to (six);
    \draw[blackarrow] (six) to (seven);
    \draw[blackarrow] (six) to (eight);

    \node at (4.25,-5.65) {\footnotesize $\mathcal{M}_{\mathrm{hard}}$};
    \filldraw[black] (4.25,-6) circle (1.2pt);
    \draw[black] (0,-6.5) -- (2.5,-6) -- (5.5, -6) -- (8.5,-6.75);
    \draw[black,decorate,decoration={coil,amplitude=1.0pt,segment length=2.5pt}] (2.5,-6) -- (1.5,-5.5);
    \draw[black,decorate,decoration={coil,amplitude=1.0pt,segment length=2.5pt}] (5.5,-6) -- (7,-5.8);
    \draw[black] (8, -5.2) -- (7,-5.8) -- (8,-6.2);

    \node at (5.0,-5.75) {\footnotesize $1$};
    \node at (3.25,-5.75) {\footnotesize $2$};
    \node at (2.15,-5.5) {\footnotesize $3$};
    \node at (0.9, -6.05) {\footnotesize $4$};
    \node at (6.5,-6.5) {\footnotesize $5$};
    \node at (6.3, -5.6) {\footnotesize $6$};
    \node at (7.5, -5.2) {\footnotesize $7$};
    \node at (7.7, -5.85) {\footnotesize $8$};
  \end{tikzpicture}
  \caption{An example of a Collins-Knowles binary tree, and a corresponding shower history.
  The root node corresponds with the hard scattering matrix element. 
  Nodes that correspond to partons that have already split store a decay matrix and a branching amplitude.
  Nodes that have not split only store a Kronecker delta decay matrix.}
  \label{fig:binary-tree}
\end{figure}

The purpose of the Collins-Knowles algorithm is to distribute the azimuthal degrees 
of freedom according to Eq.~\eqref{eq:ME-factorization-example} for an
arbitrary number of collinear branchings, while maintaining
linear complexity in the number of particles.
To that end, a binary tree is tracked, where the nodes correspond to collinear shower branchings.
Figure \ref{fig:binary-tree} shows an example of a shower history and the corresponding binary tree.

The algorithm is initialised from the hard scattering amplitude $\mathcal{M}^{\lambda_1 \lambda_2}_{\mathrm{hard}}$,
where the outgoing hard partons $1$, $2$ have spin indices $\lambda_1$, $\lambda_2$. 
The formalism is not limited to a particular number of final-state partons, but for readability we restrict ourselves to a two-body hard scattering.
Each particle $i$ (whether final or intermediate) is associated with a
so-called decay matrix, $\mathcal{D}_i^{\lambda_i\lambda_i'}$ with two
spin indices.
At the start of the shower the decay matrices are initialised to
$\mathcal{D}_1^{\lambda_1\lambda_1'} = \delta^{\lambda_1 \lambda_1'}$
and
$\mathcal{D}_2^{\lambda_2\lambda_2'} = \delta^{\lambda_2 \lambda_2'}$.

The core of the algorithm consists of the rules for generating an
azimuthal angle and adding nodes to and updating the binary tree, as well as a
subset of the decay matrices, each time the shower adds an emission to
the event.
The shower first selects a branching dipole or antenna and generates
an ordering scale and momentum-sharing variable%
\footnote{The exact definitions of which depend on the shower
  implementation at hand.} according to the regular spin-summed shower
dynamics, in which the azimuthal dependence does not
appear.\footnote{%
  This may no longer be the case when accounting for subleading colour
  effects, for example in the NODS method of
  Ref.~\cite{Hamilton:2020rcu}.
  That specific algorithm may reject an emission after its azimuth has
  been chosen.
  When combining it with our spin-correlation approach, we first
  choose the azimuth according to the spin-correlation algorithm, and
  then apply the colour rejection step.
}
The branching parton will correspond to one of the terminal nodes in
Fig.~\ref{fig:binary-tree}, as discussed in further detail below, and
we refer to the node as $a_n$, where the $n$ index labels the depth of
the node within the tree.
Then, to determine the azimuthal angle, the algorithm proceeds with the following rejection-sampling procedure:
\begin{enumerate}
\item 
Compute the spin-density matrix
$\rho_{a_n}^{\lambda_{a_n}\lambda'_{a_n}}$ as follows:
\begin{enumerate}
  \item Starting from parton $a_n$, trace the binary tree back to the root parton node $a_0 \in [1,2]$.
  This results in a sequence $a_0$...$a_n$ of parton indices, along with a sequence of complementing parton indices $b_0$...$b_n$ such 
  that $b_0$ is the other root parton and $a_i$ and $b_i$ have common parent node $a_{i - 1}$.
  \item Compute the spin-density matrix for the root node, 
  \begin{equation}
    \rho_{a_0}^{\lambda_{a_0} \lambda'_{a_0}} = \frac{1}{\textrm{tr}( \cdot)} \mathcal{D}_{b_0}^{\lambda_{b_0} \lambda'_{b_0}} 
      \mathcal{M}_{\mathrm{hard}}^{\lambda_{a_0}\lambda_{b_0}} \mathcal{M}_{\mathrm{hard}}^{*\lambda'_{a_0}\lambda'_{b_0}},
  \end{equation}
  where the denominator is the trace of the numerator.
  \item Iteratively for $i \in \{1,..,n\}$, compute 
  \begin{equation}
    \rho_{a_i}^{\lambda_{a_i} \lambda'_{a_i}} = \frac{1}{\textrm{tr}( \cdot)} \rho_{a_{i - 1}}^{\lambda_{a_{i - 1}} \lambda'_{a_{i - 1}}} 
    \mathcal{M}_{a_{i - 1}\rightarrow a_i b_i}^{\lambda_{a_{i - 1}} \lambda_{a_i} \lambda_{b_i}}
    \mathcal{M}_{a_{i - 1}\rightarrow a_i b_i}^{*\lambda'_{a_{i - 1}} \lambda'_{a_i} \lambda'_{b_i}}
    \mathcal{D}_{b_i}^{\lambda_{b_i} \lambda'_{b_i}}.
  \end{equation}
\end{enumerate}
\item Repeat the following until a value of $\phi$ is accepted:
\begin{enumerate}
  \item Sample a value of $\phi$ uniformly and, using the other shower variables, construct the post-branching momenta $p_{a_{n+1}}$ and $p_{b_{n+1}}$ using the usual kinematic mapping of the parton shower.
  \item Compute the branching amplitude $\mathcal{M}_{a_n \rightarrow a_{n+1} b_{n+1}}^{\lambda_{a_n} \lambda_{a_{n+1}} \lambda_{b_{n+1}}}$ using Eq.~\eqref{eq:branch-amp-spinor-prod}.
  \item Compute the acceptance probability 
  \begin{equation}
    p_{\mathrm{accept}} = \frac{1}{N} \rho_{a_{n}}^{\lambda_{a_n} \lambda'_{a_n}} 
    \mathcal{M}_{a_n \rightarrow a_{n+1} b_{n+1}}^{\lambda_{a_n} \lambda_{a_{n+1}} \lambda_{b_{n+1}}}
    \mathcal{M}_{a_n \rightarrow a_{n+1} b_{n+1}}^{*\lambda'_{a_n} \lambda_{a_{n+1}} \lambda_{b_{n+1}}},
  \label{eq:accept-prob}
  \end{equation}
  where $N$ is a $\phi$-independent normalisation factor to ensure $p_{\mathrm{accept}} < 1$.%
  \footnote{One can make use of the fact that the spin-density matrix is hermitian and has trace $1$ to find, for instance
  \begin{equation} 
    N = \mathcal{M}_{a_n \rightarrow a_{n+1} b_{n+1}}^{\alpha \, \lambda_{a_{n+1}} \lambda_{b_{n+1}}} 
        \mathcal{M}_{a_n \rightarrow a_{n+1} b_{n+1}}^{*\alpha \, \lambda_{a_{n+1}} \lambda_{b_{n+1}}} + 
        2 \Big|\rho_{a_{n}}^{\alpha \, (\scalebox{0.75}{-} \alpha)}\Big|
        \Big| \mathcal{M}_{a_n \rightarrow a_{n+1} b_{n+1}}^{\alpha \, \lambda_{a_{n+1}} \lambda_{b_{n+1}}} 
        \mathcal{M}_{a_n \rightarrow a_{n+1} b_{n+1}}^{*(\scalebox{0.75}{-}\alpha) \, \lambda_{a_{n+1}} \lambda_{b_{n+1}}} \Big|,
  \end{equation}
  where $\alpha = 1$ or $-1$.}
  \item Accept $\phi$ with probability $p_{\mathrm{accept}}$.
\end{enumerate}
\item Update the binary tree
\begin{enumerate}
  \item Insert new nodes $a_{n+1}$ and $b_{n+1}$ with parent node $a_n$ and initialise their decay matrices to a Kronecker delta.
  \item Store $\mathcal{M}_{a_n \rightarrow a_{n+1} b_{n+1}}^{\lambda_{a_n} \lambda_{a_{n+1}} \lambda_{b_{n+1}}}$ in node $a_n$.
  \item Iteratively for $i \in \{n,..,0\}$, recompute and store the updated decay matrices
  \begin{equation}
    \mathcal{D}_{a_i}^{\lambda_{a_{i}} \lambda'_{a_{i}}} = \frac{1}{\textrm{tr}( \cdot)} \mathcal{D}_{a_{i+1}}^{\lambda_{a_{i+1}} \lambda'_{a_{i+1}}} \mathcal{D}_{b_{i+1}}^{\lambda_{b_{i+1}} \lambda'_{b_{i+1}}}
    \mathcal{M}_{a_n \rightarrow a_{n+1} b_{n+1}}^{\lambda_{a_i} \lambda_{a_{i+1}} \lambda_{b_{i+1}}}
    \mathcal{M}_{a_n \rightarrow a_{n+1} b_{n+1}}^{*\lambda'_{a_i} \lambda'_{a_{i+1}} \lambda'_{b_{i+1}}}.
  \end{equation}
\end{enumerate}
\end{enumerate}

Note that the identification of the parton that branches is not
without subtleties.
We have assumed that it is always one of the terminal nodes in
Fig.~\ref{fig:binary-tree}.
To understand why this is a non-trivial choice, suppose that we have a
dipole stretching between particles $3$ and $7$.
When the left-hand end of the dipole emits a gluon (which we label
$9$), our spin-correlation algorithm always views this as a $g\to gg$
splitting of particle $3$.
However if the emitted gluon is at large angle relative to the $34$
splitting, i.e.\ $\theta_{93} \gg \theta_{34}$ the gluon effectively
sees the coherent charge of $3$ and $4$ and could more properly be
viewed as being emitted from $2$ (which unlike $3$ is a quark). 
Because of the interplay between the shower ordering variable and
emission kinematics, this occurs only for situations in which $9$ is
soft relative to particle $3$, and also soft relative to any of the
parents of $3$.
Inspecting Table~\ref{tab:hel-dep-fcns}, one sees that soft gluon
emission (the $z\to1$ limit) leads to splitting amplitudes that are
independent of the flavour of the parent, $a$, and that are non-zero only for
$\lambda_a = \lambda_b$, i.e.\ they are diagonal in the spin space
relating the parent and its harder offspring.
This means that in the limit where  emission $9$ could conceivably
have been emitted from $2$, it is immaterial whether we actually view
it as being emitted from $2$ or instead organise the tree as if it had
been emitted by $3$.
The latter is considerably simpler and so it is the solution that we
adopt.

%======================================================================
\section{Collinear spin correlations: expectations and measurement strategy}
\label{sec:spin-Lund-observable}

In this section, we start (section~\ref{sec:azimuthal-structure}) by
examining how the spin correlations translate into azimuthal
correlations between the planes of separate collinear branchings, both
within a single jet and across pairs of jets.
We do so at fixed order, $\order{\as^2}$, where it is trivial to define the
observables.
We then propose (section~\ref{sec:lund-azimuth-variables}) a set of
observables that are suitable for use at all orders.
They exploit a Lund diagram~\cite{Andersson:1988gp} representation of
individual jets~\cite{Dreyer:2018nbf}.
Next (section \ref{sec:EEEC-definition}), we recall the definition of
the EEEC spin-sensitive observable, which was proposed and resummed in
Ref.~\cite{Chen:2020adz}.
Finally (section~\ref{sec:lund-all-order}), we use these observables
to study the impact on the azimuthal correlations coming from the
all-order resummation of collinear spin-correlation effects.

%----------------------------------------------------------------------
\subsection{Azimuthal structure}
\label{sec:azimuthal-structure}

\begin{figure}
  \centering
  \begin{subfigure}{0.65\textwidth}
    \includegraphics[width=1.\textwidth]{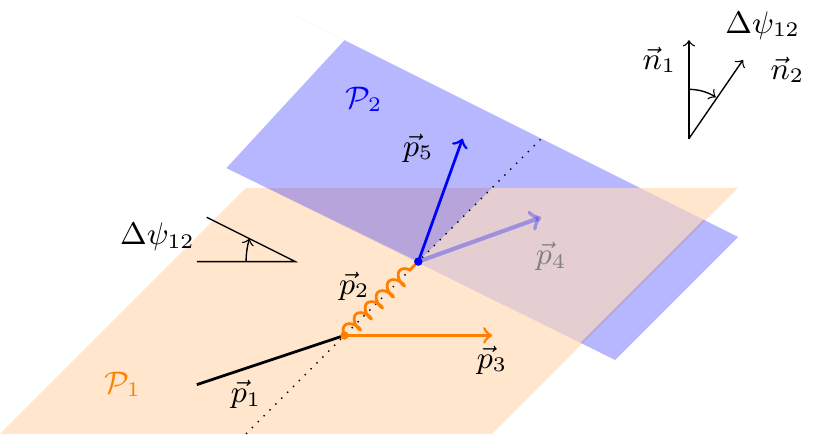}
  \end{subfigure}
  \caption{Azimuthal angles are defined between successive splitting
    planes for the $1\to23$ splitting,
    $\mathcal{P}_1 \supset \lbrace \vec{p}_{2}, \vec{p}_{3} \rbrace$
    and the $2\to45$ splitting,
    $\mathcal{P}_{2} \supset \lbrace \vec{p}_{4}, \vec{p}_{5}
    \rbrace$.
    The figure also depicts the vectors normal to the two planes,
    $\vec n_1$ and $\vec n_2$. 
  }
  \label{fig:dpsi12-Lund-definition}
\end{figure}

Each collinear branching in an event can be associated with the plane
that contains the momenta of the two offspring partons.
The simplest observable one may think of to study spin correlations is
the azimuthal difference, $\Delta\psi$, between the planes defined by
two distinct branchings.
Here we will consider two broad cases: \emph{intra-jet} correlations,
i.e.\ between the planes of two branchings within a single jet, for
example between the plane of the $1\to56$ splitting and the plane of
the $6\to 78$ splitting in Fig.~\ref{fig:binary-tree};
and \emph{inter-jet} correlation, i.e.\ between the planes of two
splittings in separate jets, for example between the plane of the
$1\to56$ splitting and the plane of the $2\to 34$ splitting in
Fig.~\ref{fig:binary-tree}.\footnote{That particular case, with a
  $q\bar q$ hard process, would have zero correlation, but the
  correlation is non-zero for a $gg$ hard process.}
We will refer to the two azimuthal differences as $\Delta \psi_{12}$
and $\Delta \psi_{11'}$ where the $1$ and $2$ labels refer to the
first and second splitting within a given jet and the $1'$ label
refers to the first splitting in a distinct jet.
The $\Delta \psi_{12}$ and $\Delta \psi_{11'}$ observables are
straightforward to define at $\order{\as^2}$ relative to the hard
scattering and it is this situation that we will concentrate on here.
In the $\Delta \psi_{12}$ case, the splitting planes and the azimuthal
angle between them are illustrated in Fig.~\ref{fig:dpsi12-Lund-definition}.

At second order in the coupling, and in the collinear limit, the cross
sections differential in the intra- and inter-jet observables
$\Delta\psi_{12}$, respectively $\Delta \psi_{11'}$, take the simple
form (see e.g.~section 2.3 in Ref~\cite{Richardson:2018pvo}, or the example calculated in Appendix~\ref{sec:app:derivation-A-B})
\begin{align}
  \frac{d\sigma}{d\Delta \psi_{ij}} \propto a_0 \left( 1 + \frac{a_2}{a_0}
\cos(2\Delta \psi_{ij}) \right) = a_0 \left( 1 + A_{i}(z_i) B_{j}(z_{j})
\cos(2\Delta \psi_{ij}) \right)\,,
\label{eq:dsig-dpsi12}
\end{align}
where the coefficients $a_0$ and $a_2$ depend on the observable, the
final state under consideration, and the momentum fractions associated
with the first ($z_i$) and second splitting ($z_j$).
If the splittings are restricted to have opening angles greater than
$e^{-|L|}$, with $|L| \gg 1$, and $\ln z_1$ and $\ln z_2$ are both of
order $1$, the $a_0$ and $a_2$ coefficients are
dominated by terms $\as^2 L^2$, i.e.\ they belong to the
single-logarithmic set of terms that we aim to control for NLL
accuracy.
For large $L$, at this order, the ratio $a_2/a_0$ is independent of
$\as L$, and so Eq.~(\ref{eq:dsig-dpsi12}) can always be written in
terms of
the functions $A(z)$ and $B(z)$ given in
Table~\ref{tab:A-B-coefficients}.

\begin{table}
  \centering
  \begin{tabularx}{.75\textwidth}{>{\hsize=0.2\hsize}C
                             >{\hsize=0.5\hsize}C
                             >{\hsize=0.3\hsize}C}
    \toprule
    \multicolumn{3}{c}{$\Delta \psi_{12}$} \\
    \midrule
    \multicolumn{3}{c}{Primary splitting} \\
    \midrule
     &&\\[-2em]
   $q \to q g$        & $A(z)= \frac{2z}{1+z^2}$ & \includegraphics[width=70pt]{{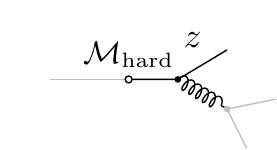}} \\[-0.8em]%\hline
    $g \to g g$        & $A(z)= \frac{z^2}{(1-z(1-z))^2}$ & \includegraphics[width=70pt]{{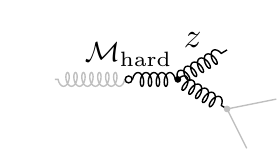}} \\[-0.4em]%\hline
    \midrule
    \multicolumn{3}{c}{Secondary splitting} \\
    \midrule
    &&\\[-2em]
    $g \to q'\bar q'$  & $B(z)= \frac{-2z(1-z)}{1-2z(1-z)}$ & \includegraphics[width=70pt]{{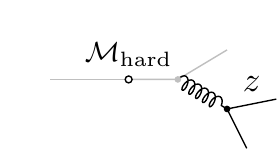}} \\[-0.8em]%\hline
    $g \to gg$  & $B(z)= \frac{z^2(1-z)^2}{(1-z(1-z))^2}$ & \includegraphics[width=70pt]{{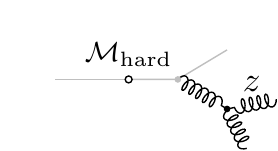}} \\[-0.4em]
%    \hline
    %    \hline
    \bottomrule
    \multicolumn{3}{c}{$\Delta \psi_{11'}$} \\
    \midrule
    &&\\[-2em]
    $g \to q'\bar q'$  & $A(z) = B(z)= \frac{-2z(1-z)}{1-2z(1-z)}$ & \includegraphics[width=70pt]{{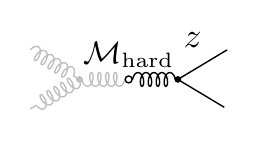}} \\[-0.8em]%\hline
    $g \to gg$  & $A(z) = B(z)= \frac{z^2(1-z)^2}{(1-z(1-z))^2}$ & \includegraphics[width=70pt]{{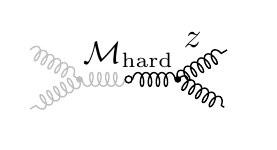}} \\[-0.8em]
    \bottomrule
  \end{tabularx}
  \caption{The functions $A(z)$ and $B(z)$ entering Eq.~(\ref{eq:dsig-dpsi12})
for the intra- and inter-jet observables $\Delta\psi_{12}$,
$\Delta\psi_{11'}$, for all channels at second order. See
e.g.~Appendix~\ref{sec:app:derivation-A-B} for an exemplified derivation. Here,
for a $1 \to 23$ splitting, the variable $z$ is the momentum fraction carried away by parton
$2$.
Representative diagrams at $\order{\as^2}$ are shown on the right.
In those
diagrams, the splitting in black is the one associated with the corresponding
function $A(z)$ or $B(z)$. The partons shown in grey serve as an example of what
the rest of the branching history can look like, but they do not matter in
choosing the function itself, since the contributions factorise in the final
result, Eq.~(\ref{eq:dsig-dpsi12}).  
Configurations that do not involve an
intermediate gluon have vanishing spin correlations.
}
\label{tab:A-B-coefficients}
\end{table}

\begin{figure}
  \centering
  \begin{subfigure}{0.48\textwidth}
    \centering
    \includegraphics[width=1.\textwidth,page=1]{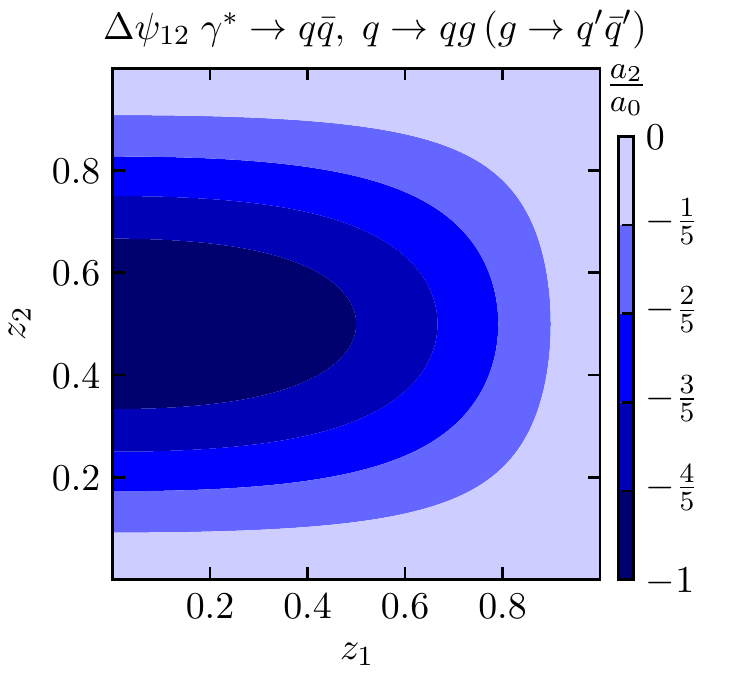}
  \end{subfigure}
  \begin{subfigure}{0.48\textwidth}
    \centering
    \includegraphics[width=1.\textwidth,page=2]{figures/z1z2-fixed-order-plots/fixed-order-z1z2-dependence.pdf}
  \end{subfigure}
  \caption{Size of the spin correlations, at fixed order
$\mathcal{O}(\alpha_s^2)$, for a quark-initiated jet. We consider the azimuthal
difference $\Delta \psi_{12}$ between the splitting planes of a primary (with
gluon momentum fraction $z_1$) and secondary ($z_2$) branching, separated by channel.
The colour scale indicates the relative size $a_2/a_0$ of the correlations,
where $\Delta \psi_{12}$ is distributed proportionally to $a_0 + a_2 \cdot {\rm
  cos}(2\Delta \psi_{12})$.
Note the use of $z_1$ as the gluon momentum fraction in the $q \to qg$
splitting, while for amplitude expressions in the text, $z$ often
refers to the quark momentum fraction.
}
\label{fig:fixedorder-z1z2-quarkjet-sameside-dpsi12}
\end{figure}
In Fig.~\ref{fig:fixedorder-z1z2-quarkjet-sameside-dpsi12} we show a
contour plot of the ratio $a_2/a_0 = A(z_1)B(z_2)$, for our intra-jet observable,
as a function of $z_1$ ($x$-axis)
and $z_2$ ($y$-axis) for a quark-initiated jet.
In this case there are only two non-trivial final states to consider,
namely $q\rightarrow qg(g\rightarrow q' \bar{q}')$ and
$q\rightarrow qg(g\rightarrow gg)$, while channels such as
$q\to q(q\to qg) g$ that do not involve an intermediate gluon have
vanishing spin correlations.
In the case of $q\rightarrow qg(g\rightarrow q'
\bar{q}')$, we see that the coefficient $a_2$ is negative,
corresponding to an enhancement when the $q\to qg$ and
$g \to q'\bar q'$ planes are perpendicular.
The ratio $a_2/a_0$ peaks at $-100\%$ around $z_1=0$ and $z_2=0.5$,
i.e.\ when the gluon is soft and the quark-antiquark pair share the
energy equally.\footnote{$z_1=0$ does not satisfy our requirement of
  finite $\ln z_1$, but it is indicative of the behaviour that we will
  see if we take moderately small values of $z_1$ while keeping finite
  $\ln z_1$. Similarly for the rest of the discussion.}
The spin correlations stay large even for moderate
values of $z_1$ and $z_2$ although they vanish completely for $z_1 \to
1$ (soft quark emerging from the $q \to qg$ splitting) and for $z_2
\to 0$ or $z_2 \to 1$ (either of the secondary $q'$ or $\bar q'$ becomes
soft). 
Similarly, for
$q\rightarrow qg(g\rightarrow gg)$, the ratio peaks at $z_1=0$ and
$z_2=0.5$ but the correlation has the opposite sign to the $g\to
q\bar q$ case, which implies an enhancement when the $q\to qg$ and 
$g\to gg$ splittings are in the same plane. 
The magnitude of $a_2/a_0$ is substantially smaller, with a peak at
$1/9$.
The spin correlations fall off more sharply than in the
$g\rightarrow q' \bar{q}'$ case, in particular as a function of $z_2$,
and also vanish for $z_1 \to 1$ and for $z_2 \to 0$ or
$z_2 \to 1$.

\begin{figure}
  \centering
  \begin{subfigure}{0.48\textwidth}
    \centering
    \includegraphics[width=1.\textwidth,page=3]{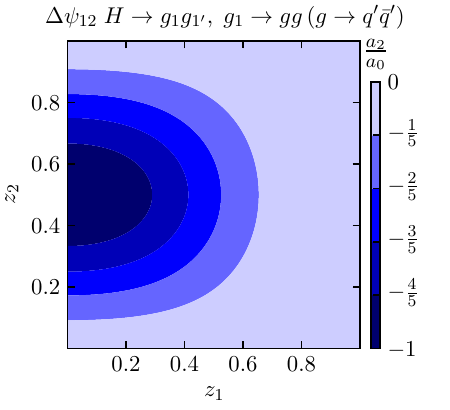}
  \end{subfigure}
  \begin{subfigure}{0.48\textwidth}
    \centering
    \includegraphics[width=1.\textwidth,page=4]{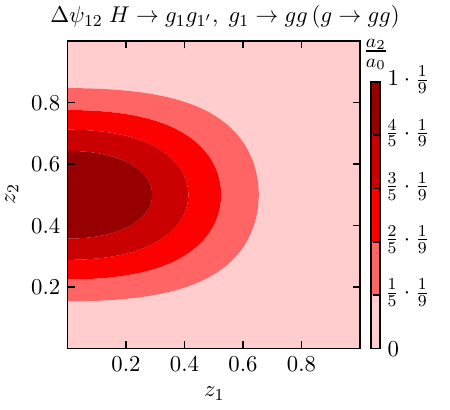}
  \end{subfigure}
  \caption{Same as Fig.~\ref{fig:fixedorder-z1z2-quarkjet-sameside-dpsi12}, for
a gluon-initiated jet. Note that the colour scale is spanning the same range as
in Fig.~\ref{fig:fixedorder-z1z2-quarkjet-sameside-dpsi12}.}
\label{fig:fixedorder-z1z2-gluonjet-sameside-dpsi12}
\end{figure}

The overall picture is very similar for a gluon-initiated jet, shown
in Fig.~\ref{fig:fixedorder-z1z2-gluonjet-sameside-dpsi12}, although
the dependence on $z_1$ is moderately stronger for both $g\to q'\bar{q} '$ and $g\to gg$ secondary splittings, which can be
understood from Table~\ref{tab:A-B-coefficients}.

Turning now to the correlations between azimuthal angles in two
distinct jets, they are zero for a $\gamma^* \to q\bar q$, but non-zero
for $H \to g_1g_{1'}$, so we consider only the latter.
In
this case we examine the azimuthal difference $\Delta\psi_{11'}$
between the splitting planes of the two primary emissions. There are
three possible final states given by $\{g_1 \rightarrow q\bar{q},
g_{1'}\rightarrow q\bar{q}\}$, $\{g_1 \rightarrow q\bar{q},
g_{1'}\rightarrow gg\}$, and $\{g_1 \rightarrow gg,
g_{1'}\rightarrow gg\}$. 
\begin{figure}
  \centering
  \begin{subfigure}{0.48\textwidth}
    \centering
    \includegraphics[width=1.\textwidth,page=5]{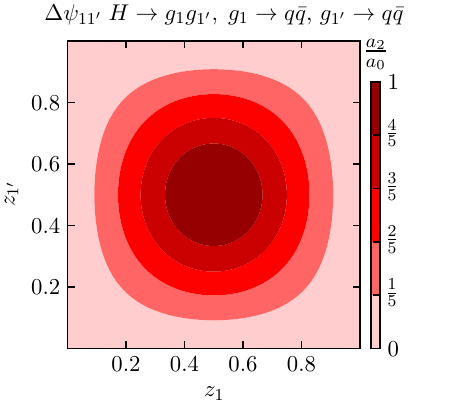}
  \end{subfigure}
  \begin{subfigure}{0.48\textwidth}
    \centering
    \includegraphics[width=1.\textwidth,page=6]{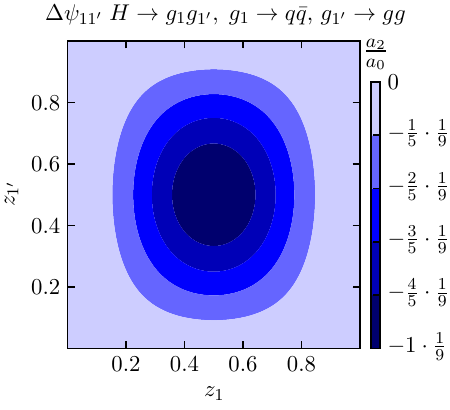}
  \end{subfigure}
  \begin{subfigure}{0.48\textwidth}
    \centering
    \includegraphics[width=1.\textwidth,page=7]{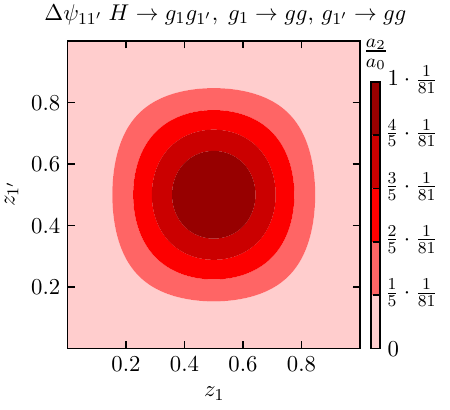}
\end{subfigure}
  \caption{Same as Figs.~\ref{fig:fixedorder-z1z2-quarkjet-sameside-dpsi12}
and~\ref{fig:fixedorder-z1z2-gluonjet-sameside-dpsi12}, for a $H \to gg$ event,
where we consider the azimuthal difference $\Delta \psi_{11'}$ between the
splitting planes of the two primary branchings on each side of the event. Note
the different scale of the colour bar for the inter-jet observable: here, the
spin correlations are largest for the quark-only final-state (top left) which
peak at $+100\%$, mid-range for the mixed quark-gluon final state
(top right) peaking at $-11.1\%$, and much smaller for the all-gluon
final state (bottom), with a peak at $+1.23\%$.}
\label{fig:fixedorder-z1z2-h2gg-oppositeside-dpsi11}
\end{figure}
In Fig.~\ref{fig:fixedorder-z1z2-h2gg-oppositeside-dpsi11} we show
the ratio $a_2/a_0$ as a function of the energy fractions $z_1$ and
$z_{1'}$. In all three cases the ratio is peaked when
$z_1=z_{1'}=0.5$ and is largest in magnitude when both gluons
split into quark-antiquark pairs. When only one gluon splits into a
quark-antiquark pair the ratio becomes negative and
$-11.1\%$ around the peak. When both gluons split into
gluons the spin correlations almost vanish. In all three cases the
ratio vanishes when either of the energy fractions approach $0$ or
$1$.

%----------------------------------------------------------------------
\subsection{Definition of spin-sensitive Lund observables}
\label{sec:lund-azimuth-variables}

The observables discussed so far start at $\mathcal{O}(\alpha_s^2)$
relative to the hard scattering, but care has to be taken in order to
define them in an infrared safe way beyond that order.
To facilitate a definition which can also be directly applied in
experimental analyses, we make use of the procedure from
Ref.~\cite{Dreyer:2018nbf} to build Lund
diagrams~\cite{Andersson:1988gp} from individual jets.\footnote{
  In Ref.~\cite{Dreyer:2018nbf} an
  azimuthal variable $\psi$ was already defined. Azimuthal differences
  computed using this $\psi$ coincide with the definition that we give
  here in the collinear limit, but turn out to have an undesirable
  dependence on the rotation of the initial hard event.}
We remind the reader that the Lund jet plane is constructed through a
declustering of a C/A-jet~\cite{Dokshitzer:1997in,Wobisch:1998wt}.
The primary Lund plane is constructed by following the harder branch
at each step of the declustering.
If at some point one instead follows a softer branch and then all its subsequent
harder branches, those subsequent branches populate a secondary Lund plane.

The procedure that we describe here can be applied at hadron colliders
and $e^+e^-$ colliders (or also in the decay of heavy particles at
hadron colliders, for example in their rest frame).
The first branching within a jet is identified by taking all
declusterings on the primary Lund plane that satisfy $z$ larger than
some $z_\text{cut}$, and selecting the one with largest relative
$k_t$.\footnote{In the $pp$ case, $z$, $k_t$ and the azimuth $\phi$ for a
  branching were defined in Ref.~\cite{Dreyer:2018nbf}; in our $e^+e^-$
  studies here, for an $i\to jk$ branching, we use $z=\min(|\vec
  p_j|,|\vec p_k|)/(|\vec p_j| + |\vec p_k|)$, $k_t =
  \min(|\vec p_j|,|\vec p_k|)\sin\theta_{jk}$ and we discuss the
  definition of the azimuths below.}
The second branching within a jet is identified by taking all
declusterings on the secondary Lund plane associated with the first
branching, and then again identifying those that satisfy $z$ larger
than some $z_\text{cut}$ and selecting the one with largest relative
$k_t$.
One may choose to use different $z_\text{cut}$ values for the first
and second branchings within a single jet.\footnote{Instead of
  selecting the highest-$k_t$ emission that satisfies
  $z > z_\text{cut}$, one could instead take the first that
  appears in the declustering.
  That would have made the procedure more similar to the modified
  mass-drop tagger~\cite{Dasgupta:2013ihk} or SoftDrop with
  $\beta=0$~\cite{Larkoski:2014wba}.
  At the logarithmic accuracy that we discuss here, both procedures
  should yield equivalent results.
  Selecting the highest-$k_t$ emission without a $z_\text{cut}$ would
  be similar to Dynamical Grooming~\cite{Mehtar-Tani:2019rrk} and would
  involve a double rather than single logarithmic resummation
  structure~\cite{Caucal:2021bae}.
  The use of two declusterings is also part of a range of top taggers,
  e.g.\ Refs.~\cite{Kaplan:2008ie,CMS:2009lxa,CMS:2014fya,Plehn:2009rk,Plehn:2010st}, 
  theoretical aspects of which are further discussed in Ref.~\cite{Dasgupta:2018emf}.
}
Given a choice of $z_\text{cut}$, one may then study the results
differentially in the two $z$ values, so as to obtain plots similar to
Figs.~\ref{fig:fixedorder-z1z2-quarkjet-sameside-dpsi12}--\ref{fig:fixedorder-z1z2-h2gg-oppositeside-dpsi11}.
Here, however, we will consider just results integrated over $z$.
Note that in section~\ref{sec:azimuthal-structure}, we investigated
the structure for $0<z<1$, while in the Lund construction we always
have $z \le 1/2$, because $z$ is the momentum carried by the softer
offspring (and the secondary Lund plane is the one associated with
further emission from the softer offspring).
So, for example, in section~\ref{sec:azimuthal-structure} we could
meaningfully discuss a $q \to qg$ splitting where the quark was soft
($z_1$ close to $1$), followed by a splitting of the gluon.
In contrast, in the Lund declustering picture, the quark in such a
case will be assigned to the secondary plane, and the second splitting
that we study will necessarily be a further $q\to qg$ splitting. 
The reason for adopting this procedure in Lund declustering is that
experimentally one cannot observe the flavour of the underlying
parton.\footnote{We will still show results below classified according
  to the flavour channel.
  This is useful in terms of diagnostics, and it is possible because
  the parton shower does contain information about flavours.
  Strictly speaking, that flavour information for massless quarks is
  infrared unsafe~\cite{Banfi:2006hf} within the Cambridge/Aachen
  algorithm.
  The infrared-safe flavour algorithm of Ref.~\cite{Banfi:2006hf}
  cannot be applied in conjunction with Lund declustering, because it
  adapts the $k_t$ jet algorithm rather than the Cambridge/Aachen
  algorithm.
  However in the limit that we will consider for our logarithmic
  tests, $\as \to 0$ with $\as L$ fixed, if the logarithm of the infrared cutoff is of
  the order of $-|L|$, the infrared unsafe contributions obtained
  with the Cambridge/Aachen algorithm vanish, because they scale as
  $\as^2 L$.
 }

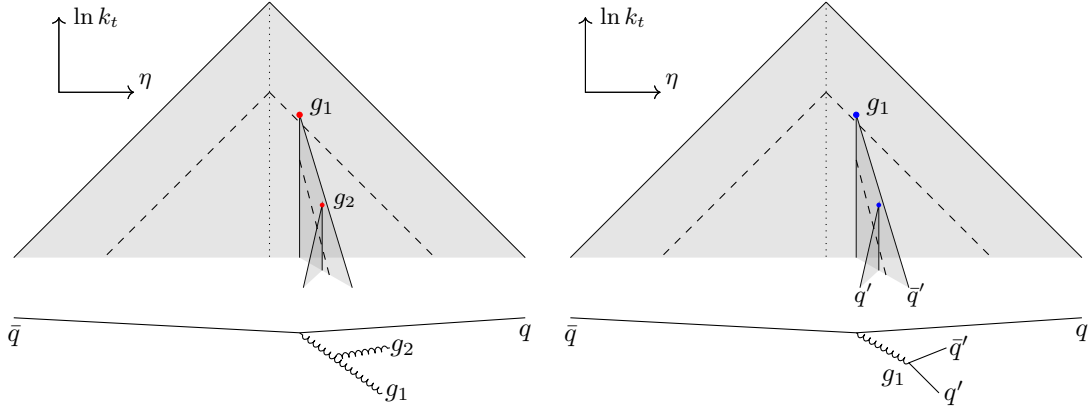
\begin{figure}
\centering
\begin{tikzpicture}
  % Lund plane
  \draw[black]  ( 0,0) -- ( 3.4,-3.4);
  \draw[black]  ( 0,0) -- (-3.4,-3.4);
  \fill [gray,opacity=0.2] (0,0) -- (-3.4,-3.4) -- (3.4,-3.4);
  \draw[black,dotted]  ( 0,0) -- (0,-3.4);
  \draw[black, <->]  (-2.8, -0.2) -- (-2.8,-1.2) -- (-1.8,-1.2);
  \draw[black, <->]  (-2.8, -0.2) -- (-2.8,-1.2) -- (-1.8,-1.2);
  \node at (-2.3,-0.2) {\footnotesize $\ln k_t$};
  \node at (-1.65,-1.05) {\footnotesize $\eta$};
  %
  % Cuts primary
  \draw[thin,black,dashed]  ( 0,-1.2) -- (2.2,-3.4);
  \draw[thin,black,dashed]  ( 0,-1.2) -- (-2.2,-3.4);
  %\fill [gray,opacity=0.3] (0,-1.3) -- (-1.7,-3.0) -- (1.7,-3.0);
  %
  % 1st emission and 2ndary plane
  \node at (0.7,-1.4) {\small  $g_1$};
  \draw [very thin,black]  (0.4,-3.4) -- (0.4,-1.5) -- (1.1,-3.8);
  \fill [gray,opacity=0.2] (0.4,-3.4) -- (0.4,-1.5) -- (1.1,-3.8);
  \filldraw [red] (0.4,-1.5) circle (1pt);
  % g->gg splitting
  \draw [very thin,black]  (0.7,-3.57) -- (0.7,-2.7) -- (0.45,-3.8);
  \fill [gray,opacity=0.2] (0.7,-3.57) -- (0.7,-2.7) -- (0.45,-3.8);
  \filldraw [red] (0.7,-2.7) circle (0.7pt);
  \node at (1.0,-2.65) {\footnotesize  $g_2$};
  % Cuts secondary
  \draw[thin,black,dashed]  ( 0.4,-2.1) -- (0.8,-3.65);
  %
  % Event
  \draw[black] (-3.4,-4.2) -- (0.4,-4.4) -- (3.4,-4.2);
  \draw[black,decorate,decoration={coil,amplitude=1.0pt,segment length=2.5pt}] (1.5,-5.2) -- (0.4,-4.4);
  \draw[black,decorate,decoration={coil,amplitude=1.0pt,segment length=2.5pt}] (0.883,-4.72) -- (1.6,-4.6);
  \node at (-3.4,-4.45) {\small $\bar q$};
  \node at (3.4,-4.4) {\small $q$};
  \node at (1.77,-4.6) {\small $g_2$};
  \node at (1.7,-5.2) {\small $g_1$};

  % Lund plane
  \draw[black]  ( 7.4,0) -- ( 10.8,-3.4);
  \draw[black]  ( 7.4,0) -- (4.0,-3.4);
  \fill [gray,opacity=0.2] (7.4,0) -- (4.,-3.4) -- (10.8,-3.4);
  \draw[black,dotted]  ( 7.4,0) -- (7.4,-3.4);
  \draw[black, <->]  (4.2, -0.2) -- (4.2,-1.2) -- (5.2,-1.2);
  \draw[black, <->]  (4.2, -0.2) -- (4.2,-1.2) -- (5.2,-1.2);
  \node at (4.7,-0.2) {\footnotesize $\ln k_t$};
  \node at (5.35,-1.05) {\footnotesize $\eta$};
  %
  % Cuts primary
  \draw[thin,black,dashed]  ( 7.4,-1.2) -- (9.6,-3.4);
  \draw[thin,black,dashed]  ( 7.4,-1.2) -- (5.2,-3.4);
  %
  % 1st emission and 2ndary plane
  \node at (8.1,-1.4) {\small  $g_1$};
  \draw [very thin,black]  (7.8,-3.4) -- (7.8,-1.5) -- (8.5,-3.8);
  \fill [gray,opacity=0.2] (7.8,-3.4) -- (7.8,-1.5) -- (8.5,-3.8);
  \filldraw [blue] (7.8,-1.5) circle (1pt);
  %
  % g->qq splitting
  \draw [very thin,black]  (8.1,-3.57) -- (8.1,-2.7) -- (7.85,-3.8);
  \fill [gray,opacity=0.2] (8.1,-3.57) -- (8.1,-2.7) -- (7.85,-3.8);
  \filldraw [blue] (8.1,-2.7) circle (0.7pt);
  \node at (7.9,-3.9)  {\footnotesize  $q'$};
  \node at (8.6,-3.9) {\footnotesize  $\bar q'$};
  %
  % Cuts secondary
  \draw[thin,black,dashed]  ( 7.8,-2.1) -- (8.2,-3.65);
  %
  % Event
  \draw[black] (4,-4.2) -- (7.8,-4.4) -- (10.8,-4.2);
  \draw[black,decorate,decoration={coil,amplitude=1.0pt,segment length=2.5pt}] (8.5,-4.8) -- (7.8,-4.4);
  \draw[black] (8.9,-5.2) -- (8.5,-4.8) -- (9,-4.6);
  \node at (4,-4.45) {\small $\bar q$};
  \node at (10.8,-4.4) {\small $q$};
  \node at (8.3,-5.0) {\small $g_1$};
  \node at (9.17,-4.6) {\small $\bar q'$};
  \node at (9.1,-5.2) {\small $q'$};
\end{tikzpicture}
\caption{For the intra-jet definition of the Lund observable $\Delta
\psi_{12}$, we consider the primary $q \to q g_1$ and secondary splittings
$g_1 \to g_1 g_2$ (left) and $g_1 \to q' \bar q'$ (right). We apply a cut (dashed lines),
$z_{\rm cut}$, on the momentum fraction used to identify the primary and
secondary splittings.  The red (blue) dots indicate that the azimuthal angles in
those splittings are correlated, with the preferred value of $\Delta \psi_{12}$
being in-plane (out-of-plane).}
\label{fig:lund-plane-z2qq}
\end{figure}

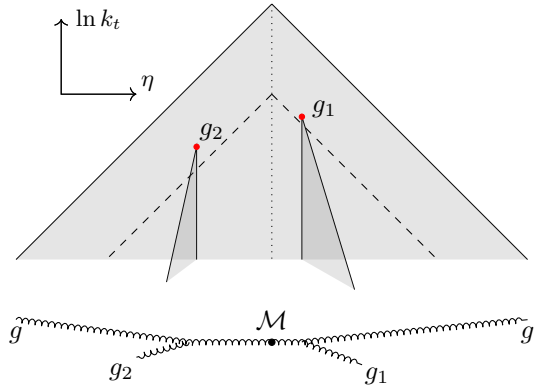
\begin{figure}
\centering
\begin{tikzpicture}
  % Lund plane
  \draw[black]  ( 0,0) -- ( 3.4,-3.4);
  \draw[black]  ( 0,0) -- (-3.4,-3.4);
  \fill [gray,opacity=0.2] (0,0) -- (-3.4,-3.4) -- (3.4,-3.4);
  \draw[black,dotted]  ( 0,0) -- (0,-3.4);
  \draw[black, <->]  (-2.8, -0.2) -- (-2.8,-1.2) -- (-1.8,-1.2);
  \draw[black, <->]  (-2.8, -0.2) -- (-2.8,-1.2) -- (-1.8,-1.2);
  \node at (-2.3,-0.2) {\footnotesize $\ln k_t$};
  \node at (-1.65,-1.05) {\footnotesize $\eta$};
  %
  % Cuts primary
  \draw[thin,black,dashed]  ( 0,-1.2) -- (2.2,-3.4);
  \draw[thin,black,dashed]  ( 0,-1.2) -- (-2.2,-3.4);
  %
  % 1st emission and 2ndary plane
  \node at (0.7,-1.4) {\small  $g_1$};
  \draw [very thin,black]  (0.4,-3.4) -- (0.4,-1.5) -- (1.1,-3.8);
  \fill [gray,opacity=0.2] (0.4,-3.4) -- (0.4,-1.5) -- (1.1,-3.8);
  \filldraw [red] (0.4,-1.5) circle (1pt);
  %
  % 2nd emission and 2ndary plane
  \node at (-0.8,-1.7) {\small  $g_2$};
  \draw [very thin, black] (-1.0,-3.4) -- (-1.0,-1.9) -- (-1.4,-3.7);
  \fill [gray,opacity=0.2] (-1.0,-3.4) -- (-1.0,-1.9) -- (-1.4,-3.7);
  \filldraw [red] (-1.0,-1.9) circle (1pt);
  %
  % Event
  \draw[black,decorate,decoration={coil,amplitude=1.0pt,segment length=2.5pt}] (-3.4,-4.2) -- (-1.1,-4.5) -- (0.4,-4.5) -- (3.4,-4.2);
  \draw[black,decorate,decoration={coil,amplitude=1.0pt,segment length=2.5pt}] (-1.1,-4.5) -- (-1.8,-4.7);
  \draw[black,decorate,decoration={coil,amplitude=1.0pt,segment length=2.5pt}] (0.4,-4.5) -- (1.2,-4.8);
  \filldraw[black] (0,-4.5) circle (1.2pt);
  \node at (0.,-4.2) {\small $\mathcal{M}$};
  \node at (-3.4,-4.4) {\small $g$};
  \node at (3.4,-4.4) {\small $g$};
  \node at (-2.,-4.9) {\small $g_2$};
  \node at (1.4,-4.95) {\small $g_1$};
\end{tikzpicture}
\caption{For the inter-jet definition of the Lund observable $\Delta
  \psi_{11'}$, to be used in $H \to g g$ events, we consider two
  primary splittings in different hemispheres.}
\label{fig:lund-plane-h2gg}
\end{figure}

To track the azimuthal angles in the $e^+e^-$ case within the Lund
declustering, we adopt a procedure that differs from that introduced
in Ref.~\cite{Dasgupta:2020fwr} (supplemental material), because we
have found that the latter results in differences in azimuthal angles that are not
invariant under rotations of the event. 
We start with an initial azimuthal angle $\psi_0 = 0$ (the particular
choice has no impact on the differences of azimuthal angles that we
eventually study).
Then, working through the declusterings in the Cambridge/Aachen
sequence, for each declustering $i$ we:
\begin{enumerate}
\item Compute the normalised cross product,
  $\hat{n}_i=\vec{p}_{i,a}\times \vec{p}_{i,b}/|\vec{p}_{i,a}\times \vec{p}_{i,b}|$, between the two pseudojets,
  $\vec{p}_{i,a}$ and $\vec{p}_{i,b}$ that result from the declustering, where
  $\vec{p}_{i,a}$ is the harder of the two.
\item Compute the signed angle $\Delta \psi_{(i-1,i)}$ between
  $\hat{n}_{i-1}$ and $\hat{n}_{i}$ where $\Delta \psi_{(i-1,i)}$ is
  positive if
  $\left(\hat{n}_{i-1}\times \hat{n}_{i}\right)\cdot \vec{p}_{i,a}>0$ and
  negative otherwise, and ${\hat n}_{i-1}$ is the normalised cross
  product obtained for the splitting that produced the parent parton.
\item Compute $\psi_i = \psi_{i-1} + \Delta \psi_{(i-1,i)}$.
\end{enumerate}
The variable $\Delta\psi_{12}$ is now defined as the difference
between the $\psi$ obtained for the primary splitting and secondary
splittings as selected above (which may not follow in immediate
sequence in the C/A declustering).
For the case of two successive splittings, this is equivalent to the
$\Delta\psi_{12}$ illustrated in
Fig.~\ref{fig:dpsi12-Lund-definition}. 
Likewise we define $\Delta\psi_{11'}$ as the $\psi$ difference between
the highest-$k_t$ primary passing the $z>z_\text{cut}$ requirement
inside each of the two different jets.
The Lund diagrams are illustrated at second order in
Figs.~\ref{fig:lund-plane-z2qq} and \ref{fig:lund-plane-h2gg}.

In practice, in our $e^+e^-$ implementation of the $\Delta\psi_{12}$
observable, we cluster the event back to two jets and analyse each jet
independently.
Our final distributions will be normalised to the number of events,
rather than the number of jets. 

%......................................................................
\subsection{Recall of 3-point energy-correlator observable} 
\label{sec:EEEC-definition}

Recently Chen, Moult and Zhu proposed 3-point energy
correlators (EEEC) as suitable observables for measuring the quantum
interference effects associated with spin
correlations~\cite{Chen:2020adz}.
For concreteness, we use the following definition for the EEEC,
\begin{equation}
\label{eq:EEEC-definition}
\frac1{\sigma_\text{tot}} \frac{d^3\Sigma}{d\Delta\psi d\theta_S d\theta_L}
  =
  \left\langle \sum_{i,j,k=1}^{N} \frac{8E_i E_j E_k}{Q^3}
    \delta\left(\Delta\psi - \phi_{(ij)k})\right)\delta\left(\theta_S - \theta_{ij}\right)\delta\left(\theta_L - \theta_{jk}\right)    
  \right\rangle
\end{equation}
where $\sigma_\text{tot}$ is the total cross section, $Q$ is
the event centre-of-mass energy, $\theta_{mn}$ is the opening angle
between two emissions $m$ and $n$ and $\phi_{(ij)k}$ is the angle
between the plane that contains the $p_{i}+p_{j}$ and $p_k$ directions
and the plane that contains the $p_i$ and $p_j$ directions.
The average is carried out across events, and in each event the sum over each
of the $i$, $j$ and $k$ runs over all particles in the event,
$1\ldots N$.
One may also apply the definition to particles in a single jet of
energy $E_\text{jet}$, in which case one would replace $8/Q^3$ with
$1/E_\text{jet}^3$.
Refs.~\cite{Chen:2019bpb,Chen:2020adz} provided techniques for
resumming such observables, and below we will compare our
resummation results for the EEEC to theirs. 

%......................................................................
\subsection{MicroJets resummation of spin correlations and comparison
  to fixed order}
\label{sec:lund-all-order}

Although the above fixed-order analysis gives a sense of the overall
structure of spin correlations in quark- and gluon-initiated jets,
there are important effects which can only be captured by all-order
resummation.  

To our knowledge, no analytical result exists for the logarithmic
structure of observables sensitive 
to spin correlations, except for the recently computed all-order result~\cite{Chen:2020adz}
for the 3-point energy correlator, reproduced in section \ref{sec:EEEC-definition}.  
In order to enable comparisons to other observables we have implemented a
numerical resummation, henceforth called the \emph{toy shower}, based
on the MicroJets code~\cite{Dasgupta:2014yra,Dasgupta:2016bnd,microjets}.

The toy shower is ordered in an angular-type evolution variable $t$,
\begin{equation}
t(\theta, p_t) = \int_{\theta}^1 \frac{d{\theta'}}{\theta'}\frac{\alpha_s(E \theta')}{\pi}\;,
\end{equation}
where $E$ is the energy of the hard parton initiating the shower,
and $\theta$ is an angular scale. For a 1-loop running of the strong coupling
$\alpha_s(p_t \theta)$, the scale $t$ is related to the opening angle of the
splitting $\theta$ by
\begin{equation}
\label{eq:t-from-theta-running}
t = \frac{1}{\beta_0} \ln
\left( \frac{1}{1+\frac{\alpha_s}{\pi}\beta_0 \ln \theta} \right) =
\frac{1}{\beta_0} \ln \left( \frac{1}{1+\frac{\lambda}{\pi}\beta_0} \right) \,,
\end{equation}
where $\beta_0=\frac{1}{6}(11C_A-4 T_R n_f)$, and where we have
introduced $\lambda = \alpha_s \ln \theta$.  Single-logarithmic terms
of the form $(\alpha_s {\ln}(1/\theta))^n$ can then be resummed by
solving corresponding DGLAP-style equations. Unlike in a real parton
shower, there is no kinematic map associated with the emissions, and
crucially it thus does not generate any spurious higher-order
terms.
Since the toy shower is angular ordered, the Collins-Knowles
algorithm can readily be applied to it, and we can make predictions
for angular observables correct at NLL in the strongly ordered
limit. The toy shower can also provide fixed order predictions at
$\mathcal{O}(\alpha_s^2)$ in the collinear limit.

\subsubsection{Results for Lund declustering observables}

We start by evaluating the impact of the single-logarithmic resummation on the
size of the spin correlations, $a_2/a_0$, for the observables $\Delta
\psi_{12}$ and $\Delta \psi_{11'}$ introduced above. In order to
do so, we consider the distributions of $\Delta \psi_{12}$ and $\Delta
\psi_{11'}$ generated by the toy shower for a value of $t_{\rm max} =
\frac{0.5}{\pi}$, which corresponds to $\lambda \approx -0.3743$ for a 1-loop
running of the strong coupling, see
Eq.~(\ref{eq:t-from-theta-running}), which is compatible with the range of $\alpha_s$
and $L$ accessible at the LHC (cf.\ table~1 of
Ref.~\cite{Hamilton:2020rcu}).
We then
apply the Lund declustering procedure: we identify a splitting as primary or
secondary only if it passes the cut $z > z_{\rm cut}$. We then determine the
coefficients $a_0$ and $a_2$ from the distribution of the observable.

\begin{figure}
  \captionsetup[subfigure]{oneside,margin={20pt,0pt}}
  \centering
  \begin{subfigure}{0.33\textwidth}
    \centering
    \includegraphics[width=1.\textwidth,page=2]{figures/zcut-all-order-plots/fixed-vs-allorder-zcut-scan.pdf}
  \caption{$q\rightarrow qg(g\rightarrow gg)$}
  \label{fig:allorder-Z2qq-zcut-1}
  \end{subfigure}%
  \begin{subfigure}{0.33\textwidth}
    \centering
    \includegraphics[width=1.\textwidth,page=3]{figures/zcut-all-order-plots/fixed-vs-allorder-zcut-scan.pdf}
  \caption{$q\rightarrow qg(g\rightarrow q'\bar{q}')$}
  \label{fig:allorder-Z2qq-zcut-2}
  \end{subfigure}%
  \centering
  \begin{subfigure}{0.33\textwidth}
    \centering
    \includegraphics[width=1.\textwidth,page=1]{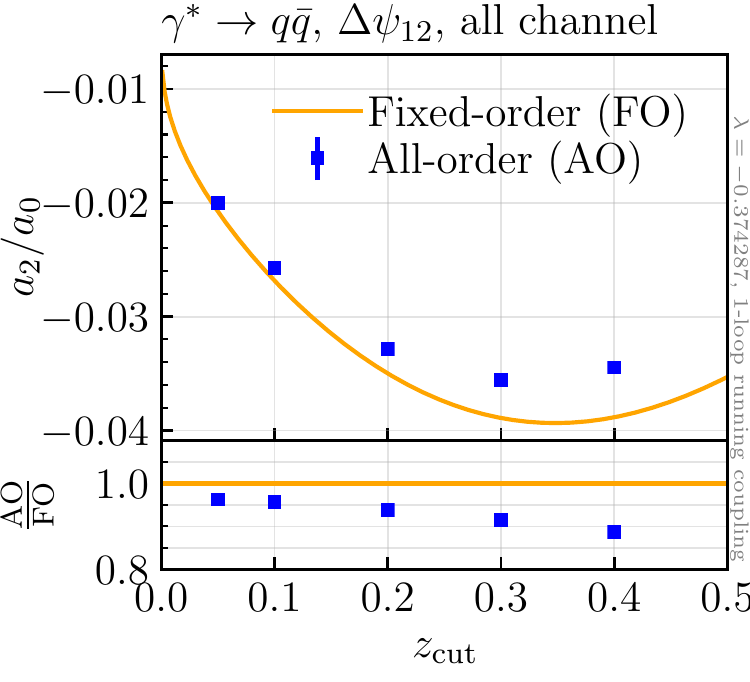}
  \caption{All channels}
  \label{fig:allorder-Z2qq-zcut-3}
  \end{subfigure}
  \caption{Relative size of the spin correlations $a_2/a_0$ for $\Delta
\psi_{12}$, in $\gamma^* \to q \bar q$ (quark-initiated jet), from a numerical
resummation (AO, blue triangles), compared to the second-order result (FO, orange
curve) for (a) the $gg$, (b) the $q' \bar q'$ and (c) all channels.
}
\label{fig:allorder-Z2qq-zcut}
\end{figure}
\begin{figure}
  \captionsetup[subfigure]{oneside,margin={20pt,0pt}}
  \centering
  \begin{subfigure}{0.33\textwidth}
    \centering
    \includegraphics[width=1.\textwidth,page=6]{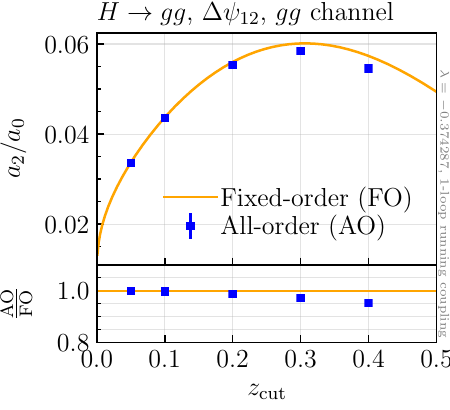}
  \caption{$q\rightarrow qg(g\rightarrow gg)$}
  \end{subfigure}%
  \begin{subfigure}{0.33\textwidth}
    \centering
    \includegraphics[width=1.\textwidth,page=7]{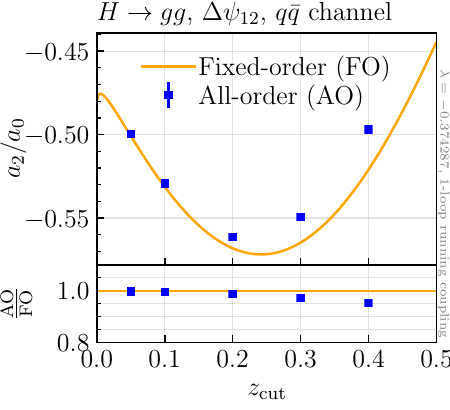}
  \caption{$q\rightarrow qg(g\rightarrow q'\bar{q}')$}
  \end{subfigure}%
  \centering
  \begin{subfigure}{0.33\textwidth}
    \centering
    \includegraphics[width=1.\textwidth,page=5]{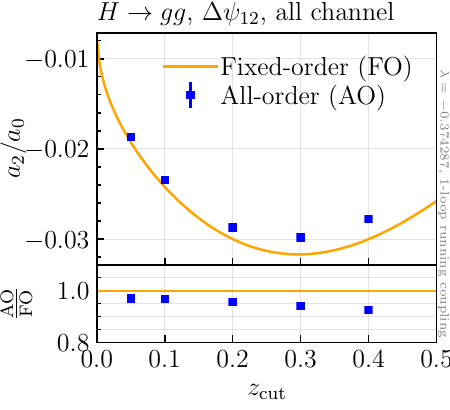}
  \caption{All channels}
  \end{subfigure}
  \caption{Same as Fig.~\ref{fig:allorder-Z2qq-zcut}, for $H \to gg$ (gluon-initiated jet).}
\label{fig:allorder-h2gg-dpsi12-zcut}
\end{figure}

Figure \ref{fig:allorder-Z2qq-zcut} depicts the ratio $a_2/a_0$ at fixed second
order (FO) and all orders (AO) for the case of $\Delta \psi_{12}$ in $\gamma^* \to q
\bar q$ events (quark-initiated intra-jet azimuthal angle), for five values of
$z_{\rm cut} \in \lbrace 0.05,0.1,0.2,0.3,0.4 \rbrace$ (the fixed
order extends to $z_\text{cut}=0$, however the resummation cannot be
extended to that region without also addressing soft logarithms).
The size of the spin
correlations is shown separately for the two different branching histories
$q\rightarrow qg(g\rightarrow gg)$ in
Fig.~\ref{fig:allorder-Z2qq-zcut-1}, and $q\rightarrow qg(g\rightarrow q'\bar{q}')$ in
Fig.~\ref{fig:allorder-Z2qq-zcut-2}. The {\it rest} channel, characterised by
the absence of an intermediate gluon (e.g. $q\rightarrow gq(q\rightarrow
gq)$ at second order, and other possible histories at all orders), is not
shown as it does not produce correlations.\footnote{Note that the rest channel
does contribute to the normalisation of the sum of all channels.} The
ratio $a_2/a_0$ is also given for the sum of all channels in
Fig.~\ref{fig:allorder-Z2qq-zcut-3}.

Turning on the IR cut, $z_{\rm cut} > 0$, first leads to an increase in the
absolute size of $|a_2/a_0|$, with the large-$z_2$ contributions in
Fig.~\ref{fig:fixedorder-z1z2-quarkjet-sameside-dpsi12} driving the spin
correlations. As $z_{\rm cut}$ continues to increase, the
intermediate gluon becomes harder and spin correlations decrease again up to
$z_{\rm cut} = 0.5$.\footnote{In phenomenological applications, one would
optimise the values of the cuts separately, $z_{{\rm cut},1}$ and $z_{{\rm
cut},2}$, to have a maximal signal-to-background ratio.}
We observe that the size of the spin correlations agrees between the fixed order
and the resummed, all-order case, for small values of $z_{\rm cut}\sim 0$, in
the separate channels. Once we turn $z_{\rm cut}$ on, the resummation starts
diluting the spin correlations, as intermediate, soft gluons can be emitted with
values of $z<z_{\rm cut}$, transporting away some of the spin information before
it is propagated to the identified secondary splitting. Thus, the size of the
correlations in the resummed prediction decreases with respect to the
fixed-order calculation, with the resummed $|a_2/a_0|$ being about $88\%$
of the fixed-order value at $z_{\rm cut} = 0.4$. 

If one considers the sum of all
channels, given in Fig.~\ref{fig:allorder-Z2qq-zcut-3}, i.e.\ without the
inclusion of any flavour-tagging, the relative size of the resummed correlations
starts at about $95\%$ of their fixed-order counterpart at small values of
$z_{\rm cut} \sim 0$.
This decrease finds its source in the higher relative
contribution from possible branching histories without correlations --- what we
call the rest channel --- in all-order events: indeed, there is only one such
possible history at second order ($q \to q(q \to qg)g) $, where the emission of
the first gluon is hard and the declustering follows the soft quark branch to
the secondary splitting). This rest channel contributes to $a_0$ but not to
$a_2$, thus the additional damping of $|a_2/a_0|$
at small values of $z_{\rm cut}$ in
\begin{equation}
\left(\frac{a_2}{a_0}\right)_\text{\!all} = \frac{a_{2,{\rm q'\bar q'}} +
a_{2,{\rm gg}}}{a_{0,{\rm q'\bar q'}} + a_{0,{\rm gg}} + a_{0,{\rm rest}}}\,.
\end{equation}

Similar results are shown in Fig.~\ref{fig:allorder-h2gg-dpsi12-zcut} for the
gluon-initiated ($H\to gg$) intra-jet observable $\Delta \psi_{12}$. While the
spin correlations are smaller when compared to the quark-initiated case
presented above, we find that they are also less sensitive to resummation: the
all-order correlations $|a_2/a_0|$ are about $\mathcal{O}(95\%)$ of
their fixed-order counterpart at $z_{\rm cut} = 0.4$ for the separate channels,
and $\mathcal{O}(92.5\%)$ for the sum of all channels.

\begin{figure}
  \captionsetup[subfigure]{oneside,margin={20pt,0pt}}
  \centering
  \begin{subfigure}{0.33\textwidth}
    \centering
    \includegraphics[width=1.\textwidth,page=10]{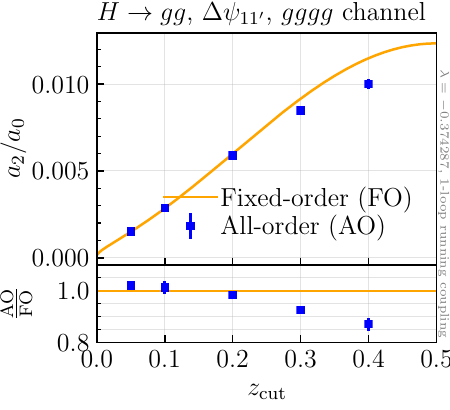}
  \caption{$\{g_1 \rightarrow gg,g_{1'}\rightarrow gg\}$}
  \end{subfigure}
  \begin{subfigure}{0.33\textwidth}
    \centering
    \includegraphics[width=1.\textwidth,page=11]{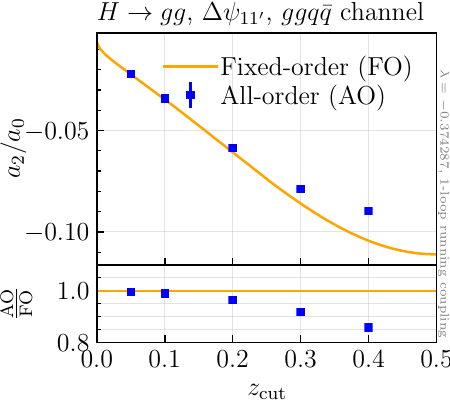}
  \caption{$\{g_1 \rightarrow q\bar{q},g_{1'}\rightarrow gg\}$}
  \end{subfigure}\\[1.5em]
  \centering
  \begin{subfigure}{0.33\textwidth}
    \centering
    \includegraphics[width=1.\textwidth,page=12]{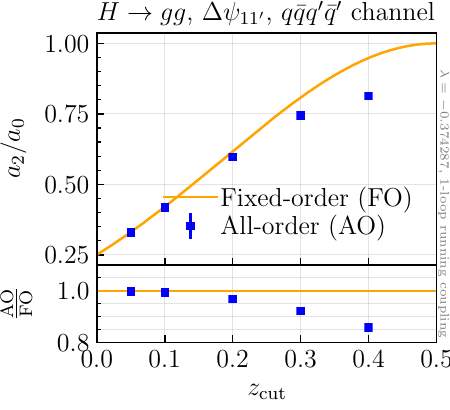}
  \caption{$\{g_1 \rightarrow q\bar{q},g_{1'}\rightarrow q\bar{q}\}$}
  \end{subfigure}
  \centering
  \begin{subfigure}{0.33\textwidth}
    \centering
    \includegraphics[width=1.\textwidth,page=9]{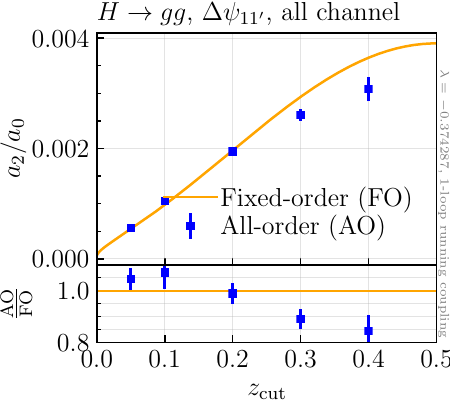}
  \caption{All channels}
  \end{subfigure}
  \caption{Size of the correlations $a_2/a_0$ in $\Delta \psi_{11'}$ for $H
\to gg$ events (inter-jet). Results are shown for (a) the $gggg$, (b) the
$ggq\bar q$ , (c)  the $q\bar{q}q\bar{q}$ and (d) all channels.}
\label{fig:allorder-h2gg-dpsi11bar-zcut}
\end{figure}

Finally, fixed- and all-order predictions are shown in
Fig.~\ref{fig:allorder-h2gg-dpsi11bar-zcut} for the inter-jet observable $\Delta
\psi_{11'}$ in $H \to g_1g_{1'}$ events. The correlations $|a_2/a_0|$
are very small in each separate channel, with the exception of the all-quark
final state $g_1 \to q \bar q, g_{1'} \to q' \bar q '$. Because the largest
cross section comes from the all-gluon final state, where the correlations
$|a_2/a_0| \lesssim 1\%$ are extremely small, and because of partial cancellations
between different channels, the correlations in the sum of all
channels are almost vanishing.
Since the effect is so small, there is a large
statistical uncertainty on the value of the coefficient $a_2$ fitted from the
Monte-Carlo runs of the toy shower. It still gives results consistent with the
observations made above.
Additionally, we note that the inter-jet spin correlations are also
somewhat more sensitive to resummation than for the intra-jet
correlations in $\Delta \psi_{12}$ studied above.

%......................................................................
\subsubsection{Validation and results for energy correlators}

As part of the validation of the MicroJets code, we also compare it
to the analytic resummation of the EEEC presented in
Ref.~\cite{Chen:2020adz}.
While in the latter reference, the spin
correlations in the EEEC were shown differentially as a function of
the opening angle $\theta_S$ of the secondary (small) splitting and
fixed opening angle $\theta_L$ of the primary (large) splitting, here we
show the results integrated over angles to enhance the statistics.
Inspired by Ref.~\cite{Chen:2020adz} we choose the following
integration bounds on the opening angles of the primary and secondary
branchings:\footnote{Cf. figure 3 in Ref.~\cite{Chen:2020adz}}
\begin{align}
\sqrt{0.1} &< \theta_L < 1\,, \notag\\
0.01 &< \theta_S < 0.1\,,
\label{eq:theta-cuts}
\end{align}
\begin{figure}
  \centering
  \includegraphics[width=0.85\textwidth]{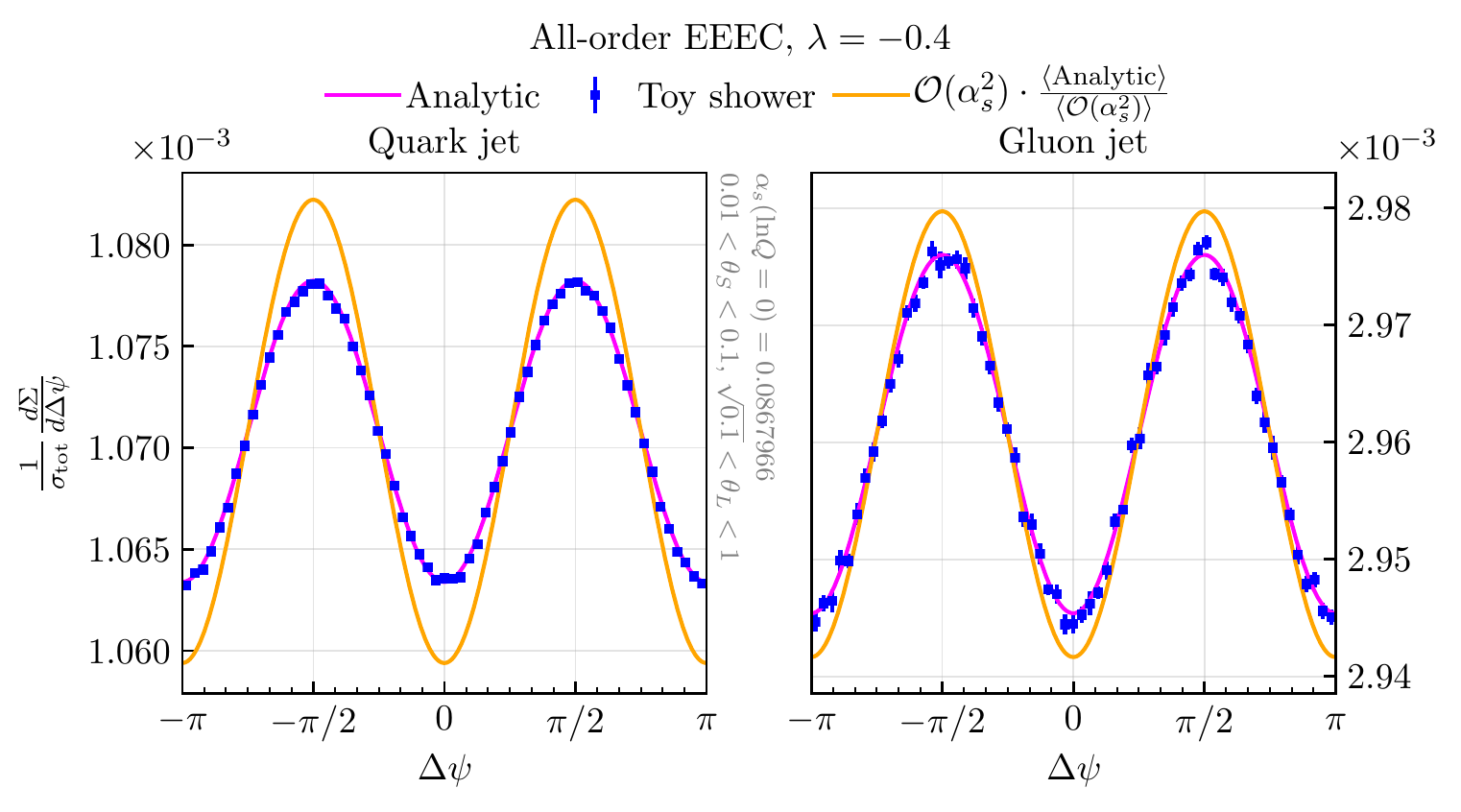}
  \caption{All-order comparison of the toy shower and the analytic
    resummation performed in Ref.~\cite{Chen:2020adz}, for a quark-initiated
    (left) and a gluon-initiated jet (right). The resummation
    is performed using $\alpha_s = 0.0868$ and restricting the opening
    angles as in Eq.~\eqref{eq:theta-cuts}. The result from a
    fixed-order expansion, normalised so that its mean coincides with
    the mean value of the analytic curve, is also shown for
    comparison.  }
\label{fig:allorder-test-EEEC-noshowers}
\end{figure}%
and take $\alpha_s = 0.0868$ corresponding to a hard scale of roughly
$1~\mathrm{TeV}$.
The toy shower and analytic resummation
results~\cite{Chen:2020adz} are shown in
Fig.~\ref{fig:allorder-test-EEEC-noshowers}, summed over all final
branching flavour channels, demonstrating good agreement.
The figure also shows the second-order expansion, normalised so
that its mean value coincides with the resummed result, illustrating a
modest reduction in the degree of spin correlations from the
resummation, which is similar to our findings above with the Lund
declustering observables.

\section{Numerical validation of spin correlations within PanScales
  showers}
\label{sec:tests}

In this section, we validate the PanScales showers against various
numerical predictions.
In particular we want to demonstrate that the algorithm described
above reproduces fixed-order matrix elements in the strongly ordered
limit and that it produces the correct NLL distributions at all orders.
Here, we first provide a brief summary of the PanScales showers. 
A comprehensive description can be found in Ref. \cite{Dasgupta:2020fwr}.

The PanScales showers fit into two categories, namely PanLocal which 
features local recoil, and PanGlobal which features global recoil.  
The PanLocal mapping for the emission of $p_k$ from a dipole $\{\tilde{p}_i, \tilde{p}_j\}$ 
is given by
\begin{subequations}
  \begin{align}
    p_k &= a_k \tilde{p}_i + b_k \tilde{p}_j + k_{\perp}\,, \\
    p_i &= a_i \tilde{p}_i + b_i \tilde{p}_j - f k_{\perp}\,, \\
    p_j &= a_j \tilde{p}_j + b_j \tilde{p}_j - (1-f) k_{\perp}\,,
  \end{align}
\end{subequations}
where $k_{\perp} = k_{\perp,1} \cos(\varphi) + k_{\perp,2}
\sin(\varphi)$ and $-k_{\perp}^2 = k_t^2$.
The coefficients of the map are
functions of the ordering variable, $v$, and an auxiliary variable
$\bar \eta$,
\begin{equation}
\label{eq:panscales-variable-defs}
v = \frac{k_t}{\rho} e^{-\beta |\bar{\eta}|},
\qquad
\rho = \left(\frac{s_{\itilde}
    s_{\jtilde}}{Q^2 s_{\itilde\jtilde}}\right)^{\frac{\beta}{2}}\,,
\qquad
  a_{k} = \sqrt{\frac{s_{\jtilde}}{s_{\itilde\jtilde}s_{\itilde}}}\,
  k_t{e}^{+\bar \eta}\,,
  \qquad
  b_{k} = \sqrt{\frac{s_{\itilde}}{s_{\itilde\jtilde}s_{\jtilde}}}\,
  k_t{e}^{-\bar \eta}\,,
\end{equation}
where
$s_{\itilde \jtilde} =2\ptilde_{i}\cdot \ptilde_{j}$,
$s_{\itilde}=2\ptilde_{i} \cdot Q$, and $Q$ is the total event momentum.
The quantity $\beta$ parameterises the choice of ordering variable.
The PanGlobal map
drops the $k_{\perp}$-contributions to $p_i$ and $p_j$, as well as the
terms $b_i$ and $a_j$, and instead boosts and rescales the full event.
The PanLocal shower comes in a dipole variant, where $f = 1$ and every
soft eikonal is reproduced by the sum of two opposite branching
kernels, and in an antenna variant, where
\begin{equation}
f = f(\bar{\eta}) = \frac{e^{2\bar{\eta}}}{1+e^{2\bar{\eta}}}\,,
\end{equation}
and every eikonal is instead
contained in a single contribution. 
We will show results for the PanGlobal shower with
$\beta=0$\footnote{We have performed validations also with the
  PanGlobal $\beta=1/2$ shower but do not show them here to avoid
  cluttering already busy plots.} and the PanLocal dipole shower with
$\beta = 1/2$ and in some places also the PanLocal antenna shower with
$\beta = 1/2$.

\subsection{Validation at second order}
\label{sec:fixed-order-tests}
The first essential test to carry out is at fixed
$\mathcal{O}(\alpha_s^2)$ since this is the first order at which spin
correlations are non-vanishing.
We compare the collinear branching amplitudes presented in
section~\ref{sec:coll-branching} against the exact
$\mathcal{O}(\alpha_s^2)$ tree-level matrix element, taking the limit
of strongly ordered angles for fixed $z_1$ and $z_2$.
We then proceed to validate our adaptation of the Collins-Knowles algorithm by applying it at
fixed order in the PanScales showers and comparing against the second
order expansion of the toy shower and the exact
$\mathcal{O}(\alpha_s^2)$ cross section.

\subsubsection{Matrix element validation of collinear branching amplitudes}
\label{sec:point-by-point}
Matrix elements constructed using collinear branching amplitudes are
expected to reproduce the exact amplitudes only in the strongly
ordered limit. In order to demonstrate that this is the case, we
consider the angular correlations $\Delta\psi_{12}$ and
$\Delta\psi_{11'}$ as a function of
\begin{align}
  \Theta_{12}&=\max\left(\theta_1,\theta_2/\theta_1\right),\notag\\
  \Theta_{11'}&=\max\left(\theta_1,\theta_{1'}\right),
\end{align}
where $\theta_i$ is the opening angle of the $i^\text{th}$ branching. The
strongly ordered limit is then approached as $\Theta\rightarrow 0$. In
practice we achieve this by fixing the energy fractions $z_1$ and
$z_2$. We then generate angles $\theta_1$, $\theta_2$, $\phi_1$, and
$\phi_2$ from which we can deduce the full kinematics of our event
using the PanLocal dipole $\beta=0.5$ map.\footnote{%
  Here, for conciseness, we show just the PanLocal dipole $\beta=0.5$
  results.
  When we come to fixed order tests with the full shower phase space
  and observables in
  section~\ref{sec:fourier-tests}, we will also show results for the
  PanGlobal shower.}
The full kinematics are then passed to
the exact matrix element, obtained using amplitudes from
Ref.~\cite{Badger:2005jv}, to be compared directly with
the Collins-Knowles weights, Eqs.~(\ref{eq:ME-factorization-example})
and \eqref{eq:branch-amp-spinor-prod}.
\begin{figure}
  \centering 
  \includegraphics{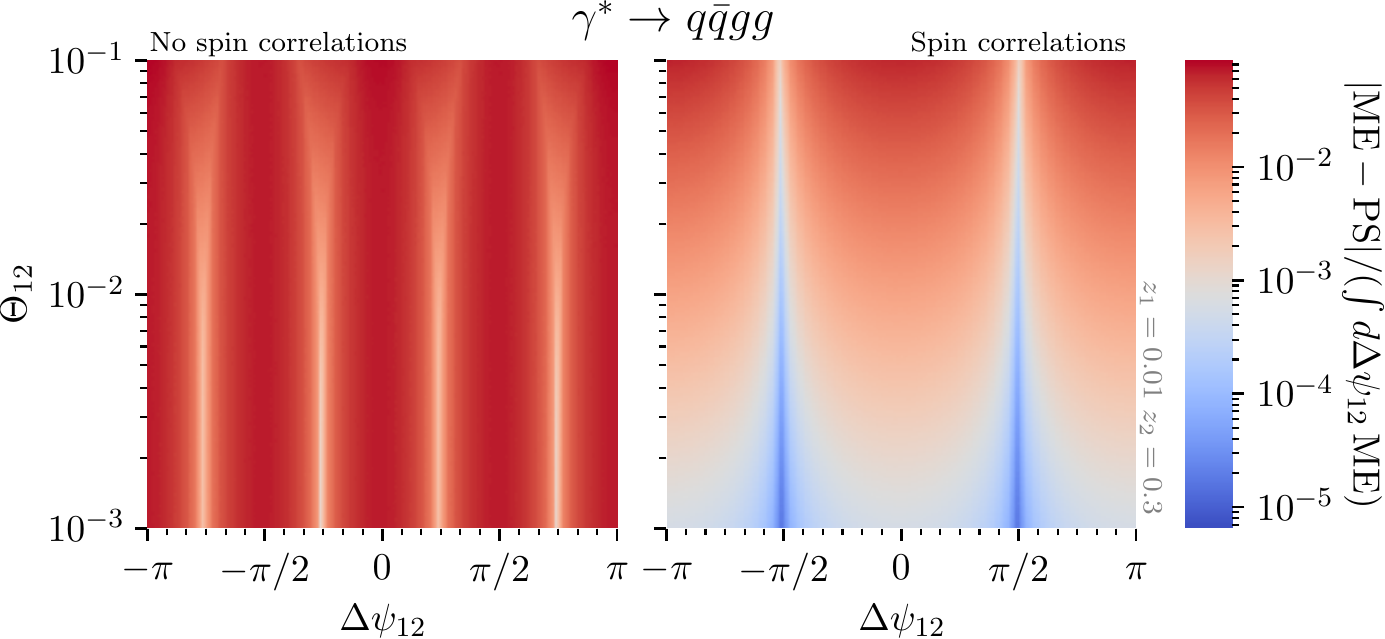}
  \includegraphics{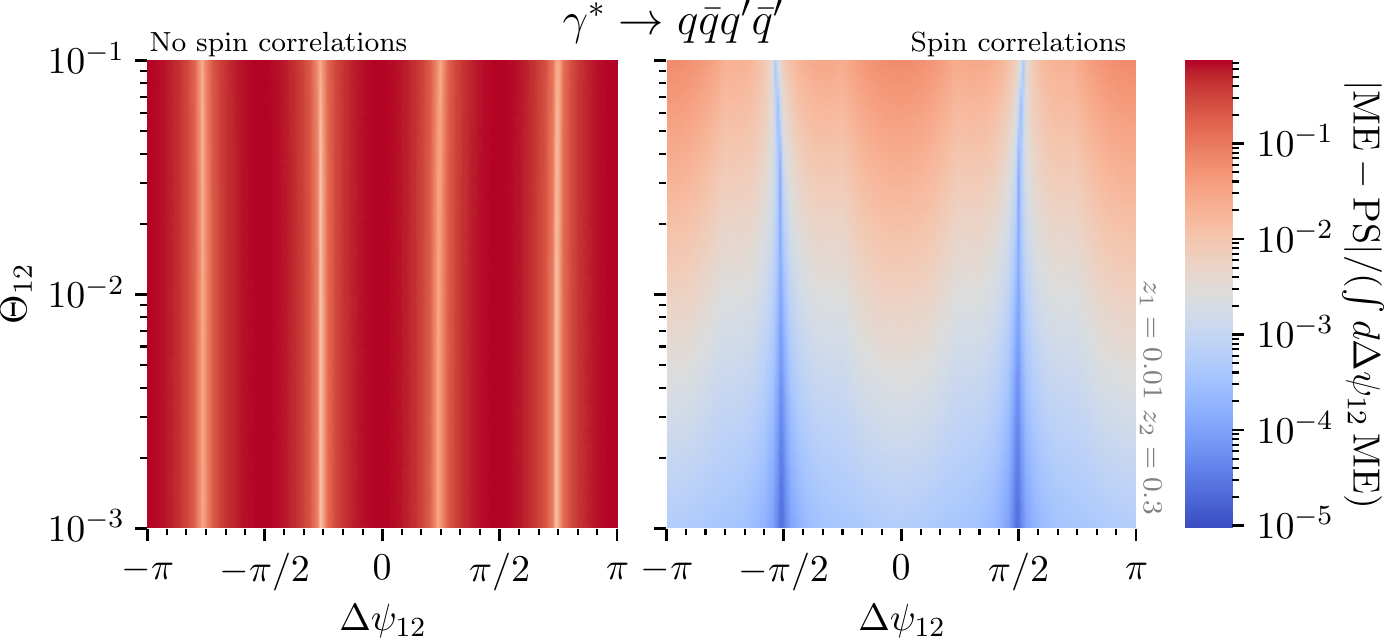}
  \caption{Normalised difference between the Collins-Knowles weight
    and the full squared matrix element for $e^+e^- \to \gamma^* \to
    q\bar q + \text{2-parton} $ events.
    Both branchings are on the same side of the event and have
    $z_1 = 0.01$ and $z_2 = 0.3$. 
    The left-hand (right-hand) column shows the results when we
    neglect (include)
    spin-correlations in the Collins-Knowles weights.
    The rows show different partonic channels.
  }
  \label{fig:point-by-point-Zgamma}
\end{figure}
In Fig.~\ref{fig:point-by-point-Zgamma} we show this comparison as a function of
$\Delta\psi_{12}$ and $\Theta_{12}$ for the processes $\gamma^*\rightarrow
q\bar{q}q'\bar{q}'$ and $\gamma^* \rightarrow q\bar{q}gg$ for
$z_1=0.01$ and $z_2=0.3$. On the left we show the comparison between
the exact matrix element and the shower matrix element without 
spin correlations, while they are included on the right. We show
the absolute difference between the exact matrix element and the
shower matrix element normalised to the exact matrix element
integrated in a slice of $\Delta\psi_{12}$. When spin correlations are
not included in the branching amplitudes the exact matrix element and
the shower matrix element disagree for all values of $\Theta_{12}$ and
$\Delta\psi_{12}$ in both processes.
The band structure that shows up for
$\Delta\psi_{12}=\{\frac{-3\pi}{4},\frac{-\pi}{4},\frac{\pi}{4},\frac{3\pi}{4}\}$
corresponds to the points in $\Delta\psi_{12}$ where the
azimuthal modulation of the full matrix element intersects
the prediction by the shower, which is flat in the collinear
limit.
When spin correlations are included we see that the discrepancy
between the exact matrix element and the shower matrix element only
persists away from the collinear limit and we find that the residual
differences vanish as $\Theta_{12}$ becomes small.
\begin{figure}
  \includegraphics{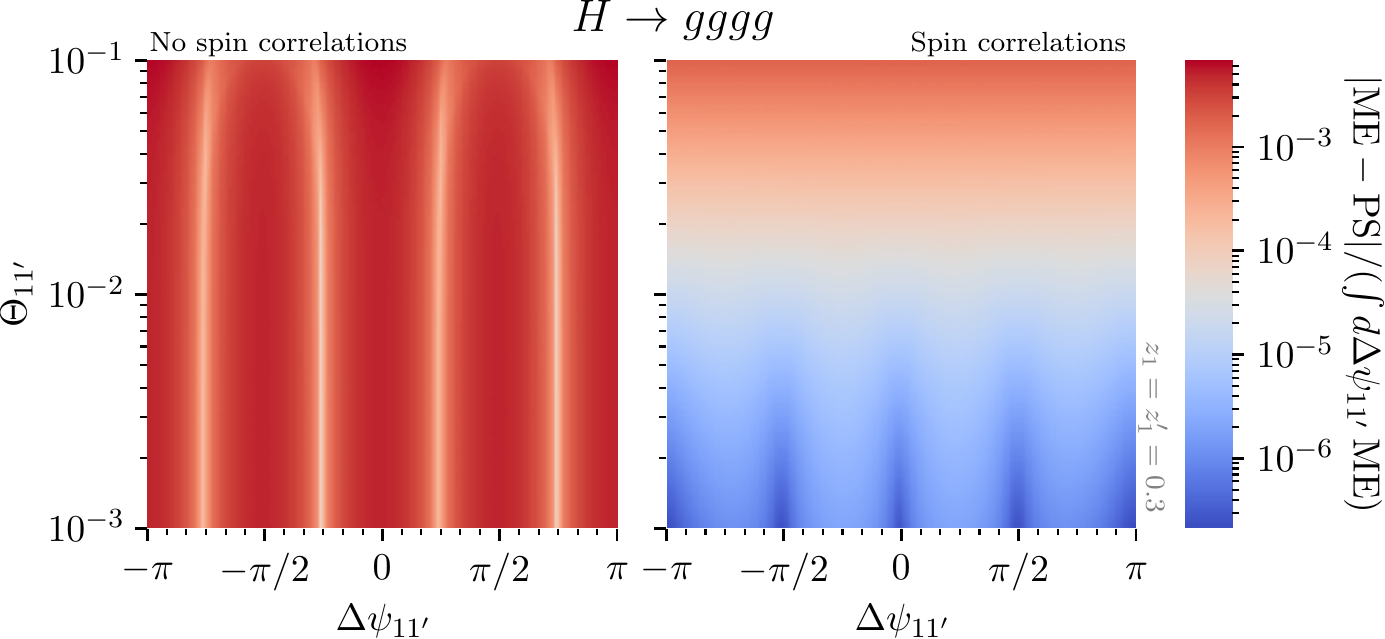}
  \includegraphics{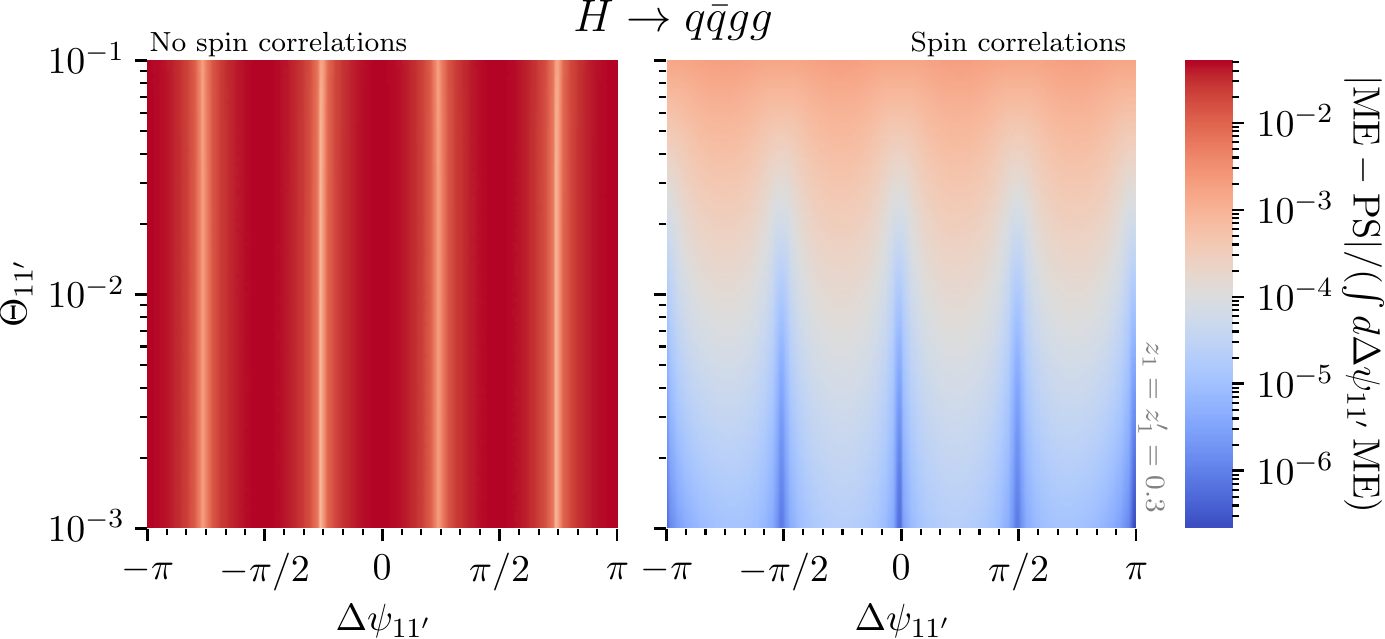}
  \includegraphics{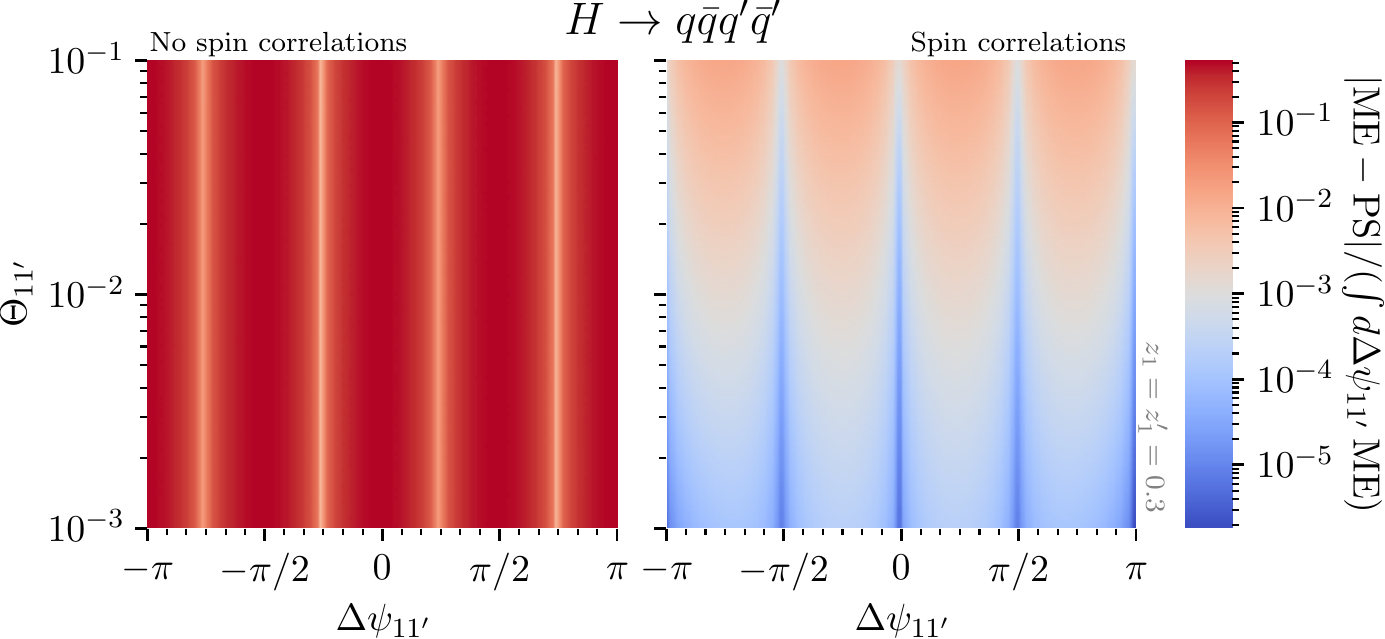}
  \caption{Same as Fig.~\ref{fig:point-by-point-Zgamma} but for $H \to
    gg + \text{2-parton}$ events.
    The two branchings are on opposite sides of the event and have
    $z_1 = z_{1'} = 0.3$.
  }
  \label{fig:point-by-point-H}
\end{figure}
The picture is very similar if we instead focus on $H\rightarrow gg$
processes, as can be seen in Fig.~\ref{fig:point-by-point-H}. In
this case we show the three subprocesses
$H\rightarrow gggg$, $H\rightarrow ggq\bar{q}$, and $H\rightarrow
q\bar{q}q'\bar{q}'$ as a function of $\Delta\psi_{11'}$ and $\Theta_{11'}$, using $z_1=z_2=0.3$. 
When spin correlations are not included in the shower branching amplitudes, the shower matrix element
and the exact matrix element again disagree for all values of $\Theta_{11'}$ and
$\Delta\psi_{11'}$. The same band structure can be seen as in
Fig.~\ref{fig:point-by-point-Zgamma} which is again due to the
intersection of the cosine and an almost flat curve. When spin
correlations are included in the shower we again find excellent
agreement as we approach the strongly ordered limit. 
Some band structure remains, associated with the existence of azimuths
for which the exact and parton shower results happen to coincide
exactly.

\subsubsection{Tests with full shower phase space and observables}
\label{sec:fourier-tests}

\begin{figure}
  \centering
  \begin{subfigure}{0.49\textwidth}
    \centering
    \includegraphics[width=1.\textwidth,page=3]{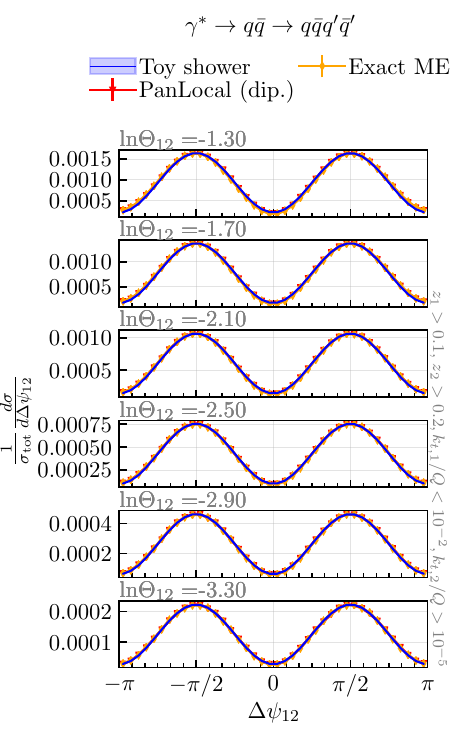}
    \caption{$q\bar q q' \bar{q}'$}
	\label{fig:fixedorder-fourier-dpsi-qqqq}
  \end{subfigure}
  \begin{subfigure}{0.49\textwidth}
    \centering
    \includegraphics[width=1.\textwidth,page=4]{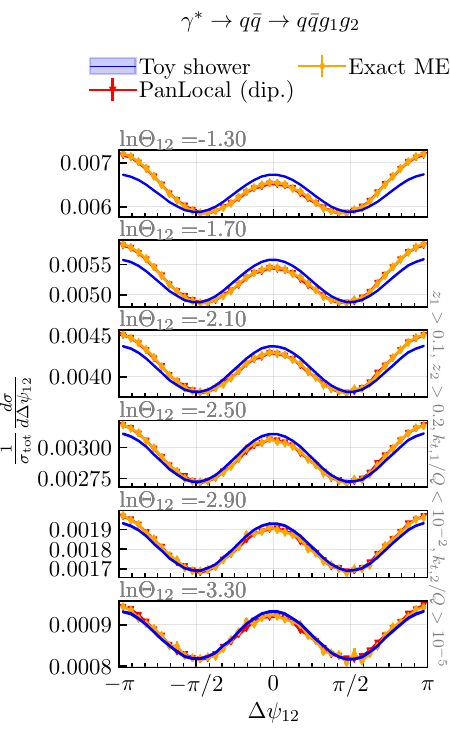}
    \caption{$q\bar q g_1 g_2$}
	\label{fig:fixedorder-fourier-dpsi-qqgg}
  \end{subfigure}
  \caption{The second-order distribution of $\Delta \psi_{12}$ from the toy
shower, the dipole PanLocal shower, and the tree-level matrix element,
shown for increasingly collinear configurations, ${\ln} \Theta_{12}
\in \lbrace -1.3,-1.7,-2.1,-2.5,-2.9,-3.3 \rbrace$. Results are
given for (a) quark-only and (b) two quarks-two gluons final
states.
}

\label{fig:fixedorder-fourier-dpsi}
\end{figure}

We now consider integrated results over a portion of the Lund plane,
i.e.\ using the analysis approach discussed in
section~\ref{sec:lund-azimuth-variables}. 
This serves to test the full combination of shower phase space
generation, branching probabilities and observable implementation
(which comes in independent variants for the toy shower and the full
shower). 
We still work at
fixed (second) order, and compare the resulting distribution of $\Delta
\psi_{12}$ for several of the PanScales showers, the toy shower, and the exact
tree-level matrix element.
We produce results for a fixed strong coupling at
$Q = 100 \,{\rm TeV}$, setting the following cuts on the primary and
secondary splittings:
\begin{subequations}
  \begin{align}
    &z_1 > 0.1,\quad k_{t,1} < 1 \,{\rm TeV}\,, \\
    &z_2 > 0.2,\quad k_{t,2} > 1 \,{\rm GeV}\,.
  \end{align}
\end{subequations}
Fig.~\ref{fig:fixedorder-fourier-dpsi} depicts the second-order (normalised) differential
cross-section
\begin{align}
  \frac{1}{\sigma_\text{tot}} \frac{d \sigma}{d \Delta\psi_{12}} = \frac{1}{\int d|{\rm
  ME}_{\gamma^* \to q\bar q}|^2}
  \frac{d|{\rm ME}_{\gamma^*\to q\bar{q} + X}|^2}{d \Delta \psi_{12}}\,,
\end{align}
for different values of $\Theta_{12}$,
where $X$ may be $q\bar q$ or $gg$.
As in the previous section, the latter
variable serves as a measure of ordered collinearity for the two successive
splittings. We show the distributions obtained by the toy shower, by the
PanLocal shower (run with a value of $\beta=0.5$, see
Eq.~(\ref{eq:panscales-variable-defs})) in its dipole formulation, and by the
exact tree-level matrix element. In the bulk of the phase space, where opening
angles are not strongly restricted, the distribution of $\Delta \psi_{12}$ is
not expected to agree across all considered setups: the parton shower will
feature non-negligible recoil effects, and the toy shower is applying the
strictly-collinear limit of the Collins algorithm irrespective of the opening
angles. However, we do expect agreement between those setups, and the exact
matrix element, in the limit where the angles are small and strongly ordered.

To quantify the agreement between the different setups in the
strongly-ordered collinear limit,
${\rm lim}_{\Theta_{12} \to 0} \left( \frac{d \sigma}{d \Delta
    \psi_{12}}\right)$, we apply a Discrete Cosine Transform (DCT) to
the binned distribution of $\Delta \psi_{12}$, where we integrate over
all configurations that fulfil the strong angular ordering
requirement for a given value of $\Theta_{12}$.
Specifically, for a distribution with $\Delta \psi_{12}$ bins running
from $i=0,\ldots,n-1$ we define the $k^\text{th}$ DCT coefficient as follows
\begin{equation}
  \label{eq:Ak-def}
  A_k(\Theta_{12}) =
  \frac{1}{n} \sum_{i=0}^{n-1} \cos \left(\frac{2k\pi}{n}\left(
      i+\frac{1}{2} \right)\right)
  \int_{\text{bin } i} \!\!\!\! d\Delta \psi_{12}
  \int_{ \substack{\theta_1 < \Theta_{12}
      \hphantom{\theta_1} \\ \theta_2 < \Theta_{12} \theta_1 }}
  d\theta_2 d\theta_1
  \frac{d\sigma}{d\theta_1 d\theta_2 d\Delta \psi_{12}} \,.
\end{equation}
The $A_0$ and $A_2$ coefficients can be related to the Fourier
coefficients $a_0$ and $a_2$ used above,
\begin{equation}
a_0 = A_0\,, \qquad
a_2 = 2A_2\,.
\end{equation}
This definition exclusively picks out the even (cosine-like) Fourier components,
whereas the odd (sine-like) modes are zero by definition.

\begin{figure}
  \centering
  \begin{subfigure}{0.49\textwidth}
    \centering
    \includegraphics[page=1]{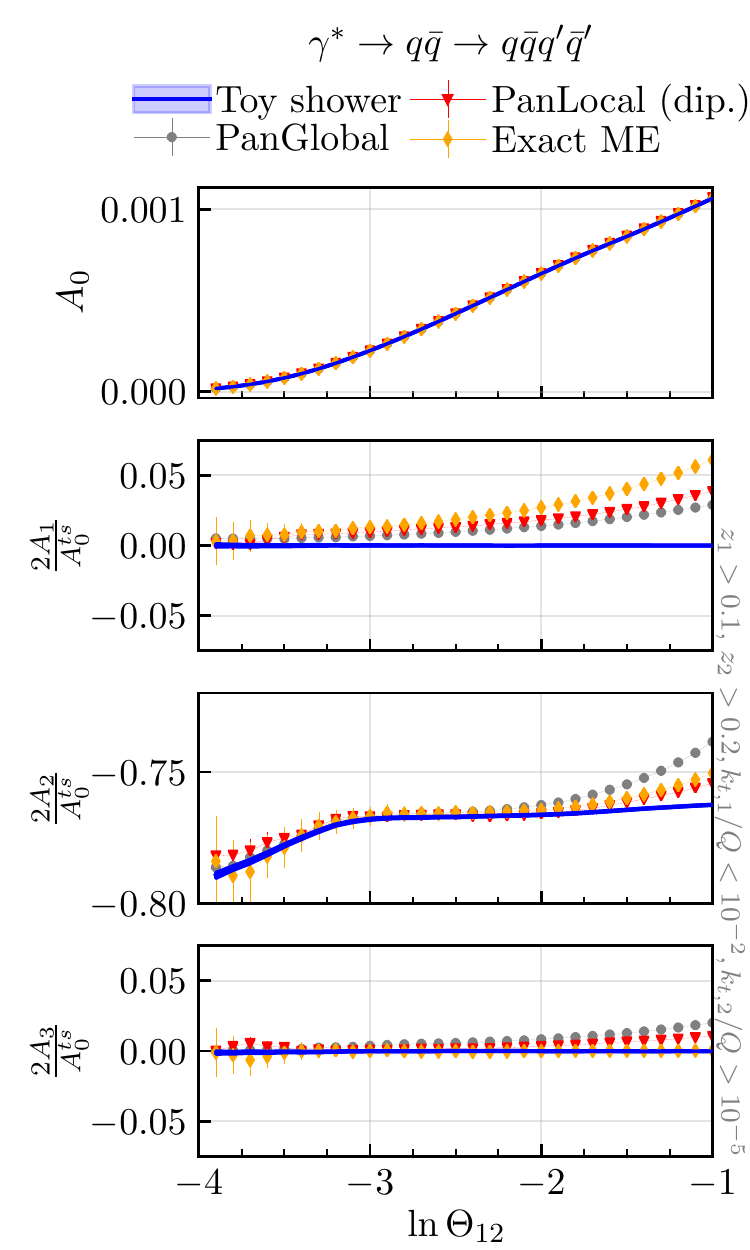}
  \end{subfigure}
  \begin{subfigure}{0.49\textwidth}
    \centering
    \includegraphics[page=2]{figures/fixed-order-dct-coeffs.pdf}
  \end{subfigure}
  \caption{Fixed-order Fourier analysis of the toy shower, the PanGlobal
$(\beta=0)$, and the dipole PanLocal ($\beta=0.5$) showers, as compared to the
exact tree-level matrix element. The first four modes are shown here, and their
associated coefficients suitably converge to identical values in the strongly
angular-ordered limit $\Theta_{12} \to 0$.}
\label{fig:fixedorder-fourier-fits}
\end{figure}

The results are summarised in Fig.~\ref{fig:fixedorder-fourier-fits}, for
$\gamma^* \to q\bar q q' \bar q '$ (left column) and $\gamma^* \to q \bar q gg$
(right column). We show the first four DCT coefficients, normalised by the toy
shower $A_0^{ts}(\Theta_{12})$, for two PanScales showers, the exact matrix
element, and the toy shower. For the latter, since the toy shower
is free of recoil effects, the correlations introduced by the spin algorithm take
the form given by Eq.~(\ref{eq:dsig-dpsi12}) irrespective of the value of
$\Theta_{12}$. Thus, for the toy shower, $A_0$ and $A_2$ are the only non-zero coefficients,
and $2A_2 / A_0 = a_2 / a_0$ gives the relative size of the integrated spin
correlations, which are of the order $\mathcal{O}(-77\%)$ for the $g \to q'\bar
q'$ channel, and $\mathcal{O}(+7\%)$ for the $g \to gg$ channel in most of the
phase space.\footnote{There is an interplay between the imposed phase-space cuts on
$k_t$, and the integration bound on $\Theta_{12}$, which we believe
may be responsible for the kink at the left-hand end of the spectrum.}
The parton showers and the matrix
element partly generate non-zero coefficients at large values of $\Theta_{12}$,
all of which become consistent with zero in the strongly-ordered limit at $\ln
\Theta_{12} \sim -4$, with the exception of the coefficient $A_2$ encoding the
spin correlations. All setups produce compatible values of $A_2$ at small
$\Theta_{12}$. We also observe that the PanScales showers
display distortions that are of same sign, and similar in size, as the exact
matrix element at larger values of $\Theta_{12}$.
The difference between the PanLocal dipole and the PanGlobal showers
is a consequence of the different kinematic map rather than the
difference in $\beta$ values that we have used ($0.5$ and $0.0$
respectively).
\logbook{c5ea9da8}{see 2020-eeshower/analyses/spin-correlation-toyshower/toyshower-vs-panscales-fixed-order/fixed-order-dct-coeffs-wPanGlobal-beta0.5.pdf}

\subsection{Validation at all orders in $\alpha_s$} 
\label{sec:all-order-tests}

We now turn to the validation of our spin-correlation algorithm in the PanScales parton showers.
To test specifically the single logarithmic (NLL),  $\as^n
L^n$, terms generated by the shower, 
we run the PanGlobal ($\beta=0$) shower and the
dipole and antenna versions of the PanLocal ($\beta = 0.5$) showers with
asymptotically small $\alpha_s \to 0$, keeping $\lambda = \alpha_s L$
fixed.
For the Lund declustering observable (cf.\
section~\ref{sec:lund-azimuth-variables}), $e^{-|L|}$ is the smallest
value that we allow for $k_{t,2}/Q$, or equivalently at
$\as^n L^n$ accuracy, the smallest value for $\theta_2$, while
$\theta_1$ and $k_{t,1}$ are allowed to take on any value (typically,
the Lund declustering procedure ensures $\theta_1 > \theta_2$ and
$k_{t,1} > k_{t,2}$).
For the EEEC variable (cf.\ section~\ref{sec:EEEC-definition}),
$e^{-|L|}$ is the smallest value that we allow for $\theta_S$, while
$\theta_L$ is allowed to take on any value larger than $\theta_S$.
For the purpose of the following tests we choose a value of
$\lambda=-0.5$, which corresponds roughly to the range of $\alpha_s$
and $L$ accessible at the LHC.
In practice, we set $\alpha_s = 10^{-7}$, and the value of the shower cutoff to
$\ln v_{\rm min} = (1+\beta) \lambda / \alpha_s = (1+\beta) \cdot (-0.5) \cdot
10^7$. For results generated by the toy shower, this translates to setting the
toy shower's cutoff scale to $t_{\rm max} = t(\lambda)$ as given in
Eq.~(\ref{eq:t-from-theta-running}).
We use a 1-loop running of $\alpha_s$ for all results shown below
(in our $\alpha_s \to 0$ but fixed $\lambda = \alpha_s L$ limit,
higher-loop running leaves the results unchanged).

In the following, we consider both $\gamma^* \to q \bar q$ and $H \to gg$
initial setups.
A variety of issues arise from the use of small values of $\alpha_s$
and large values of $L$.
The approaches that we take to dealing with them include truncating
the shower to avoid large numbers of soft gluons, tracking differences
in angles between particles as well as their $4$-momenta, the use of a
numerical type that extends the available exponent beyond that
normally accessible in double precision, and a stand-in analysis using
the shower tree structure rather than the full Lund declustering
analysis.
Some of the techniques were introduced in
Refs.~\cite{Dasgupta:2020fwr,Hamilton:2020rcu} and
Appendix~\ref{sec:app:small-alphas-techniques} gives further details
about the techniques that are new here, including a number of tests to
verify that our conclusions remain robust even with these techniques.
We apply a cut $z_{\rm
cut} = 0.1$ in the identification of the primary and secondary (respectively,
both primary) splittings used in reconstructing the Lund observable $\Delta
\psi_{12}$ (respectively, $\Delta \psi_{11'}$).
We also use a stand-in analysis for the EEEC observable, presented in
Appendix~\ref{sec:app:proxy-analysis}, which again is valid in the
extreme collinear limit.

\begin{figure}
  \centering
  \includegraphics[page=1]{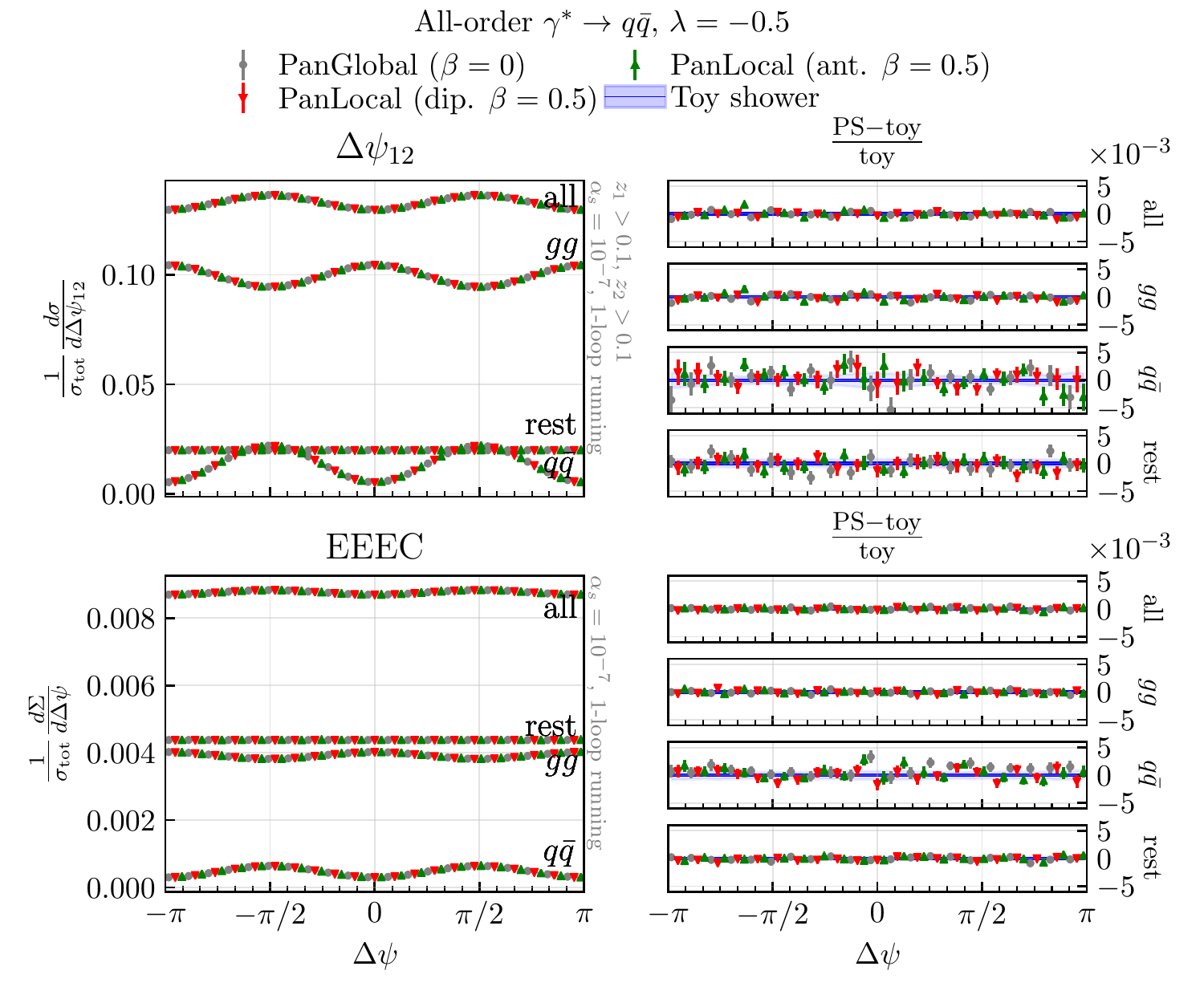}
  \caption{All-order comparison of the toy shower and different PanScales
    showers, for $\gamma^* \to q\bar q$ events.
    The two observables shown are the azimuthal angle,
    $\Delta \psi_{12}$, between a primary and secondary splitting
    planes in Lund declustering, and the difference in angle
    $\Delta \psi$ between the $(ij)k$ and $ij$ planes in the EEEC
    (Eq.~(\ref{eq:EEEC-definition})).
    The results are obtained in the limit $\as \to 0$ for fixed
    $\lambda = \as L=-0.5$.
    For the Lund declustering
    $\Delta \psi_{12}$ we consider events with
    $k_{t,2}/Q > e^{-|L|}$ and for the EEEC $\Delta\psi$ we consider events
    with $\theta_S > e^{-|L|}$.
  }
\label{fig:allorder-test-Ztoqq}
\end{figure}

Figures \ref{fig:allorder-test-Ztoqq} and~\ref{fig:allorder-test-H2gg} depict the
distribution of our new Lund observables, $\Delta \psi_{12}$ and $\Delta
\psi_{11'}$, as well as the EEEC, for $\gamma^* \to q \bar q$, respectively
for $H \to gg$ initial events. The three showers PanGlobal ($\beta=0$),
PanLocal ($\beta=0.5$, dipole and antenna versions) are compared to the
numerically resummed result obtained from the toy shower. In all cases, we show
the contributions stemming from the different channels to the full observable.
The relative deviation between the PanScales showers and the toy shower is
shown on the right, separately for each channel, and is compatible with zero
with statistical uncertainties below the 5 permille level.

\begin{figure}
  \centering
  \includegraphics[width=0.99\textwidth,page=2]{{figures/allorder-toy-vs-panscales-lambda-0.5}.pdf}
  \caption{All-order comparison of the toy shower and different PanScales
showers, for $H \to gg$ events. The three observables shown here are the azimuthal
angle between a primary and secondary splitting plane $\Delta \psi_{12}$, the EEEC, and the
azimuthal angle between two primary splittings (on opposite sides of the event) $\Delta \psi_{11'}$.}
\label{fig:allorder-test-H2gg}
\end{figure}

%======================================================================

\subsection{Phenomenological remarks}
\label{sec:pheno-remarks}

We comment on three aspects here that are potentially relevant for
phenomenological applications.

\begin{table}
  \centering
  \begin{tabular}{lccc}
    \toprule
    flavour channel for 2$^\text{nd}$ splitting
    & $g\to q\bar q$  & $g\to gg$  & all \\\midrule

    EEEC                                        & -0.36 & 0.026 &-0.008\\
    $\Delta \psi_{12}$, $z_1,z_2 > 0.1$         & -0.61 & 0.050 &-0.025\\
    $\Delta \psi_{12}$, $z_1>0.1$, $z_2 > 0.3$  & -0.81 & 0.086 &-0.042 \\
    \bottomrule
  \end{tabular}
  \caption{The relative magnitude of the azimuthal modulation,
    $a_2/a_0$ (cf.\ Eq.~(\ref{eq:dsig-dpsi12})), for the EEEC and Lund intra-jet $\Delta \psi_{12}$
    observables, the latter for two sets of cuts on $z_1$ and $z_2$.
    The results are shown for $\gamma^*\to q\bar q$ events for
    $n_f=5$, separately for two specific flavour channels, as well as
    the sum over all flavour channels (including the channel without
    spin correlations, $q\to qg$).
    As in Fig.~\ref{fig:allorder-test-Ztoqq}, the results are obtained
    in the limit $\as \to 0$ for fixed 
    $\lambda = \as L=-0.5$ and
    for the Lund declustering $\Delta \psi_{12}$ we consider events
    with $k_{t,2}/Q > e^{-|L|}$, while for the EEEC $\Delta\psi$ we
    consider events with $\theta_S > e^{-|L|}$.  }
  \label{tab:a2a0-various-obs}
\end{table}

Our first comment concerns the relative size of spin correlations in the EEEC
and the $\Delta \psi_{12}$ Lund declustering observable.
The EEEC has the advantage of not requiring a $z_\text{cut}$, reducing
the number of parameters that need to be chosen for the observable.
However its weighting with the energies in Eq.~(\ref{eq:EEEC-definition})
tends to favour configurations where a $q\to qg(g\to xy)$ splitting
shares energy equally between the three final particles.
In the notation of Figs.~\ref{fig:fixedorder-z1z2-quarkjet-sameside-dpsi12} and
\ref{fig:fixedorder-z1z2-gluonjet-sameside-dpsi12}, this corresponds
to $z_1 \simeq 2/3$ and $z_2\simeq 1/2$.
While $z_2\simeq1/2$ acts to enhance the spin correlations,
$z_1\simeq 2/3$ tends to reduce them.
In contrast, with the Lund declustering $\Delta \psi_{12}$ one can
adjust the cuts on the $z_1$ and $z_2$ values so as to maximise the
azimuthal modulations.\footnote{Too tight a cut on $z_1$ and $z_2$
  would reduce the available statistics, so one might want to optimise
  the cuts  to maximise a combination of statistical accuracy
  and degree of modulation.}
Table~\ref{tab:a2a0-various-obs} summarises the degree of azimuthal
modulation for different observables in $\gamma^* \to q\bar q$ events.
With our default (non-optimised) cuts of $z_1$ and $z_2 > 0.1$, we see
substantially larger azimuthal modulations than in the EEEC variables,
both in individual flavour channels and in their sum.
The potential for further enhancement of the modulations is made
evident by the results obtained with the $z_2>0.3$ requirement.

Our second comment concerns the sum over all flavour
channels.
The results shown here have been obtained with $n_f=5$ light flavours.
The final magnitude of the spin correlations after the sum over
flavour channels is quite sensitive to the cancellation between
$g \to q\bar q$ and $g \to gg$ splittings and the degree of
cancellation is strongly influenced by the value of $n_f$.
At the scales where one might aim to probe spin correlations, the $c$-
and especially $b$-quark masses are not entirely negligible.
A full phenomenological study of the flavour-summed structure of
azimuthal correlations might, therefore, needs to take into account
finite quark-mass effects.
Note that effects related to $k_t$ values in the neighbourhood of a
heavy-quark threshold are formally suppressed by a logarithm.
For a complete understanding of phenomenological expectations one
would also want to examine the impact of other subleading
logarithmic effects, as well as contributions suppressed by powers of
$k_t/Q$, and possibly also non-perturbative corrections. 
It would clearly also be of interest to find ways of carrying out
measurements with flavour tagging, given the strong effects to be seen
with $g\to q\bar q$ splittings.
While $b$ and $c$ flavour tagging are the most obviously robust
starting points in this respect, one may also wish to consider $s$
tagging~\cite{Erdmann:2020ovh} and generic light-quark versus gluon
discrimination observables.

Our final comment concerns the inter-jet $\Delta \psi_{11'}$
observable.
If Higgs boson decay to two gluons is eventually observed and if one
can carry out flavour tagging so as to have visible net modulation,
this observable could provide an interesting example of an EPR type
measurement at colliders (a constraint on a gluon's splitting angle
would then translate to a constraint on the distance travelled by the
gluon before it splits).
%

%======================================================================
\section{Conclusions}
\label{sec:concl}

The developments shown in this paper are among the last remaining
aspects needed in order to claim complete NLL accuracy in the sense of
Ref.~\cite{Dasgupta:2020fwr} for the PanScales family of parton
showers.
The one component that now remains (at leading colour) is a treatment of
the spin correlations for soft emissions, an aspect that we leave
to future work.

The approach we have taken is largely based on that proposed by
Collins and Knowles, and in part also on the extension to dipole
showers proposed by Richardson and Webster.
The main difference relative to the Richardson and Webster work is our
use of spinor products, avoiding the need for correlating branching
variables across different steps and associated boosts.
The spin correlation treatment is then expected to be correct at
single-logarithmic accuracy for showers that adhere to the general
requirements set out in Ref.~\cite{Dasgupta:2020fwr}.

Relative to earlier work on spin correlations in parton showers, one
of the main novelties of this paper is the framework for validating
the implementation of spin correlations.
As part of our testing framework we extended the MicroJets code to
enable single-logarithmic resummation for a range of spin-correlation
observables.
Figures \ref{fig:point-by-point-Zgamma}--\ref{fig:fixedorder-fourier-fits}
show clear agreement between the collinear limits of fixed-order
matrix elements and a fixed-order expansion of the shower, while
Figs.~\ref{fig:allorder-test-Ztoqq} and \ref{fig:allorder-test-H2gg}
show equally good agreement between the logarithmic structure of the
full shower and the single-logarithmic resummed expectations. 

To carry out those tests we introduced a set of new observables based
on Lund declustering that are sensitive to spin correlations.
These complement the recently proposed EEEC spin-sensitive
observables, providing information that is more differential in
momentum fractions of the different branchings.
This makes it possible to enhance the relative magnitude of the
spin-correlation signal, cf.\ Table~\ref{tab:a2a0-various-obs}.
Spin-correlation effects can be large for $g \to q\qbar$ splittings,
while they are smaller, with an opposite sign for $g \to gg$
splittings (cf.\
Figs.~\ref{fig:fixedorder-z1z2-quarkjet-sameside-dpsi12}--\ref{fig:fixedorder-z1z2-h2gg-oppositeside-dpsi11}).
All-order resummation has a modest effect on them (cf.\
Figs.~\ref{fig:allorder-Z2qq-zcut}--\ref{fig:allorder-h2gg-dpsi11bar-zcut}).
Phenomenologically, the opposite signs for $g \to gg$ and
$g \to q\qbar$ lead to a partial cancellation of the spin correlation
effects in flavour-insensitive observables, although our studies
indicate that some effect remains visible.
Further study of the potential for measuring these effects, possibly
in combination with flavour tagging, would clearly be of interest.

%======================================================================
\section*{Acknowledgements}

We thank Fabrizio Caola and Frederic Dreyer for access to the
matrix-element code used in the tests in
section~\ref{sec:fixed-order-tests}, Keith Hamilton for
discussions on spinor-product conventions and Gregory Soyez for work
on the use of direction differences within observables.
Additionally we thank Hao Chen for communications regarding the
numerical details of the analytical resummation in
Ref.~\cite{Chen:2020adz}.
We are grateful to our PanScales collaborators (Melissa van Beekveld,
Mrinal Dasgupta, Frederic Dreyer, Basem El-Menoufi, Silvia Ferrario
Ravasio, Keith Hamilton, Rok Medves, Pier Monni, Gregory Soyez and
Alba Soto Ontoso), for their work on the code, the underlying
philosophy of the approach and comments on this manuscript.

This work was supported
by a Royal Society Research Professorship
(RP$\backslash$R1$\backslash$180112) (GPS, LS),
by the European Research Council (ERC) under the European Union’s
Horizon 2020 research and innovation programme (grant agreement No.\
788223, PanScales) (AK, GPS, RV), 
by the Science and Technology Facilities Council (STFC) under
grants ST/P000274/1 (RV) and ST/T000864/1 (GPS),
by Linacre College (AK) and by Somerville College (LS).

%======================================================================
\appendix
\section{Deriving the branching amplitudes}
\label{sec:app:spinor-product-der}

In this appendix we collect some details regarding the computation of branching amplitudes 
in terms of spinor products and their numerical evaluation. 
The calculations here are done following the conventions of Ref.~\cite{Kleiss:1985yh}.
For two light-like momenta $p$ and $q$, we define the spinor product 
\begin{equation}
S_{\lambda}(p, q) = \bar{u}_{\lambda}(p) u_{-\lambda}(q),
\end{equation} 
where $\lambda = \pm 1$ is the Dirac spinor helicity.
Spinor products have the properties
\begin{subequations}
  \begin{align}
    &S_{\lambda}(p, q) = - S_{\lambda}(q, p) = -S_{-\lambda}(p,q)^*, \nonumber \\
    &|S_{\lambda}(p,q)|^2 = S_{\lambda}(p,q) S_{-\lambda}(q,p) = 2 \, p{\cdot}q.
  \end{align}
\end{subequations}
Following the HELAS conventions \cite{Murayama:1992gi}, the polarisation vector of
a gluon with momentum $p$ is defined as 
\begin{equation}
\epsilon^{*\mu}_{\lambda}(p) = \frac{1}{\sqrt{2}} \frac{1}{S_{-\lambda}(r,p)} 
\bar{u}_{\lambda}(p) \gamma^{\mu} u_{\lambda}(r),
\end{equation} 
where $r$ is a light-like vector specifying the gauge.
For calculational purposes, the Chisholm identity \cite{Kleiss:1985yh}
\begin{equation}
\slashed{\epsilon}^*_{\lambda}(p) = \frac{\sqrt{2}}{S_{-\lambda}(r, p)} \bigg[u_{\lambda}(r) \bar{u}_{\lambda}(p) + u_{-\lambda}(p) \bar{u}_{-\lambda}(r) \bigg] 
\end{equation} 
is often useful.
We define the momenta of a collinear splitting as $p_a \rightarrow p_b + p_c$, 
such that $p_b = z p_a$ and $p_c = (1-z) p_a$.
In this limit, the gauge vector $r$ cancels from the collinear branching amplitudes,
and they may be written in terms of a single spinor product 
$S_{\lambda}(p_b, p_c)$ using the identities
\begin{equation}
S_{\lambda}(p_b, p_a) = \sqrt{1-z} S_{\lambda}(p_b, p_c)
\text{ and }
S_{\lambda}(p_c, p_a) = -\sqrt{z} S_{\lambda}(p_b, p_c).
\end{equation} 
Next, we compute the branching amplitudes for all colour-stripped collinear QCD branchings, dropping any overall phase factors for convenience. 

\section*{$ q \rightarrow q g$}
The branching amplitude is 
\begin{subequations}
  \begin{align}
    \mathcal{M}_{a \to bc}^{\lambda_a \lambda_b \lambda_c} &= \frac{g_s}{2 p_b{\cdot}p_c} \bar{u}_{\lambda_b}(p_b) \slashed{\epsilon}^*_{\lambda_c}(p_c) u_{\lambda_a}(p_a) \\
    &= \frac{1}{\sqrt{2}} \frac{g_s}{p_b{\cdot}p_c} \frac{1}{S_{-\lambda_c}(r, p_c)} \bar{u}_{\lambda_b}(p_b) \Big[ u_{\lambda_j}(r) \bar{u}_{\lambda_c}(p_c) + u_{-\lambda_c}(p_c) \bar{u}_{-\lambda_j}(r) \Big] u_{\lambda_a}(p_a).
  \end{align} 
\end{subequations}
The non-zero helicity configurations are
\begin{subequations}
  \begin{align}
    \mathcal{M}_{a \to bc}^{\lambda, \lambda, \lambda} &= \frac{1}{\sqrt{2}} \frac{g_s}{p_b{\cdot}p_c} \frac{1}{\sqrt{1-z}} S_{\lambda}(p_b, p_c), \\
    \mathcal{M}_{a \to bc}^{\lambda, \lambda, -\lambda} &= \frac{1}{\sqrt{2}} \frac{g_s}{p_b{\cdot}p_c} \frac{z}{\sqrt{1-z}} S_{-\lambda}(p_b, p_c).
  \end{align} 
\end{subequations}
\section*{$g \rightarrow q \bar{q}$}
The branching amplitude is
\begin{subequations}
  \begin{align}
    \mathcal{M}_{a \to bc}^{\lambda_a \lambda_b \lambda_c} &= \frac{g_s}{2 p_b{\cdot}p_c} \bar{u}_{\lambda_b}(p_b) \slashed{\epsilon}^*_{-\lambda_a} (p_a) u_{-\lambda_c}(p_c) \\
    &= \frac{1}{\sqrt{2}} \frac{g_s}{p_b{\cdot}p_c} \frac{1}{S_{\lambda_a} (r, p_a)} \bar{u}_{\lambda_b}(p_b) \bigg[u_{-\lambda_a} (r) \bar{u}_{-\lambda_a}(p_a) + u_{\lambda_a}(p_a) \bar{u}_{\lambda_a}(r) \bigg] u_{-\lambda_c}(p_c).
  \end{align} 
\end{subequations}
The non-zero helicity configurations are
\begin{subequations}
  \begin{align}
    \mathcal{M}_{a \to bc}^{\lambda, \lambda, -\lambda} &= -\frac{1}{\sqrt{2}} \frac{g_s}{p_b{\cdot}p_c} z S_{-\lambda}(p_b, p_c), \\
    \mathcal{M}_{a \to bc}^{\lambda, -\lambda, \lambda} &=  \frac{1}{\sqrt{2}} \frac{g_s}{p_b{\cdot}p_c} (1-z) S_{-\lambda}(p_b, p_c).
  \end{align} 
\end{subequations}

\section*{$g \rightarrow gg$}
For this case, we first derive some general properties. 
If we consider the collinear limit and pick all gluons to have the same gauge vector $r$, we find 
\begin{subequations}
  \begin{align}
    \epsilon^*_{\lambda}(p_i) {\cdot} \epsilon^*_{\lambda}(p_j) &= 0,  \\
    \epsilon^*_{\lambda}(p_i) {\cdot} \epsilon^*_{-\lambda}(p_j) &= 1,\\
    \epsilon^*_{\lambda}(p_i){\cdot}p_j &= \frac{1}{\sqrt{2}} \frac{S_{-\lambda}(p_j, r)}{S_{-\lambda}(r, p_i)} S_{\lambda}(p_i, p_j),
  \end{align}
\end{subequations}
where $p_i, p_j \in \{ p_a, p_b, p_c\}$.
The branching amplitude is 
\begin{align}
\mathcal{M}_{a \to bc}^{\lambda_a \lambda_b \lambda_c} &= -\frac{g_s}{p_b{\cdot}p_c} \Big( \epsilon^*_{\lambda_b}(p_b){\cdot}p_c \, \epsilon^*_{\lambda_c}(p_c){\cdot}\epsilon^*_{-\lambda_a}(p_a) 
- \epsilon^*_{\lambda_c}(p_c){\cdot}p_b \, \epsilon^*_{\lambda_b}(p_b){\cdot}\epsilon^*_{-\lambda_a}(p_a) \nonumber \\
&+ \epsilon^*_{-\lambda_a}(p_a){\cdot}p_b \, \epsilon^*_{\lambda_{b}} (p_b) {\cdot} \epsilon^*_{\lambda_c}(p_c)\Big).
\end{align} 
The non-zero helicity configurations are
\begin{subequations}
  \begin{align}
    \mathcal{M}_{a \to bc}^{\lambda, \lambda, \lambda}  &= \frac{1}{\sqrt{2}} \frac{g_s}{p_b{\cdot}p_c}  \frac{1}{\sqrt{z(1-z)}}  S_{\lambda}(p_b, p_c),  \\
    \mathcal{M}_{a \to bc}^{\lambda, \lambda, -\lambda} &= \frac{1}{\sqrt{2}} \frac{g_s}{p_b{\cdot}p_c}  \frac{z^{3/2}}{\sqrt{1-z}} S_{-\lambda}(p_b, p_c),   \\
    \mathcal{M}_{a \to bc}^{\lambda, -\lambda, \lambda} &= \frac{1}{\sqrt{2}} \frac{g_s}{p_b{\cdot}p_c}  \frac{(1-z)^{3/2}}{\sqrt{z}} S_{-\lambda}(p_b, p_c). 
  \end{align}  
\end{subequations}

\section*{Numerical evaluation of spinor products}
To find an expression for the spinor product that can be evaluated numerically,
we may introduce arbitrary reference vectors $k_0$ and $k_1$ which obey 
$k_0^2 = 0$, $k_1^2 = -1$ and $k_0{\cdot}k_1 = 0$.
Then, without loss of generality, we may write 
\begin{equation}
u_+(p) = \frac{1}{\sqrt{2 p{\cdot}k_0}} \slashed{p} u_-(k_0) \text{ and } 
u_-(p) = \frac{1}{\sqrt{2 p{\cdot}k_0}} \slashed{p} \slashed{k}_1 u_-(k_0).
\end{equation}
The spinor product may then be expressed as 
\begin{subequations}
  \begin{align}
    S_{+}(p_b, p_c) &= \frac{1}{\sqrt{2 p_b{\cdot}k_0} \sqrt{2 p_c{\cdot}k_0}} \bar{u}_-(k_0) \slashed{k}_1 \slashed{p}_b \slashed{p}_c u_-(k_0) \\
    &= \frac{1}{2} \frac{1}{\sqrt{2 p_b{\cdot}k_0} \sqrt{2 p_c{\cdot}k_0}} \textrm{tr}\left((1 - \gamma^5) \slashed{k}_0 \slashed{k}_1 \slashed{p}_b \slashed{p}_c  \right) \\
    &= \frac{1}{\sqrt{p_b{\cdot}k_0} \sqrt{p_c{\cdot}k_0}} \bigg[ (p_c{\cdot}k_0) (p_b{\cdot}k_1) - (p_b{\cdot}k_0)(p_c{\cdot}k_1) - i \epsilon_{\mu \nu \alpha \beta} k_0^{\mu} k_1^{\nu} p_b^{\alpha} p_c^{\beta} \bigg].
  \end{align}
\end{subequations}
While this expression remains independent of a choice of representation of the Dirac algebra, 
an explicit choice for the reference vectors $k_0$ and $k_1$ must be made for the purposes of numerical evaluation.
For instance, if we choose $k_0 = (1,0,0,-1)$ and $k_1 = (0,1,0,0)$, we find
\begin{align} \label{eq:spinor-product-explicit-form}
S_+(p_b, p_c) &= \sqrt{\frac{p_c^0 + p_c^3}{p_b^0 + p_b^3}}(p_b^1 + i p_b^2) - \sqrt{\frac{p_b^0 + p_b^3}{p_c^0 + p_c^3}} (p_c^1 + i p_c^2).
\end{align} 
In the shower implementation, reference vectors are selected on an
event-by-event basis, where care is taken not to select a direction
that aligns with the initial momenta to avoid numerical instabilities.

\section{An example of calculating the functions $A(z)$, $B(z)$}
\label{sec:app:derivation-A-B}

We illustrate the computation of the spin-correlated matrix element squared,
taking the example of the observable $\Delta \psi_{12}$ at second order for a
quark-initiated jet, $q \to qg (g \to q'\bar q')$, of
Fig.~\ref{fig:intro-example}. We then cast the result into the form of
Eq.~(\ref{eq:dsig-dpsi12}). The matrix element squared, summing over
all helicities, is given by
\begin{align}
|M|^2 &= \mathcal{M}_{0\to 12}^{\lambda_0
\lambda_1 \lambda_2} \mathcal{M}_{0 \to 12}^{* \lambda_0 \lambda_1 \lambda_2'} \mathcal{M}_{2 \to 3
4}^{\lambda_2 \lambda_3 \lambda_4} \mathcal{M}_{2 \to 3 4}^{* \lambda_2' \lambda_3
\lambda_4} \notag\\
&= \mathcal{M}_{0\to 12}^{+++} \mathcal{M}_{0\to 12}^{* +++} (\mathcal{M}_{2\to 34}^{++-} \mathcal{M}_{2\to 34}^{* ++-} + \mathcal{M}_{2\to 34}^{+-+} \mathcal{M}_{2\to 34}^{* +-+}) \notag \\
&+ \mathcal{M}_{0\to 12}^{+++} \mathcal{M}_{0\to 12}^{* ++-} (\mathcal{M}_{2\to 34}^{+-+} \mathcal{M}_{2\to 34}^{* --+} + \mathcal{M}_{2\to 34}^{++-} \mathcal{M}_{2\to 34}^{* -+-}) \notag \\
&+ \mathcal{M}_{0\to 12}^{++-} \mathcal{M}_{0\to 12}^{* +++} (\mathcal{M}_{2\to 34}^{--+} \mathcal{M}_{2\to 34}^{* +-+} + \mathcal{M}_{2\to 34}^{-+-} \mathcal{M}_{2\to 34}^{* ++-}) \notag \\
&+ \mathcal{M}_{0\to 12}^{++-} \mathcal{M}_{0\to 12}^{* ++-} (\mathcal{M}_{2\to 34}^{-+-} \mathcal{M}_{2\to 34}^{* -+-} + \mathcal{M}_{2\to 34}^{--+} \mathcal{M}_{2\to 34}^{* --+}) \notag \\
&+ (+ \leftrightarrow -)\,,
\label{eq:calculation-A-B-0}
\end{align}
where the last line includes the flipping of all helicities, and we have already
excluded forbidden helicities. Using Eq.~(\ref{eq:branch-amp-phase}) and
inserting the functions $\mathcal{F}_{a\to bc}^{\lambda_a\lambda_b\lambda_c}(z)$
given in Table~\ref{tab:hel-dep-fcns}, the above reduces to
\begin{subequations}
\begin{align}
|M|^2 &= 2\frac{2 g_s^2}{\theta_{12}^2 E_0^2 z_1 (1-z_1)} \frac{2g_s^2}{\theta_{34}^2 E_2^2 z_2 (1-z_2)} \notag \\
&\times \Big( \frac{z_2^2 + (1-z_2)^2}{1-z_1} - \frac{2 z_1 z_2 (1-z_2)}{1-z_1} e^{2 i (\phi_1-\phi_2)} \notag \\
&-  \frac{2 z_1 z_2 (1-z_2)}{1-z_1} e^{-2 i (\phi_1-\phi_2)} + \frac{z_1^2((1-z_2)^2 + z_2^2)}{1-z_1}  \Big) \label{eq:calculation-A-B-1}\\
&= \frac{8g_s^4 (1+z_1^2)(z_2^2 + (1-z_2)^2)}{\theta_{12}^2 \theta_{34}^2 E_0^4 z_1(1-z_1)^4 z_2 (1-z_2)}\Big( 1 + \frac{2z_1}{1+z_1^2} \frac{-2z_2(1-z_2)}{z_2^2+(1-z_2)^2} \cos (2 \Delta \phi)\Big)\,,\label{eq:calculation-A-B-2}
\end{align}
\end{subequations}
where each of the four terms in the parenthesis of
Eq.~(\ref{eq:calculation-A-B-1}) comes from one of the four lines in
Eq.~(\ref{eq:calculation-A-B-0}). Finally, Eq.~(\ref{eq:calculation-A-B-2})
makes the form of $A(z_1)$ and $B(z_2)$ explicit.

\section{Rotational invariance in the collinear limit}
\label{sec:app:rotated-axis-check}
As part of the numerical evaluation of spinor products in
Eq.~\eqref{eq:spinor-product-explicit-form}, a set of reference
vectors must be selected. 
Any dependence on these reference vectors vanishes in the calculation of a full squared matrix element, 
but the Collins-Knowles algorithm only accounts for the contributions that dominate the matrix element in the collinear limit.
Consequently, any dependence on the reference vectors should also vanish from the Collins-Knowles result as long as the shower emissions are strongly ordered and sufficiently collinear.

\begin{figure}
  \begin{subfigure}{0.48\textwidth}
    \centering
    \includegraphics{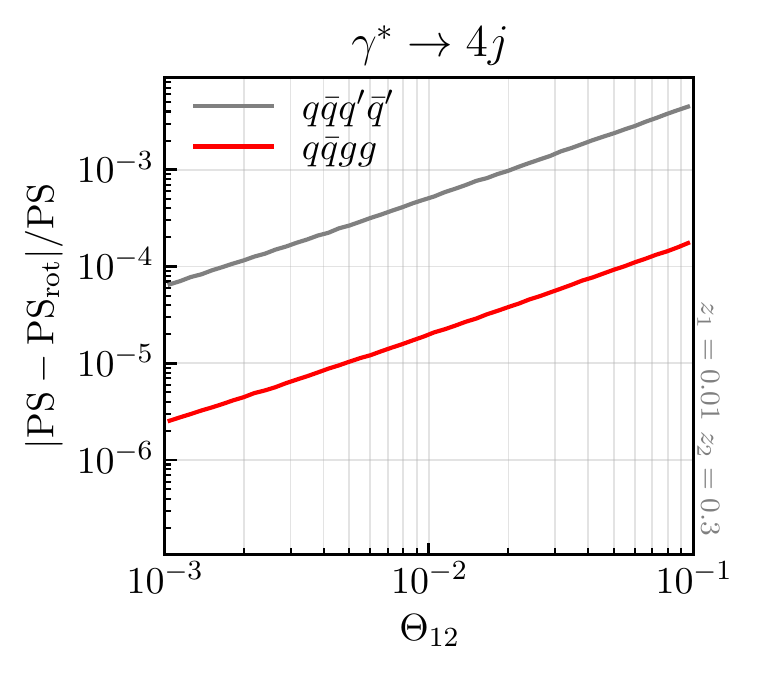}
  \end{subfigure}
  \begin{subfigure}{0.48\textwidth}
    \centering
    \includegraphics{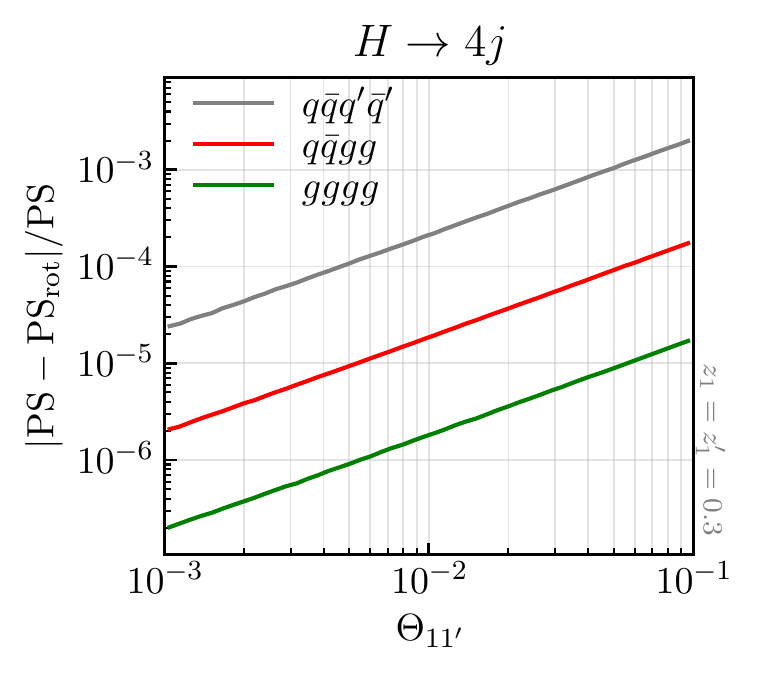}
  \end{subfigure}
  \caption{Comparison of the Collins-Knowles weights of two-emission
    events and their randomly rotated counterpart.
    The hard scattering is either $\gamma^* \rightarrow q \bar{q}$
    (left), where a single primary ($z_1 = 0.01$) and secondary emission ($z_2 = 0.3$) are
    performed, or $H \rightarrow gg$ (right), where two opposite-side
    primary emissions are performed ($z_1 = z_{1'} = 0.3$).}
  \label{fig:axis-check}
\end{figure}

In this appendix, we validate this property in the PanScales implementation of the Collins-Knowles algorithm.
To that end, like in section~\ref{sec:point-by-point}, the Collins-Knowles weight as a
function of an angular resolution is considered.
Rather than comparing to a matrix element directly, the weight is
computed for an event, and for the same event that is rotated
randomly.
As the reference vectors stay fixed, rotating the event is equivalent to rotating the reference vectors in the original event.
Figure \ref{fig:axis-check} shows the relative difference of these weights for two shower emissions off $\gamma^* \rightarrow q \bar{q}$ and $H \rightarrow gg$ hard scatterings.
The relative difference in weights decreases as a power of the angular resolution, and is thus a power correction that vanishes in the collinear limit. 
Numerically, these effects are also small compared to the size of the
spin correlations themselves. 

\section{Technical details of the all-order comparisons}
\label{sec:app:small-alphas-techniques}
To facilitate the isolation of the NLL structure of the parton shower, the all-order comparisons of section \ref{sec:all-order-tests} are performed at 
extremely small values of $\alpha_s$ and extremely large values of the logarithm. 
Several techniques need to be employed to maintain numerically feasible analyses in these extreme regimes. 
In this appendix we detail these techniques. 

\subsection{Removal of soft radiation}
\label{sec:collinear-cuts}

\begin{figure}
\centering
\begin{tikzpicture} 
  % Lund plane
  \draw[black] (0,0) -- ( 3.4,-3.4 );
  \draw[black] (0,0) -- (-3.4,-3.4);
  \fill[gray,opacity=0.2] (0,0) -- (-3.4,-3.4) -- (-1.8,-3.4) -- (0,-1.6) -- (1.6,-3.4) -- (3.4,-3.4);
  \fill[red,opacity=0.4] (-1.8,-3.4) -- (0,-1.6) -- (1.6,-3.4);
  \draw[black,dotted] (0,0) -- (0,-3.4);
  \draw[black, <->]  (-2.8, -0.2) -- (-2.8,-1.2) -- (-1.8,-1.2);
  \draw[black, <->]  (-2.8, -0.2) -- (-2.8,-1.2) -- (-1.8,-1.2);
  \node at (-2.3,-0.2) {\footnotesize $\ln k_t$};
  \node at (-1.65,-1.05) {\footnotesize $\eta$};
  %
  % Cuts primary
  \draw[thin,black,dashed]  ( 0,-0.55) -- (2.85,-3.4);
  \draw[thin,black,dashed]  ( 0,-0.55) -- (-2.85,-3.4);
  %
  % 1st emission and 2ndary plane
  \draw [very thin,black]  (0.5,-3.4) -- (0.5,-0.8) -- (1.6,-4.4);
  \fill [gray,opacity=0.2] (0.5,-2.14) -- (0.5,-0.8) -- (1.6,-4.4) -- (0.88, -3.75);
  \fill [red,opacity=0.4] (0.5,-2.14) -- (0.88, -3.75) -- (0.5,-3.4);
  \filldraw [black] (0.5,-0.8) circle (1pt);
  %
  % Cuts secondary
  \draw[thin,black,dashed]  ( 0.5,-1.35) -- (1.30,-4.05);
\end{tikzpicture}
\caption{An example of the shower Lund plane after the first
  branching, with a cut on soft emissions.
  The red shaded area
  indicates the part of phase space removed by the soft-emission cut,
  which is applied relative to the total event energy.
  The black dashed line represents a possible observable cut on $z$,
  such as that applied by the Lund declustering observables.
  It is applied relative to the parent momentum.
  To
  avoid removing parts of the phase space above the dashed line
  due to recoil effects, the soft-emission cut is applied well below the
  observable cut.
}
\label{fig:collinear-cuts-lund-diagram}
\end{figure}
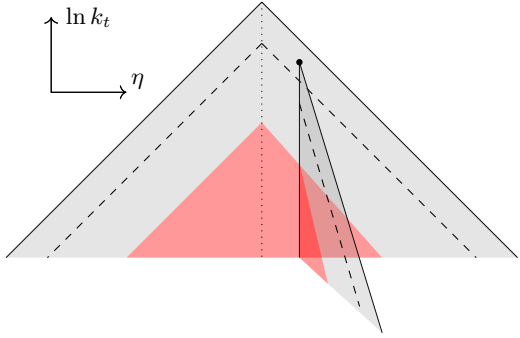

The particle multiplicity generated by the parton shower scales like $\sqrt{\alpha_s L^2}$.
This means that, because the product $\alpha_s L$ is kept constant as $\alpha_s$ decreases and $L$ increases, the multiplicity also increases.
At the logarithmic values used in section \ref{sec:all-order-tests}, the multiplicity has increased to levels that cause the event generation to become numerically unfeasible. 
However, the all-order observables considered here are insensitive to soft gluon radiation, and the spin correlations incorporated by the Collins-Knowles algorithm are also unaffected.
As a result, as long as any radiation that is removed is sufficiently
soft that its recoil has no noticeable impact on the momenta of the
partons that dominate the Lund declustering and energy-correlator
observables, the all-order tests should remain unaffected by the
removal of soft gluon radiation.
Such a cut is implemented in the PanScales shower by limiting the sampling range of the auxiliary parton shower variable that controls the collinear momentum fraction.
This strategy is more efficient than the alternative of vetoing soft branchings, as the shower would spend much of its time generating and vetoing soft emissions at low scales.
An illustration of this procedure is shown in Fig.~\ref{fig:collinear-cuts-lund-diagram}.

\begin{figure}
  \centering 
  \includegraphics{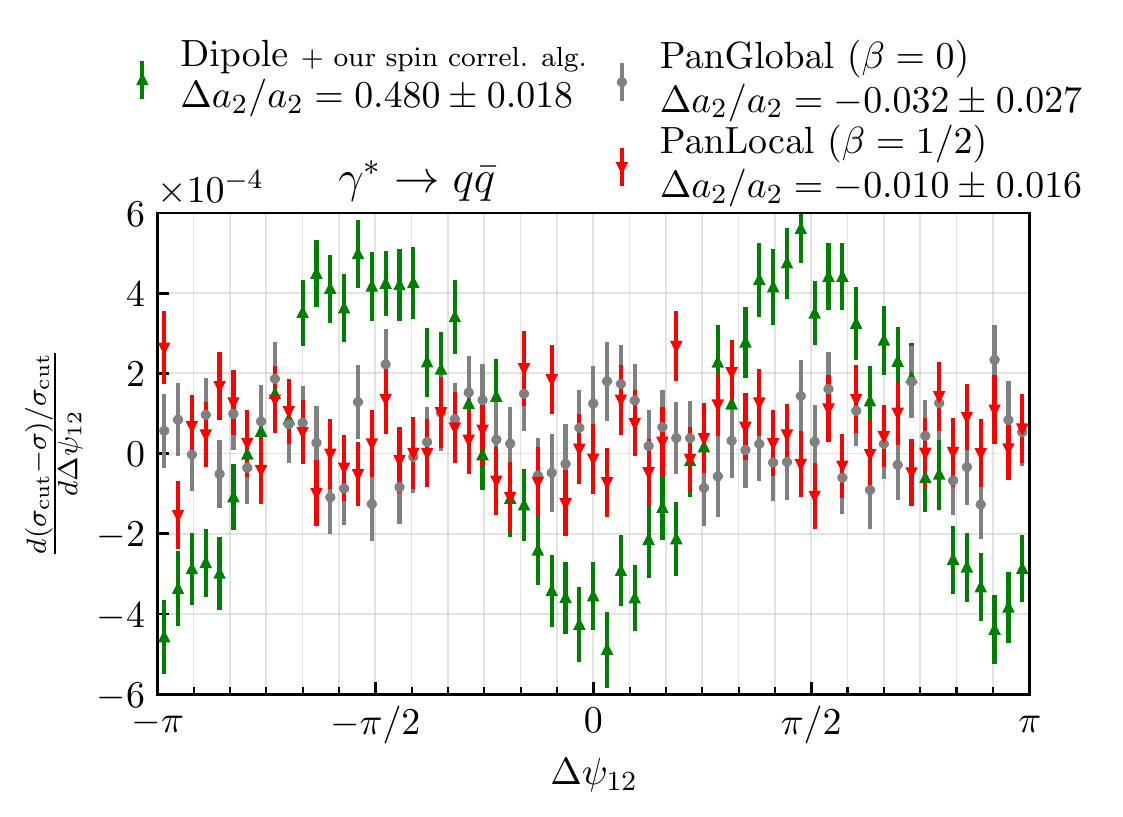}
  \caption{
    Difference in the $\Delta \psi_{12}$ distribution between a run with
    a collinear cut $\ln z_\text{cut}^{\text{PS}}=-10$ and a run without.
    The results are shown for finite
    $\alpha_s = 0.01$ and $L = -27.5$, and three showers.
    The key also shows the result for $\Delta a_2/a_2$ where $\Delta
    a_2$ is the difference between the $a_2$ Fourier coefficient as
    obtained with and without the cut.
    The result labelled ``Dipole'' corresponds to  our implementation
    of the Dire~v1~\cite{Hoche:2015sya} 
    shower algorithm supplemented with our spin-correlation
    algorithm and we expect it to be representative of a variety of
    standard dipole-type showers.
    It indicates that the use of a $\ln z_\text{cut}$ would not be
    safe for testing spin correlations in such standard dipole-type
    showers.}
  \label{fig:collinear-cuts}
\end{figure}

To validate the legitimacy of the application of this cut on soft
radiation,
we compare the predictions of the PanScales showers and our implementation of Dire v1 for $\Delta\psi_{12}$ with $z_{\text{cut}} = 0.1$ at $\alpha_s = 0.01$ and $L = -27.5$.
Figure \ref{fig:collinear-cuts} shows the difference between the
distributions with and without the application of the collinear cut
$\ln z > \ln z_\text{cut}^{\text{PS}} = -10$.
This value of $z_\text{cut}^\text{PS}$ is significantly below the $z$
values that dominate in our observables.
For the PanScales showers, we see that the cut has no statistically
significant effect on the azimuthal distribution and thus we
conclude that we can safely use a $z_\text{cut}^\text{PS}$ in our
logarithmic accuracy tests.
For other showers, the cut can have an impact on the results, as
illustrated with the Dire~v1 dipole shower, where the cut induces a
change in $a_2$ that is a significant fraction of the actual
$a_2$.
We attribute this to transverse recoil effects between gluons of
commensurate $k_t$ values, of the kind discussed in
Refs.~\cite{Dasgupta:2018nvj,Dasgupta:2020fwr}.
For such showers, a test of the logarithmic structure of the spin
correlations would need to be carried out without a
$z_\text{cut}^\text{PS}$.
We suspect that there are observables for which such tests would
reveal problems in the $\as^n L^n$ logarithmic structure, associated
with the following type of configuration: consider a hard-collinear
$g \to q \bar q$ splitting with a transverse momentum $k_{t,1}$,
followed by a soft-collinear emission with transverse momentum
$k_{t,2}\lesssim k_{t,1}$ from the dipole that contains the $q$.
Within standard dipole-shower recoil schemes, the soft-collinear
emission can take its recoil from the quark, altering its
2-dimensional vector transverse momentum, effectively smearing out the
azimuthal angle of the $q\bar q$ pair.\footnote{In PanGlobal showers,
  the problem is avoided because the transverse recoil is assigned
  through a boost, which leaves the azimuthal structure of the
  $q\bar q$ pair unchanged.
  In the PanLocal shower, where it is only valid to run with $\beta>0$
  in the definition of the ordering variable,
  Eq.~(\ref{eq:panscales-variable-defs}), when two emissions are at
  commensurate $k_t$ values, the hard-collinear one always takes place
  later, which ensures that a soft-collinear gluon does not induce
  recoil in a hard-collinear gluon of commensurate $k_t$.
}
In certain circumstances (for example if one considers events where
the $g\to q\bar q$ splitting is the highest-$k_t$ splitting in the
event), the product of squared-matrix element and phase space where
the recoil can occur will lead to a factor $\as L$ for this smearing
to occur, thus spoiling single-logarithmic accuracy.
Note, however, that in the observables that we actually consider, we
do not require either of the collinear splittings to be the hardest in
the event, and we expect that feature to bring further complications,
of the kind discussed in section~3 of the supplemental material of
Ref.~\cite{Dasgupta:2020fwr}, whereby additional double-logarithmic
effects would be expected to (at least) partially mask these issues in
the all-order tests, potentially associated with super-leading
logarithms, $\as^n L^m$ with $m>n$, for observables that should only
have $m\le n$.
Based on these arguments, we consider the presence of any effect from
emissions below $z_{\text{cut}}$, as seen in
Fig.~\ref{fig:collinear-cuts}, to be a sign of potential danger.

\subsection{Stand-in observables}
\label{sec:app:proxy-analysis}
While performing tests at small $\alpha_s$ and large logarithms, it
quickly becomes challenging to accurately evaluate the tiny angles
between collinear momenta.
To avoid large numerical cancellations in the evaluation of inner
products, the PanScales showers have the option to explicitly track the
difference in direction between dipole components.
This technique, as well as the use of a new floating-point type
(\texttt{double\_exp}) that increases the maximum size of the exponent
of a regular double-precision type, was already used in
\cite{Hamilton:2020rcu}, and was equally crucial for the all-order
tests performed here.
While it allows the shower to be run at asymptotically small values of
the coupling, the evaluation of the observables is not as
straightforward.

In the case of Lund-declustering observables, a regular analysis is
technically limited by the absence of directional difference
information for parton pairs that are not in the same dipoles.
%(
In events with at most a Born $q\bar q$ pair, we have implemented
functionality whereby we use the full dipole chain and
in-dipole direction differences to construct a look-up table for the
direction differences between any pair of partons.
For $N$ particles this has an $\order{N^2}$ time and memory cost,
while normal $e^+e^-$ clustering in FastJet also carries an $\order{N^2}$ time cost,
but an $\order{N}$ memory cost.
Used together with a winner-takes-all recombination scheme~\cite{Bertolini:2013iqa,Larkoski:2014uqa,salam:WTA}
in a version of the FastJet~\cite{Cacciari:2011ma}
code adapted to work with the \texttt{double\_exp} type, we can carry
out the Lund declustering analysis at asymptotic values of the
logarithm.
However the current technical limitation of having at most the Born
$q\bar q$ pair in the event means that we need an alternative strategy
for complete events.

Accordingly, we have developed a stand-in analysis that uses the
binary tree generated by the Collins-Knowles algorithm, even though it
is normally unobservable.
During the showering, the Lund structure is determined using
explicit shower information, where at every collinear branching that
appears in the binary tree, the appropriate azimuthal information is
stored if the collinear momentum fraction cuts are passed.
This procedure is equivalent to the evaluation of the exact observable
if the jet clustering produces the same binary tree (for the hard
emissions that we ultimately consider in the observable) as in our
implementation of the Collins-Knowles algorithm and if recoil effects
that can affect the momentum fraction cuts are absent.
At asymptotically small $\alpha_s$, for the subset of PanScales
showers that passed the NLL tests in Ref.\ \cite{Dasgupta:2020fwr},
and with the condition that one works with a hardness cut
$z > z_\text{cut}$ such that $\as \ln z_\text{cut} \ll 1$, these are
both valid assumptions.
We have verified this by considering events without $g \to q\bar q$
splittings and establishing that the full Lund declustering and the
stand-in analysis produced event-by-event identical results.
Note that the stand-in analysis also allows easy access to the flavour
separation of the spin-correlation effects, without the need to
consider flavoured jet clustering.

In the case of the EEEC, a major issue it its $\mathcal{O}(N^3)$ time
complexity, which quickly becomes prohibitive at the multiplicities under
consideration.  The following procedure, which again makes direct use
of the binary tree, is again equivalent to the full observable in the
asymptotic limit. 
For all nodes $a_n$ that are not terminal, and where $n$ is the depth of that node, do as follows:
\begin{enumerate}
\item Retrieve the opening angle $\theta_S$ and the normal vector $\vec n_S$ of the branching of node $a_n$.
\item Retrieve the global momentum fractions $z_1 = 2E_1/Q$, $z_2 =
  2E_2/Q$ of the outgoing momenta $1$ and $2$ of node
  $a_n$. 
\item Iteratively for $i \in \{1,..,n-1\}$:
\begin{itemize}
  \item Retrieve the opening angle $\theta_L$ and the normal vector $\vec n_L$ of the branching of node $a_i$.
  \item Retrieve the global momentum fraction $z_3 = E_3/2E_{\mathrm{CM}}$, where $3$ is the outgoing momentum of $a_{i}$ that is not $a_{i+1}$.
  \item Compute the signed angle $\Delta \psi$ between $\vec n_S$ and $\vec n_L$.
  \item Add to the $\Delta\psi$-$\theta_S$-$\theta_L$-bin with weight $2z_1 z_2 z_3$.
\end{itemize}
\end{enumerate}

This procedure has time-complexity $\mathcal{O} (N \log N)$ and is
again equivalent to the exact observable if $\theta_S \ll
\theta_L$ and in the absence of recoil effects, which are both the
case for asymptotically small $\alpha_s$ for the showers we study.
The above algorithm was validated by verifying that it yields
identical results to the exact observable evaluation for a
representative number of events.

%======================================================================
\bibliographystyle{utphys.bst}
\bibliography{MC}

\providecommand{\href}[2]{#2}\begingroup\raggedright\begin{thebibliography}{10}

\bibitem{Einstein:1935rr}
A.~Einstein, B.~Podolsky, and N.~Rosen, {\em {Can quantum mechanical
  description of physical reality be considered complete?}}
  \href{http://dx.doi.org/10.1103/PhysRev.47.777}{Phys. Rev. {\bf 47} (1935)
  777--780}.

\bibitem{Mahlon:1995zn}
G.~Mahlon and S.~J. Parke, {\em {Angular correlations in top quark pair
  production and decay at hadron colliders}}.
  \href{http://dx.doi.org/10.1103/PhysRevD.53.4886}{Phys. Rev. D {\bf 53}
  (1996)  4886--4896}, \href{http://arxiv.org/abs/hep-ph/9512264}{{\tt
  arXiv:hep-ph/9512264}}.

\bibitem{Bernreuther:2001rq}
W.~Bernreuther, A.~Brandenburg, Z.~G. Si, and P.~Uwer, {\em {Top quark spin
  correlations at hadron colliders: Predictions at next-to-leading order QCD}}.
  \href{http://dx.doi.org/10.1103/PhysRevLett.87.242002}{Phys. Rev. Lett. {\bf
  87} (2001)  242002}, \href{http://arxiv.org/abs/hep-ph/0107086}{{\tt
  arXiv:hep-ph/0107086}}.

\bibitem{Bernreuther:2010ny}
W.~Bernreuther and Z.-G. Si, {\em {Distributions and correlations for top quark
  pair production and decay at the Tevatron and LHC.}}
  \href{http://dx.doi.org/10.1016/j.nuclphysb.2010.05.001}{Nucl. Phys. B {\bf
  837} (2010)  90--121}, \href{http://arxiv.org/abs/1003.3926}{{\tt
  arXiv:1003.3926 [hep-ph]}}.

\bibitem{Mahlon:2010gw}
G.~Mahlon and S.~J. Parke, {\em {Spin Correlation Effects in Top Quark Pair
  Production at the LHC}}.
  \href{http://dx.doi.org/10.1103/PhysRevD.81.074024}{Phys. Rev. D {\bf 81}
  (2010)  074024}, \href{http://arxiv.org/abs/1001.3422}{{\tt arXiv:1001.3422
  [hep-ph]}}.

\bibitem{Melnikov:2011ai}
K.~Melnikov and M.~Schulze, {\em {Top quark spin correlations at the Tevatron
  and the LHC}}. \href{http://dx.doi.org/10.1016/j.physletb.2011.04.043}{Phys.
  Lett. B {\bf 700} (2011)  17--20}, \href{http://arxiv.org/abs/1103.2122}{{\tt
  arXiv:1103.2122 [hep-ph]}}.

\bibitem{Artoisenet:2012st}
P.~Artoisenet, R.~Frederix, O.~Mattelaer, and R.~Rietkerk, {\em {Automatic
  spin-entangled decays of heavy resonances in Monte Carlo simulations}}.
  \href{http://dx.doi.org/10.1007/JHEP03(2013)015}{JHEP {\bf 03} (2013)  015},
  \href{http://arxiv.org/abs/1212.3460}{{\tt arXiv:1212.3460 [hep-ph]}}.

\bibitem{ATLAS:2012ao}
{\bf ATLAS} Collaboration, G.~Aad {\em et al.}, {\em {Observation of spin
  correlation in $t \bar{t}$ events from pp collisions at sqrt(s) = 7 TeV using
  the ATLAS detector}}.
  \href{http://dx.doi.org/10.1103/PhysRevLett.108.212001}{Phys. Rev. Lett. {\bf
  108} (2012)  212001}, \href{http://arxiv.org/abs/1203.4081}{{\tt
  arXiv:1203.4081 [hep-ex]}}.

\bibitem{Aad:2013ksa}
{\bf ATLAS} Collaboration, G.~Aad {\em et al.}, {\em {Measurement of Top Quark
  Polarization in Top-Antitop Events from Proton-Proton Collisions at
  $\sqrt{s}$ = 7 TeV Using the ATLAS Detector}}.
  \href{http://dx.doi.org/10.1103/PhysRevLett.111.232002}{Phys. Rev. Lett. {\bf
  111} (2013) no.~23, 232002}, \href{http://arxiv.org/abs/1307.6511}{{\tt
  arXiv:1307.6511 [hep-ex]}}.

\bibitem{Chatrchyan:2013wua}
{\bf CMS} Collaboration, S.~Chatrchyan {\em et al.}, {\em {Measurements of
  $t\bar{t}$ Spin Correlations and Top-Quark Polarization Using Dilepton Final
  States in $pp$ Collisions at $\sqrt{s}$ = 7 TeV}}.
  \href{http://dx.doi.org/10.1103/PhysRevLett.112.182001}{Phys. Rev. Lett. {\bf
  112} (2014) no.~18, 182001}, \href{http://arxiv.org/abs/1311.3924}{{\tt
  arXiv:1311.3924 [hep-ex]}}.

\bibitem{Aad:2014pwa}
{\bf ATLAS} Collaboration, G.~Aad {\em et al.}, {\em {Measurements of spin
  correlation in top-antitop quark events from proton-proton collisions at
  $\sqrt{s}=7$ TeV using the ATLAS detector}}.
  \href{http://dx.doi.org/10.1103/PhysRevD.90.112016}{Phys. Rev. D {\bf 90}
  (2014) no.~11, 112016}, \href{http://arxiv.org/abs/1407.4314}{{\tt
  arXiv:1407.4314 [hep-ex]}}.

\bibitem{Aad:2014mfk}
{\bf ATLAS} Collaboration, G.~Aad {\em et al.}, {\em {Measurement of Spin
  Correlation in Top-Antitop Quark Events and Search for Top Squark Pair
  Production in pp Collisions at $\sqrt{s}=8$ TeV Using the ATLAS Detector}}.
  \href{http://dx.doi.org/10.1103/PhysRevLett.114.142001}{Phys. Rev. Lett. {\bf
  114} (2015) no.~14, 142001}, \href{http://arxiv.org/abs/1412.4742}{{\tt
  arXiv:1412.4742 [hep-ex]}}.

\bibitem{Bernreuther:2015yna}
W.~Bernreuther, D.~Heisler, and Z.-G. Si, {\em {A set of top quark spin
  correlation and polarization observables for the LHC: Standard Model
  predictions and new physics contributions}}.
  \href{http://dx.doi.org/10.1007/JHEP12(2015)026}{JHEP {\bf 12} (2015)  026},
  \href{http://arxiv.org/abs/1508.05271}{{\tt arXiv:1508.05271 [hep-ph]}}.

\bibitem{CMS:2018jcg}
{\bf CMS} Collaboration, A.~M. Sirunyan {\em et al.}, {\em {Measurement of the
  top quark polarization and $\mathrm{t\bar{t}}$ spin correlations using
  dilepton final states in proton-proton collisions at $\sqrt{s} =$ 13 TeV}}.
  \href{http://dx.doi.org/10.1103/PhysRevD.100.072002}{Phys. Rev. D {\bf 100}
  (2019) no.~7, 072002}, \href{http://arxiv.org/abs/1907.03729}{{\tt
  arXiv:1907.03729 [hep-ex]}}.

\bibitem{Behring:2019iiv}
A.~Behring, M.~Czakon, A.~Mitov, A.~S. Papanastasiou, and R.~Poncelet, {\em
  {Higher order corrections to spin correlations in top quark pair production
  at the LHC}}. \href{http://dx.doi.org/10.1103/PhysRevLett.123.082001}{Phys.
  Rev. Lett. {\bf 123} (2019) no.~8, 082001},
  \href{http://arxiv.org/abs/1901.05407}{{\tt arXiv:1901.05407 [hep-ph]}}.

\bibitem{Afik:2020onf}
Y.~Afik and J.~R.~M. de~Nova, {\em {Quantum information and entanglement with
  top quarks at the LHC}}. \href{http://arxiv.org/abs/2003.02280}{{\tt
  arXiv:2003.02280 [quant-ph]}}.

\bibitem{Fabbrichesi:2021npl}
M.~Fabbrichesi, R.~Floreanini, and G.~Panizzo, {\em {Testing Bell inequalities
  at the LHC with top-quark pairs}}.
  \href{http://arxiv.org/abs/2102.11883}{{\tt arXiv:2102.11883 [hep-ph]}}.

\bibitem{Barate:1997ha}
{\bf ALEPH} Collaboration, R.~Barate {\em et al.}, {\em {A Measurement of the
  QCD color factors and a limit on the light gluino}}.
  \href{http://dx.doi.org/10.1007/s002880050522}{Z. Phys. C {\bf 76} (1997)
  1--14}.

\bibitem{Moretti:1998zc}
S.~Moretti and W.~J. Stirling, {\em {Spin correlations in e+ e-
  ---\ensuremath{>} four jets}}.
  \href{http://dx.doi.org/10.1007/s100529900016}{Eur. Phys. J. C {\bf 9} (1999)
   81--93}, \href{http://arxiv.org/abs/hep-ph/9808429}{{\tt
  arXiv:hep-ph/9808429}}.

\bibitem{Webber:1986mc}
B.~R. Webber, {\em {Monte Carlo Simulation of Hard Hadronic Processes}}.
\href{http://dx.doi.org/10.1146/annurev.ns.36.120186.001345}{Ann. Rev. Nucl.
  Part. Sci. {\bf 36} (1986)  253--286}.
%%CITATION = ARNUA,36,253;%%.

\bibitem{Collins:1987cp}
J.~C. Collins, {\em {Spin Correlations in Monte Carlo Event Generators}}.
\href{http://dx.doi.org/10.1016/0550-3213(88)90654-2}{Nucl. Phys. {\bf B304}
  (1988)  794--804}.
%%CITATION = NUPHA,B304,794;%%.

\bibitem{Knowles:1987cu}
I.~Knowles, {\em {Angular Correlations in QCD}}.
  \href{http://dx.doi.org/10.1016/0550-3213(88)90653-0}{Nucl. Phys. B {\bf 304}
  (1988)  767--793}.

\bibitem{Dasgupta:2018nvj}
M.~Dasgupta, F.~A. Dreyer, K.~Hamilton, P.~F. Monni, and G.~P. Salam, {\em
  {Logarithmic accuracy of parton showers: a fixed-order study}}.
  \href{http://dx.doi.org/10.1007/JHEP09(2018)033}{JHEP {\bf 09} (2018)  033},
\href{http://arxiv.org/abs/1805.09327}{{\tt arXiv:1805.09327 [hep-ph]}}.
%%CITATION = ARXIV:1805.09327;%%.

\bibitem{Dasgupta:2020fwr}
M.~Dasgupta, F.~A. Dreyer, K.~Hamilton, P.~F. Monni, G.~P. Salam, and G.~Soyez,
  {\em {Parton showers beyond leading logarithmic accuracy}}.
  \href{http://dx.doi.org/10.1103/PhysRevLett.125.052002}{Phys. Rev. Lett. {\bf
  125} (2020) no.~5, 052002}, \href{http://arxiv.org/abs/2002.11114}{{\tt
  arXiv:2002.11114 [hep-ph]}}.

\bibitem{Hamilton:2020rcu}
K.~Hamilton, R.~Medves, G.~P. Salam, L.~Scyboz, and G.~Soyez, {\em {Colour and
  logarithmic accuracy in final-state parton showers}}.
  \href{http://dx.doi.org/10.1007/JHEP03(2021)041}{JHEP {\bf 03} (2021)  041},
  \href{http://arxiv.org/abs/2011.10054}{{\tt arXiv:2011.10054 [hep-ph]}}.

\bibitem{Andersson:1988gp}
B.~Andersson, G.~Gustafson, L.~Lonnblad, and U.~Pettersson, {\em {Coherence
  Effects in Deep Inelastic Scattering}}.
\href{http://dx.doi.org/10.1007/BF01550942}{Z. Phys. {\bf C43} (1989)  625}.
%%CITATION = ZEPYA,C43,625;%%.

\bibitem{Catani:1992ua}
S.~Catani, L.~Trentadue, G.~Turnock, and B.~R. Webber, {\em {Resummation of
  large logarithms in e+ e- event shape distributions}}.
\href{http://dx.doi.org/10.1016/0550-3213(93)90271-P}{Nucl. Phys. {\bf B407}
  (1993)  3--42}.
%%CITATION = NUPHA,B407,3;%%.

\bibitem{Corcella:2000bw}
G.~Corcella, I.~G. Knowles, G.~Marchesini, S.~Moretti, K.~Odagiri,
  P.~Richardson, M.~H. Seymour, and B.~R. Webber, {\em {HERWIG 6: An Event
  generator for hadron emission reactions with interfering gluons (including
  supersymmetric processes)}}.
  \href{http://dx.doi.org/10.1088/1126-6708/2001/01/010}{JHEP {\bf 01} (2001)
  010},
\href{http://arxiv.org/abs/hep-ph/0011363}{{\tt arXiv:hep-ph/0011363
  [hep-ph]}}.
%%CITATION = HEP-PH/0011363;%%.

\bibitem{Bahr:2008pv}
M.~Bahr {\em et al.}, {\em {Herwig++ Physics and Manual}}.
  \href{http://dx.doi.org/10.1140/epjc/s10052-008-0798-9}{Eur. Phys. J. {\bf
  C58} (2008)  639--707},
\href{http://arxiv.org/abs/0803.0883}{{\tt arXiv:0803.0883 [hep-ph]}}.
%%CITATION = ARXIV:0803.0883;%%.

\bibitem{Bellm:2015jjp}
J.~Bellm {\em et al.}, {\em {Herwig 7.0/Herwig++ 3.0 release note}}.
  \href{http://dx.doi.org/10.1140/epjc/s10052-016-4018-8}{Eur. Phys. J. {\bf
  C76} (2016) no.~4, 196},
\href{http://arxiv.org/abs/1512.01178}{{\tt arXiv:1512.01178 [hep-ph]}}.
%%CITATION = ARXIV:1512.01178;%%.

\bibitem{Bellm:2019zci}
J.~Bellm {\em et al.}, {\em {Herwig 7.2 release note}}.
  \href{http://dx.doi.org/10.1140/epjc/s10052-020-8011-x}{Eur. Phys. J. C {\bf
  80} (2020) no.~5, 452}, \href{http://arxiv.org/abs/1912.06509}{{\tt
  arXiv:1912.06509 [hep-ph]}}.

\bibitem{Richardson:2018pvo}
P.~Richardson and S.~Webster, {\em {Spin Correlations in Parton Shower
  Simulations}}. \href{http://dx.doi.org/10.1140/epjc/s10052-019-7429-5}{Eur.
  Phys. J. {\bf C80} (2020) no.~2, 83},
\href{http://arxiv.org/abs/1807.01955}{{\tt arXiv:1807.01955 [hep-ph]}}.
%%CITATION = ARXIV:1807.01955;%%.

\bibitem{Webster:2019cwq}
S.~J. Webster, {\em {Improved Monte Carlo Simulations of Massive Quarks}}.
\newblock PhD thesis, Durham U., 2019.

\bibitem{Nagy:2007ty}
Z.~Nagy and D.~E. Soper, {\em {Parton showers with quantum interference}}.
  \href{http://dx.doi.org/10.1088/1126-6708/2007/09/114}{JHEP {\bf 09} (2007)
  114},
\href{http://arxiv.org/abs/0706.0017}{{\tt arXiv:0706.0017 [hep-ph]}}.
%%CITATION = ARXIV:0706.0017;%%.

\bibitem{Nagy:2008eq}
Z.~Nagy and D.~E. Soper, {\em {Parton showers with quantum interference:
  Leading color, with spin}}.
  \href{http://dx.doi.org/10.1088/1126-6708/2008/07/025}{JHEP {\bf 07} (2008)
  025},
\href{http://arxiv.org/abs/0805.0216}{{\tt arXiv:0805.0216 [hep-ph]}}.
%%CITATION = ARXIV:0805.0216;%%.

\bibitem{Forshaw:2019ver}
J.~R. Forshaw, J.~Holguin, and S.~Pl\"atzer, {\em {Parton branching at
  amplitude level}}. \href{http://dx.doi.org/10.1007/JHEP08(2019)145}{JHEP {\bf
  08} (2019)  145}, \href{http://arxiv.org/abs/1905.08686}{{\tt
  arXiv:1905.08686 [hep-ph]}}.

\bibitem{Forshaw:2020wrq}
J.~R. Forshaw, J.~Holguin, and S.~Pl\"atzer, {\em {Building a consistent parton
  shower}}. \href{http://dx.doi.org/10.1007/JHEP09(2020)014}{JHEP {\bf 09}
  (2020)  014}, \href{http://arxiv.org/abs/2003.06400}{{\tt arXiv:2003.06400
  [hep-ph]}}.

\bibitem{Chen:2020adz}
H.~Chen, I.~Moult, and H.~X. Zhu, {\em {Quantum Interference in Jet
  Substructure from Spinning Gluons}}.
  \href{http://dx.doi.org/10.1103/PhysRevLett.126.112003}{Phys. Rev. Lett. {\bf
  126} (2021) no.~11, 112003}, \href{http://arxiv.org/abs/2011.02492}{{\tt
  arXiv:2011.02492 [hep-ph]}}.

\bibitem{Dasgupta:2014yra}
M.~Dasgupta, F.~Dreyer, G.~P. Salam, and G.~Soyez, {\em {Small-radius jets to
  all orders in QCD}}. \href{http://dx.doi.org/10.1007/JHEP04(2015)039}{JHEP
  {\bf 04} (2015)  039}, \href{http://arxiv.org/abs/1411.5182}{{\tt
  arXiv:1411.5182 [hep-ph]}}.

\bibitem{Dasgupta:2016bnd}
M.~Dasgupta, F.~A. Dreyer, G.~P. Salam, and G.~Soyez, {\em {Inclusive jet
  spectrum for small-radius jets}}.
  \href{http://dx.doi.org/10.1007/JHEP06(2016)057}{JHEP {\bf 06} (2016)  057},
  \href{http://arxiv.org/abs/1602.01110}{{\tt arXiv:1602.01110 [hep-ph]}}.

\bibitem{Knowles:1988vs}
I.~Knowles, {\em {Spin Correlations in Parton - Parton Scattering}}.
  \href{http://dx.doi.org/10.1016/0550-3213(88)90092-2}{Nucl. Phys. B {\bf 310}
  (1988)  571--588}.

\bibitem{Knowles:1988hu}
I.~G. Knowles, {\em {A Linear Algorithm for Calculating Spin Correlations in
  Hadronic Collisions}}.
\href{http://dx.doi.org/10.1016/0010-4655(90)90063-7}{Comput. Phys. Commun.
  {\bf 58} (1990)  271--284}.
%%CITATION = CPHCB,58,271;%%.

\bibitem{Richardson:2001df}
P.~Richardson, {\em {Spin correlations in Monte Carlo simulations}}.
  \href{http://dx.doi.org/10.1088/1126-6708/2001/11/029}{JHEP {\bf 11} (2001)
  029}, \href{http://arxiv.org/abs/hep-ph/0110108}{{\tt arXiv:hep-ph/0110108}}.

\bibitem{Dreyer:2018nbf}
F.~A. Dreyer, G.~P. Salam, and G.~Soyez, {\em {The Lund Jet Plane}}.
  \href{http://dx.doi.org/10.1007/JHEP12(2018)064}{JHEP {\bf 12} (2018)  064},
\href{http://arxiv.org/abs/1807.04758}{{\tt arXiv:1807.04758 [hep-ph]}}.
%%CITATION = ARXIV:1807.04758;%%.

\bibitem{Dokshitzer:1997in}
Y.~L. Dokshitzer, G.~D. Leder, S.~Moretti, and B.~R. Webber, {\em {Better jet
  clustering algorithms}}.
  \href{http://dx.doi.org/10.1088/1126-6708/1997/08/001}{JHEP {\bf 08} (1997)
  001},
\href{http://arxiv.org/abs/hep-ph/9707323}{{\tt arXiv:hep-ph/9707323
  [hep-ph]}}.
%%CITATION = HEP-PH/9707323;%%.

\bibitem{Wobisch:1998wt}
M.~Wobisch and T.~Wengler, ``{Hadronization corrections to jet cross-sections
  in deep inelastic scattering},'' in {\em {Workshop on Monte Carlo Generators
  for HERA Physics (Plenary Starting Meeting)}}, pp.~270--279.
\newblock 1998.
\newblock \href{http://arxiv.org/abs/hep-ph/9907280}{{\tt
  arXiv:hep-ph/9907280}}.

\bibitem{Dasgupta:2013ihk}
M.~Dasgupta, A.~Fregoso, S.~Marzani, and G.~P. Salam, {\em {Towards an
  understanding of jet substructure}}.
  \href{http://dx.doi.org/10.1007/JHEP09(2013)029}{JHEP {\bf 09} (2013)  029},
  \href{http://arxiv.org/abs/1307.0007}{{\tt arXiv:1307.0007 [hep-ph]}}.

\bibitem{Larkoski:2014wba}
A.~J. Larkoski, S.~Marzani, G.~Soyez, and J.~Thaler, {\em {Soft Drop}}.
  \href{http://dx.doi.org/10.1007/JHEP05(2014)146}{JHEP {\bf 05} (2014)  146},
  \href{http://arxiv.org/abs/1402.2657}{{\tt arXiv:1402.2657 [hep-ph]}}.

\bibitem{Mehtar-Tani:2019rrk}
Y.~Mehtar-Tani, A.~Soto-Ontoso, and K.~Tywoniuk, {\em {Dynamical grooming of
  QCD jets}}. \href{http://dx.doi.org/10.1103/PhysRevD.101.034004}{Phys. Rev. D
  {\bf 101} (2020) no.~3, 034004}, \href{http://arxiv.org/abs/1911.00375}{{\tt
  arXiv:1911.00375 [hep-ph]}}.

\bibitem{Caucal:2021bae}
P.~Caucal, A.~Soto-Ontoso, and A.~Takacs, {\em {Dynamical Grooming meets LHC
  data}}. \href{http://dx.doi.org/10.1007/JHEP07(2021)020}{JHEP {\bf 07} (2021)
   020}, \href{http://arxiv.org/abs/2103.06566}{{\tt arXiv:2103.06566
  [hep-ph]}}.

\bibitem{Kaplan:2008ie}
D.~E. Kaplan, K.~Rehermann, M.~D. Schwartz, and B.~Tweedie, {\em {Top Tagging:
  A Method for Identifying Boosted Hadronically Decaying Top Quarks}}.
  \href{http://dx.doi.org/10.1103/PhysRevLett.101.142001}{Phys. Rev. Lett. {\bf
  101} (2008)  142001}, \href{http://arxiv.org/abs/0806.0848}{{\tt
  arXiv:0806.0848 [hep-ph]}}.

\bibitem{CMS:2009lxa}
{\bf CMS} Collaboration, {\em {A Cambridge-Aachen (C-A) based Jet Algorithm for
  boosted top-jet tagging}} Tech. Rep. CMS-PAS-JME-09-001, 2009.

\bibitem{CMS:2014fya}
{\bf CMS} Collaboration, {\em {Boosted Top Jet Tagging at CMS}} Tech. Rep.
  CMS-PAS-JME-13-007, 2014.

\bibitem{Plehn:2009rk}
T.~Plehn, G.~P. Salam, and M.~Spannowsky, {\em {Fat Jets for a Light Higgs}}.
  \href{http://dx.doi.org/10.1103/PhysRevLett.104.111801}{Phys. Rev. Lett. {\bf
  104} (2010)  111801}, \href{http://arxiv.org/abs/0910.5472}{{\tt
  arXiv:0910.5472 [hep-ph]}}.

\bibitem{Plehn:2010st}
T.~Plehn, M.~Spannowsky, M.~Takeuchi, and D.~Zerwas, {\em {Stop Reconstruction
  with Tagged Tops}}. \href{http://dx.doi.org/10.1007/JHEP10(2010)078}{JHEP
  {\bf 10} (2010)  078}, \href{http://arxiv.org/abs/1006.2833}{{\tt
  arXiv:1006.2833 [hep-ph]}}.

\bibitem{Dasgupta:2018emf}
M.~Dasgupta, M.~Guzzi, J.~Rawling, and G.~Soyez, {\em {Top tagging : an
  analytical perspective}}.
  \href{http://dx.doi.org/10.1007/JHEP09(2018)170}{JHEP {\bf 09} (2018)  170},
  \href{http://arxiv.org/abs/1807.04767}{{\tt arXiv:1807.04767 [hep-ph]}}.

\bibitem{Banfi:2006hf}
A.~Banfi, G.~P. Salam, and G.~Zanderighi, {\em {Infrared safe definition of jet
  flavor}}. \href{http://dx.doi.org/10.1140/epjc/s2006-02552-4}{Eur. Phys. J. C
  {\bf 47} (2006)  113--124}, \href{http://arxiv.org/abs/hep-ph/0601139}{{\tt
  arXiv:hep-ph/0601139}}.

\bibitem{Chen:2019bpb}
H.~Chen, M.-X. Luo, I.~Moult, T.-Z. Yang, X.~Zhang, and H.~X. Zhu, {\em {Three
  point energy correlators in the collinear limit: symmetries, dualities and
  analytic results}}. \href{http://dx.doi.org/10.1007/JHEP08(2020)028}{JHEP
  {\bf 08} (2020) no.~08, 028}, \href{http://arxiv.org/abs/1912.11050}{{\tt
  arXiv:1912.11050 [hep-ph]}}.

\bibitem{microjets}
M.~Dasgupta, F.~Dreyer, G.~P. Salam, and G.~Soyez,
  \url{https://microjets.hepforge.org/}.

\bibitem{Badger:2005jv}
S.~D. Badger, E.~W.~N. Glover, and V.~V. Khoze, {\em {Recursion relations for
  gauge theory amplitudes with massive vector bosons and fermions}}.
  \href{http://dx.doi.org/10.1088/1126-6708/2006/01/066}{JHEP {\bf 01} (2006)
  066}, \href{http://arxiv.org/abs/hep-th/0507161}{{\tt arXiv:hep-th/0507161}}.

\bibitem{Erdmann:2020ovh}
J.~Erdmann, O.~Nackenhorst, and S.~V. Zei\ss{}ner, {\em {Maximum performance of
  strange-jet tagging at hadron colliders}}.
  \href{http://arxiv.org/abs/2011.10736}{{\tt arXiv:2011.10736 [hep-ex]}}.

\bibitem{Kleiss:1985yh}
R.~Kleiss and W.~Stirling, {\em {Spinor Techniques for Calculating p anti-p
  ---\ensuremath{>} W+- / Z0 + Jets}}.
  \href{http://dx.doi.org/10.1016/0550-3213(85)90285-8}{Nucl. Phys. B {\bf 262}
  (1985)  235--262}.

\bibitem{Murayama:1992gi}
H.~Murayama, I.~Watanabe, and K.~Hagiwara, {\em {HELAS: HELicity amplitude
  subroutines for Feynman diagram evaluations}} Tech. Rep. KEK-91-11, 1992.
\newblock
  \url{https://cp3.irmp.ucl.ac.be/projects/madgraph/attachment/wiki/ManualAndHelp/HELAS_reference.pdf}.

\bibitem{Hoche:2015sya}
S.~Hoeche and S.~Prestel, {\em {The midpoint between dipole and parton
  showers}}. \href{http://dx.doi.org/10.1140/epjc/s10052-015-3684-2}{Eur. Phys.
  J. {\bf C75} (2015) no.~9, 461},
\href{http://arxiv.org/abs/1506.05057}{{\tt arXiv:1506.05057 [hep-ph]}}.
%%CITATION = ARXIV:1506.05057;%%.

\bibitem{Bertolini:2013iqa}
D.~Bertolini, T.~Chan, and J.~Thaler, {\em {Jet Observables Without Jet
  Algorithms}}. \href{http://dx.doi.org/10.1007/JHEP04(2014)013}{JHEP {\bf 04}
  (2014)  013}, \href{http://arxiv.org/abs/1310.7584}{{\tt arXiv:1310.7584
  [hep-ph]}}.

\bibitem{Larkoski:2014uqa}
A.~J. Larkoski, D.~Neill, and J.~Thaler, {\em {Jet Shapes with the Broadening
  Axis}}. \href{http://dx.doi.org/10.1007/JHEP04(2014)017}{JHEP {\bf 04} (2014)
   017}, \href{http://arxiv.org/abs/1401.2158}{{\tt arXiv:1401.2158 [hep-ph]}}.

\bibitem{salam:WTA}
G.~P. Salam , unpublished.

\bibitem{Cacciari:2011ma}
M.~Cacciari, G.~P. Salam, and G.~Soyez, {\em {FastJet User Manual}}.
  \href{http://dx.doi.org/10.1140/epjc/s10052-012-1896-2}{Eur. Phys. J. {\bf
  C72} (2012)  1896},
\href{http://arxiv.org/abs/1111.6097}{{\tt arXiv:1111.6097 [hep-ph]}}.
%%CITATION = ARXIV:1111.6097;%%.

\end{thebibliography}\endgroup

\end{document}